\documentstyle[bbm,amsfonts,epsfig,12pt,here]{article}

\newcommand {\eqref} [1] {(\ref {#1})}
\newcommand {\slsh} [1] {\not{\hbox{\kern-2pt${#1}$}}}
\newcommand {\state} [1] {\mid \! \! {#1} \rangle}

\newcommand{\drawsquare}[2]{\hbox{%
\rule{#2pt}{#1pt}\hskip-#2pt
\rule{#1pt}{#2pt}\hskip-#1pt
\rule[#1pt]{#1pt}{#2pt}}\rule[#1pt]{#2pt}{#2pt}\hskip-#2pt
\rule{#2pt}{#1pt}}
\newcommand{\Yfund}{\raisebox{-.5pt}{\drawsquare{6.5}{0.4}}}
\newcommand{\Ysymm}{\Yfund\hskip-0.4pt%
                     \Yfund}
\def\symm{\Ysymm}
\def\bsymm{\overline{\Ysymm}}
\def\drawbox#1#2{\hrule height#2pt
         \hbox{\vrule width#2pt height#1pt \kern#1pt
               \vrule width#2pt}
               \hrule height#2pt}

\def\Asym#1#2{\vcenter{\vbox{\drawbox{#1}{#2}
               \kern-#2pt       
               \drawbox{#1}{#2}}}}

\def\asymm{\Asym{6.4}{0.3}}

\def\basymm{\overline{\asymm}}

\newcommand{\gsim}{\lower.7ex\hbox{$
\;\stackrel{\textstyle>}{\sim}\;$}}
\newcommand{\lsim}{\lower.7ex\hbox{$
\;\stackrel{\textstyle<}{\sim}\;$}}
\newcommand {\beq} {\begin{equation}}
\newcommand {\eeq} {\end{equation}}
\newcommand {\ber}{\begin{eqnarray*}}
\newcommand {\eer} {\end{eqnarray*}}
\newcommand {\bea}{\begin{eqnarray}}
\newcommand {\eea} {\end{eqnarray}}
\newcommand{\Nfour} {${\cal N}=4\ $}
\newcommand{\Ntwo}{${\cal N}=2\ $}
\newcommand{\None}{${\cal N}=1\ $}

\begin{document}
\begin{titlepage}

\begin{flushright}{CERN-PH-TH/2004-022\\
FTPI-MINN-04/01, UMN-TH-2228/04}
\end{flushright}
\vskip 0.2cm

\centerline{{\Large \bf  From Super-Yang--Mills Theory to QCD:}}
\vskip 0.1cm
\centerline{{\Large \bf  Planar Equivalence and its Implications }}
\vskip 0.7cm
\centerline{\large A. Armoni ${}^a$, M. Shifman ${}^{b}$, G. 
Veneziano ${}^{a}$}

\vskip 0.3cm

\centerline{${}^a$ Department of Physics, Theory Division,}
\centerline{CERN, CH-1211 Geneva 23, Switzerland}
\vskip 0.2cm
\centerline{${}^b$ William I. Fine Theoretical Physics Institute, 
University
of Minnesota,}
\centerline{Minneapolis, MN 55455, USA}

\vskip 0.2cm
\begin{center}
\noindent
{\small\em To be published in the Ian Kogan Memorial Collection ``From Fields to Strings: Circumnavigating Theoretical Physics," World Scientific, 2004}
\end{center}

\vskip 0.5cm

\begin{abstract}

We review and extend our recent work on the planar (large-$N$) equivalence
between gauge theories with  varying degree of supersymmetry. The main emphasis
is put on the  planar  equivalence between \mbox{${\cal N}=1$}
gluodynamics (super-Yang--Mills theory) and a 
non-supersym\-metric ``orientifold field theory.'' 
We outline an ``orientifold" large-$N$ expansion,
 analyze its possible phenomenological consequences
in one-flavor massless QCD, and make  a first attempt 
at extending the correspondence to three massless flavors.
An analytic calculation of the quark condensate in  one-flavor QCD 
starting from the gluino condensate in \None 
gluodynamics is thoroughly discussed.  We also comment
on  a planar equivalence involving  ${\cal N} =2$ supersymmetry,  on
 ``chiral rings" in non-supersymmetric theories, and
on the origin of  planar equivalence
 from an underlying, non-tachyonic type-0 string theory.
Finally,  possible other directions of investigation,
 such as the gauge/gravity correspondence 
in  large-$N$ orientifold field theory, are   briefly discussed.
\end{abstract}

\end{titlepage}

\newpage

\tableofcontents

\newpage
 
\section*{Dictionary and some notation}
\addcontentsline{toc}{section}{Dictionary and some notations}
 
In the bulk of this review the gauge group is assumed to be SU($N$),
with the exception of Sects.~\ref{orbifoldization} 
and \ref{npo}.  Different theories that we consider include:
 
\vspace{1.5mm}
 
{\bf Orientifold theory A} (one two-index Dirac fermion in the 
antisymmetric representation);

\vspace{1.5mm}
 
{\bf Orientifold theory S} (one two-index Dirac fermion in the 
symmetric representation);

\vspace{1.5mm}
 
General version:  {\bf Orientifold theory} (one of the two above);

\vspace{1.5mm}
 
{\bf Orienti/A}{\boldmath{$_n$}} {\bf theory} ($n$ antisymmetric Dirac fermions);

\vspace{1.5mm}
 
{\bf Orienti/S}{\boldmath{$_n$}} ($n$ symmetric Dirac fermions);

\vspace{1.5mm}
 
{\bf Adjoint}{\boldmath{$_n$}} {\bf theory} ($n$ adjoint Majorana fermions);

\vspace{1.5mm}

{\boldmath{$N^2\,$}}-{\bf QCD} ($N_f=N_c=N$ Dirac flavors);

\vspace{1.5mm}
 
{\bf Orienti/nf theory} (one two-index Dirac fermion in the 
antisymmetric representation plus $n$ fundamental Dirac quarks);

Version: {\bf Orienti/2f theory};

\vspace{1.5mm}
 
{\bf kf-QCD} (QCD with $k$ Dirac fundamental flavors);

Version: {\bf 1f-QCD} (QCD with one Dirac fundamental flavor);

\vspace{1.5mm}
 
{\bf Common sector}  (for a definition see the discussion
preceding Sect.~\ref{pert-theory}).

\begin{center}
$\star\star\star$
\end{center}
\vspace{1.5mm}
 
Note that, from the string theory perspective, only
 some of the theories listed above deserve the name ``orientifold theories"
--- not all appear as a result of {\em bona fide}  orientifoldization.
Keeping in mind, however, that our prime emphasis is
on  field theory, we will take the liberty of naming them
in a way that is convenient
for our purposes, albeit not quite justified from the string theory standpoint.
\begin{center}
$\star\star\star$
\end{center}
\vspace{1.5mm}
 
We normalize the Grassmann integration as follows:
\beq
\int \theta^2\, d^2\theta =2\,.
\eeq
The rest of the notation for superfield/superspace
variables is essentially the same as in the classical textbook
of Bagger and Wess \cite{BagWes}, except that
our metric is $g_{\mu\nu} ={\rm diag} (1,-1,-1,-1)$
while in Bagger and Wess's book  it is 
$g_{\mu\nu}^{\rm BW} ={\rm diag} (-1,1,1,1)$.

If we have two Weyl (right-handed) spinors
$\xi^\alpha$ and $\eta_\beta$, transforming in the
representations $R$ and $\bar R$ of the gauge group, respectively,
then the Dirac spinor $\Psi$ can be formed as
\beq
\Psi = \left(\begin{array}{l}
\xi \\
\bar\eta\end{array}\right)\,.
\eeq
The Dirac spinor $\Psi$ has four components, while
$\xi^\alpha$ and $\eta_\beta$ have two components each.

\begin{center}
$\star\star\star$
\end{center}
\vspace{2.5mm}

The gluon field-strength tensor is denoted by
$G_{\mu\nu}^a$. We use the abbreviation 
$G^2$  for
\beq
G^2\equiv G_{\mu\nu}^a\, G^{\mu\nu\,\, a}\,,
\eeq
while
\beq
G\tilde G \equiv G_{\mu\nu}^a\, \tilde{G}^{\mu\nu\,\, a} 
\equiv \frac{1}{2} \epsilon^{\mu\nu\rho\sigma} G_{\mu\nu}^a G_{\rho\sigma}^a \,.
\eeq

\begin{center}
$\star\star\star$
\end{center}
\vspace{2.5mm}

\noindent 
\underline{\bf Group-theory coefficients}
\vspace{2mm}
  
As was mentioned, the gauge group is assumed to be SU($N$). For a given
representation $R$ of SU($N$), the definitions of the
Casimir operators to be used below are
\beq
{\rm Tr} (T^a T^b )_R= T(R) \delta^{ab}\,,\qquad (T^aT^a)_R = C(R) \,I\,, 
\label{casop}    
\eeq
where  $I$ is the unit
matrix in this representation. It is quite obvious that
\beq
C(R) = T(R)\,\,\frac{{\rm dim} (G)}{{\rm dim} (R)}\,,
\label{gripp}
\eeq
where ${\rm dim} (G)$ is the dimension of the group
(= the  dimension of the adjoint representation). 
Note that $T(R)$ is also known as (one half of) the Dynkin index,
or the dual Coxeter number.
For the adjoint representation, $T(R)$ is denoted by $T(G)$. Moreover,
 $T({\rm SU}(N)) = N$.
 
 \vspace{2.5mm}

\noindent 
\underline{\bf Renormalization-group conventions}
\vspace{2mm} 
 
We use the following definition of the $\beta$ function and anomalous dimensions:
\beq
\mu\,\frac{\partial\alpha}{\partial \mu}\equiv \beta (\alpha)
=-\frac{\beta_0}{2\pi } \alpha^2 -\frac{\beta_1}{4\pi^2 } \alpha^3 +...
\eeq
while 
\beq
\gamma  = -d \ln Z (\mu ) /d\ln\mu \,.
\label{defgin}
\eeq
In supersymmetric theories
\beq
\beta (\alpha) = -\frac{\alpha^2}{2\pi}\left[3\,T(G) -\sum_i T(R_i)(1-\gamma_i )
\right]\left(1-\frac{T(G)\,\alpha}{2\pi} \right)^{-1}
\, ,
\eeq
where the sum runs over all matter supermultiplets.
The anomalous dimension of the $i$-th  matter superfield is
\beq
\gamma_i =  -2 C(R_i)\,\frac{\alpha}{2\pi } + ...
\eeq
Sometimes, when one-loop anomalous dimensions
are discussed,  the coefficient in front of
$-\alpha/(2\pi )$ in (\ref{defgin})  is also
referred to as  an ``anomalous dimension."

\newpage
 
\section{Introduction}
\label{introduction}
\renewcommand{\theequation}{\thesection.\arabic{equation}}
\setcounter{equation}{0}

Gauge field theories at strong coupling are obviously of great 
importance in particle physics. 
Needless to say, exact results in such theories have a special weight.
Supersymmetry proved to be an extremely powerful tool, allowing one
to solve some otherwise intractable  problems 
(for a review see e.g. \cite{srev}).
During the last twenty years or so, a significant number of results were obtained in this direction,  starting from the gluino
condensate and exact $\beta$ functions, to, perhaps, the most famous
example --- the Seiberg--Witten solution \cite{SW1} of \Ntwo supersymmetric theory,
which (upon a small perturbation that breaks \Ntwo to \None)
explicitly exhibits the dual Mei{\ss}ner mechanism of color confinement
conjectured by Mandelstam \cite{mandelstam} and 't~Hooft \cite{thooft}  
in the 1970's and 80's. This review is devoted to a recent addition to 
the supersymmetry-based kit,
which goes under the name of {\em planar equivalence}. Some 
{\em non}-supersymmetric gauge theories --- quite close relatives of QCD ---
were shown \cite{adisvone,adisvtwo,adisvtwop} to be equivalent to supersymmetric 
gluodynamics in a bosonic sub-sector,
provided the number of colors $N\to\infty$. Needless to say that, if true, this statement
has far-reaching consequences. A large number of supersymmetry-based predictions, 
such as spectral degeneracies, low-energy theorems, and so on, hold 
(to leading order in $1/N$) in strongly coupled non-supersymmetric gauge theories!

We will start this review by summarizing the basic features of the
``main" parent theory --- supersymmetric
gluodynamics, also known as supersymmetric Yang--Mills (SYM) theory.
We review the genesis of the idea of planar equivalence
and its physical foundation, both at the perturbative and non-perturbative
levels. After a brief introduction on perturbative planar equivalence,
we proceed to the {\em orientifold field theory}, whose bosonic
sector was argued \cite{adisvone} to be  fully 
(i.e. perturbatively {\em and} non-perturbatively) equivalent to SYM theory
at $N\to\infty$. At $N=3$ the daughter theory is also equivalent
to {\em one-flavor} QCD. We explore the detailed
consequences of these equivalences.

Among other aspects, our review will demonstrate cross-fertilization
between field theory and string theory. The idea of planar equivalence,
born in the depths of string theory, was reformulated and adapted to
field theory, where it found a life of its own and was developed
to a point where practical applications are looming. 

Instead of summarizing here the main results and topics to be covered ---
a usual practice in an introductory section --- we refer the reader to the
Table of Contents and to Sect.~\ref{conclu}.

\section{Genesis of the idea}
\label{genesis}
\renewcommand{\theequation}{\thesection.\arabic{equation}}
\setcounter{equation}{0}

{\it This section can be omitted in a first reading. Subsequent sections 
explain much of its material.}
\\ \\
Kachru and Silverstein studied \cite{kachru}
various orbifolds of $R^6$ within the framework of the 
AdS/CFT correspondence. Starting from ${\cal N}=4$, they obtained distinct
four-dimensional (daughter) gauge field theories with matter, with varying degree of supersymmetry, ${\cal N}=2,1,0$, all with vanishing $\beta$ functions. In the latter case, ${\cal N}=0$, Kachru and Silverstein predicted 
the first coefficient of the Gell-Mann--Low function, $\beta_0$,
to vanish  in the large-$N$ (planar) limit in the daughter theories, while for ${\cal N}=2,1$  they predicted $\beta_0 =0$ even for finite $N$.
Shortly after their study, this analysis was expanded in \cite{Lawrence:1998ja},
where generic orbifold projections of ${\cal N}=4$ were considered,
and those preserving  conformal invariance 
of the resulting --- less supersymmetric --- gauge theories (at two loops) were identified.

A decisive step was made \cite{K} by Bershadsky {\em et al.} These authors
abandoned the AdS/CFT limit of large 't Hooft coupling.
They considered the string perturbative expansion in the
presence of $D$ branes embedded in orbifolded space-time in the limit
$\alpha '\to 0$. In this limit the string perturbative expansion coincides 
with the 't Hooft $1/N$ expansion \cite{thooft1n}.
The genus-zero string graphs can be identified with the leading (planar)
term in the 't Hooft expansion, the genus-one is the next-to-leading correction, and so on. In \cite{K} one can find a proof that in the large-$N$
limit the $\beta$ functions of all orbifold
theories considered previously \cite{kachru, Lawrence:1998ja}
vanish to any finite order in the gauge coupling constant.
Moreover, a remarkable theorem was derived in \cite{K}.
The authors showed that, not only the
$\beta$ functions, but a variety of amplitudes
which can be considered in the orbifold theories
coincide in the large-$N$ limit (upon an appropriate rescaling of the
gauge coupling)
with the corresponding amplitudes of the parent ${\cal N}=4$ theory, 
order by order in the gauge coupling.

In a week or so, Bershadsky and Johansen abandoned
the string theory set-up altogether. They proved \cite{KK}
the above theorem in the framework of field theory {\em per se}.

In a parallel development, Kakushadze suggested \cite{Kakushadze:1998tr}
considering orientifold rather than orbifold daughters of ${\cal N}=4$ theory.
His construction was close to that of Ref.~\cite{K}, except that
instead of the string expansion in the presence of $D$ branes he added
orientifold planes.
The main emphasis was on the search of ${\cal N}=2,1$ daughters with  
gauge groups other than SU$(N)$ and matter
fields other than bifundamentals.

The first attempt to apply the idea of orbifoldization to non-conformal
field theories was carried out by Schmaltz who suggested \cite{schmaltz}
a version of Seiberg duality between a pair of non-supersymmetric large-$N$
orbifold field theories. Later on, the $Z_2$ version of Schmaltz'
suggestion was realized in a brane configuration of type 0 string
theory \cite{Armoni:1999gc}.

After a few years of a relative oblivion, the interest in the
issue of planar equivalence was revived by Strassler \cite{strassler}.
In the inspiring paper entitled
``On methods for extracting exact non-perturbative results in 
non-supersymmetric gauge theories," he shifted the emphasis away from the search of the conformal daughters, towards engineering QCD-like daughters.
Similarly to Schmaltz, Strassler noted that one does not need to limit oneself to an ${\cal N}=4$ parent; \None gluodynamics, which is much closer
to actual QCD, gives rise to orbifold planar-equivalent daughters too.
Although this equivalence was proved only at the perturbative level,
Strassler formulated a ``non-perturbative orbifold conjecture" (NPO).
According to this conjecture,
for ``those vacua which appear in both theories (i.e. parent and orbifold daughter) the Green's functions which exist in both theories
are identical in the limit $N\to\infty$ after an appropriate rescaling of the
coupling constant."

Unfortunately, it turned out \cite{gorshi,dt} that the NPO
conjecture could not be  valid. The orbifold daughter theories ``remember" that they
have fewer vacua than the parent one, which results in a mismatch
\cite{gorshi} in low-energy theorems. In  string theory language, 
the killing factor is the presence of tachyons in
the twisted sector. This is clearly seen in the light of the calculation
presented in Ref.~\cite{dt}.

The appeal of the idea was so strong, however, that
the searches continued. They culminated in the discovery of the
orientifold field theory \cite{adisvone} whose planar equivalence to 
SYM theory, both perturbative and non-perturbative,
rests on a solid footing. The orientifold daughter has no twisted sector.
The SU$(N)$ orientifold daughter and its parent,  SU$(N)$ \None 
gluodynamics, have  the same numbers of vacua.

\section{The basic parent theory}
\label{parent}
\renewcommand{\theequation}{\thesection.\arabic{equation}}
\setcounter{equation}{0}

Let us describe a prototypical parent theory of this review. It is a
generic  supersymmetric gauge theory of the following form:
\beq
{\cal L} = \left\{ \frac{1}{4g^2}\, 
\int \! {\rm d}^2 \theta\, \mbox{Tr}\,W^2 +{\rm h.c.}\right\} 
+{\cal L}_{\rm matter}\,,
\label{genlagr}
\eeq
where the spinorial superfield $W_\alpha$ in the Wess-Zumino gauge is
\begin{equation}
{W}_{\alpha} = \frac{1}{8}\;\bar{D}^2\, \left( {e}^{-V}\! D_{\alpha } 
{e}^V\right) =
i\left( \lambda_{\alpha} + i\theta_{\alpha}D - \theta^{\beta}\, 
G_{\alpha\beta} - 
i\theta^2{\cal D}_{\alpha\dot\alpha}\bar{\lambda}^{\dot\alpha} 
\right)\, , 
\label{sgfst}
\end{equation}
and ${\cal L}_{\rm matter}$ is the matter part of the Lagrangian,
\beq
{\cal L}_{\rm matter} =\sum_{\rm matter}
\left\{ \frac{1}{4} \int \! {\rm d}^2\theta {\rm d}^2\bar\theta\,
\bar Q e^V Q +
\left( \frac{1}{2}\int\! {\rm d}^2\theta\, {\cal W}(Q)
+{\rm h. c.}\right)\right\}\,,
\eeq
where $Q$ is a generic notation for the matter chiral superfields and
${\cal W}(Q)$ is their superpotential. As is seen from Eq.~(\ref{sgfst}),
${W}_{\alpha}$ is nothing but the superspace field strength.

The gauge group is assumed to be SU($N$). The matter fields $Q$ can belong to 
various representations $R$ of the gauge group.  The definition of  
various group-theoretical quantities is given at the
very  beginning of the review.
For the representations used in  this paper the relevant Casimir operators  are
collected in Table \ref{table1}.

\vspace{3mm}

\begin{table}
\begin{center}
\begin{tabular}{|c|c|c|c|c|}
\hline
$\frac{\mbox{ Representations} \to  }{\mbox{  Casimirs} \downarrow} $ \rule{0cm}{0.7cm} &{\rm Fundamental}&{\rm Adjoint} & {\rm 2-index A} &{\rm 2-index S }\\[3mm]
\hline\hline
\rule{0cm}{0.7cm}$T(R)$  & $\frac{1}{2}$ &$N$ & $\frac{N-2}{2}$
 & $\frac{N+2}{2}$\\[2mm]
\hline
\rule{0cm}{0.7cm}$C(R)$ &$\frac{N^2-1}{2N}$ &$N$& $\frac{(N-2)(N+1)}{N}$ & $\frac{(N+2)(N-1)}{N}$ \\[2mm]
\hline
\end{tabular}
\end{center}
\caption{The Casimir coefficients for relevant
representations of SU($N$).}
\label{table1}
\end{table}

Most of this review will deal with \None SYM theory, 
also known as supersymmetric gluodynamics --- the theory of gluons and
gluinos. The Lagrangian of the theory is obtained
from Eq.~(\ref{genlagr}) by omitting the matter part.
In  component form
\beq
{\cal L} = -\frac{1}{4g^2}G_{\mu\nu}^aG_{\mu\nu}^a
+\frac{i}{g^2}\lambda^{a\alpha}{\cal D}_{\alpha\dot\beta}
\bar \lambda^{\alpha\dot\beta}\,,
\label{one}
\eeq
where 
$\lambda$ is the Weyl (Majorana) fermion in the adjoint representation,
and  spinorial notation is used in the fermion sector. 
One can add, if one wishes, a $\theta$ term,
\beq
{\cal L}_\theta = \frac{\theta}{32\pi^2}\, G_{\mu\nu}^a\widetilde G_{\mu\nu}^a\,.
\label{thetaterm}
\eeq
For the time being we will  set $\theta =0$. 

\vspace{1mm}

Later on we will let $N\to\infty$, following  't Hooft, i.e. keeping\,\footnote{Here 
and below we follow the standard 
convention $g^2 =4\pi\alpha$.}
\beq
\lambda \equiv \frac{g^2 \,  N}{8\pi^2 }
\equiv \frac{N\alpha}{2\pi} ={\rm const}.
\label{thocoup}
\eeq
The distinction between QCD and  SUSY gluodynamics lies 
in the fermion sector ---  the QCD quarks
are the Dirac fermions in the fundamental 
representation of the gauge group.

The theory (\ref{one}) is supersymmetric.
The conserved spin-3/2 current $J^\mu_\beta$ has the form
(in  spinorial notation)
\beq
J_{\beta\alpha\dot\alpha} 
\equiv (\sigma_\mu)_{\alpha\dot\alpha}J^\mu_\beta 
=\frac{2i}{g^2}\, G^a_{\alpha\beta}\bar\lambda^a_{\dot\alpha}\,.
\label{two}
\eeq
Moreover, the theory  is believed to be confining, with a mass gap.  
Some new semi-theoretical--semi-empirical 
arguments substantiating this point will be provided later, 
see Sect.~\ref{sc1fqcd}.

\begin{figure}
\epsfxsize=7cm
\centerline{\epsfbox{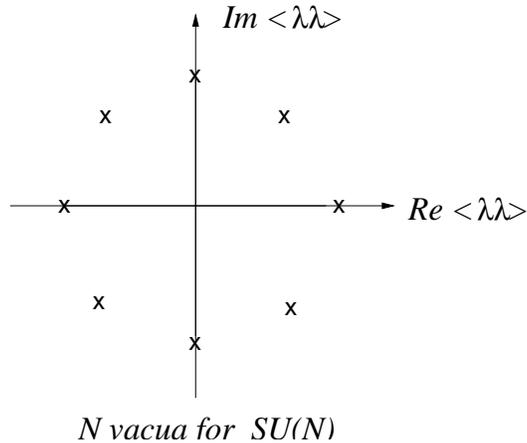}}
\caption{
The gluino condensate $\langle \lambda\lambda\rangle$
is the  order parameter labeling distinct vacua in supersymmetric gluodynamics.
For the SU$(N)$ gauge group there are $N$ discrete degenerate vacua.}
\label{gc}
\end{figure}

The spectrum of supersymmetric gluodynamics
consists of composite (color-singlet) hadrons which enter 
in degenerate supermultiplets.  
Note that there is no conserved fermion-number current
in the theory at hand. The only U(1) symmetry of the classical Lagrangian
(\ref{one}),
$$ 
\lambda \to e^{i\alpha}\lambda\,,\quad \bar\lambda \to
e^{-i\alpha}\bar\lambda\,,
$$
is broken at the quantum level by the chiral
anomaly. A discrete $Z_{2N}$ subgroup,
$ \lambda \to e^{\pi ik / N}\lambda$,
is non-anomalous. It is known to be dynamically broken down to $Z_2$.
The order parameter, the gluino condensate\,\footnote{The gluino 
condensate in supersymmetric gluodynamics was first conjectured, 
on the basis of the value of his index, by E.~Witten \cite{wittenindex}.
It was confirmed in an effective Lagrangian approach by 
G.~Veneziano and S.~Yankielowicz \cite{vy}, and exactly
calculated (by using holomorphy and analytic continuations
in mass parameters \cite{holm}) by  M.~A.~Shifman and A.~I.~Vainshtein \cite{SV}.
The exact value of the coefficient $6N$ in Eq.~(\ref{gluco})
can be extracted from several sources.
All numerical factors are carefully collected for SU(2) in the review paper
\cite{Shifman:1999mv}. A weak-coupling calculation for SU($N$)
with arbitrary $N$ was carried out in \cite{Davies:1999uw}.
Note, however, that an unconventional definition 
of the scale parameter $\Lambda$ is used in Ref.~\cite{Davies:1999uw}. 
One can pass to
the conventional definition of $\Lambda$ either by normalizing the result
to the SU(2) case \cite{Shifman:1999mv} or by analyzing the context 
of Ref.~\cite{Davies:1999uw}. Both methods give one and the same result.}
 $\langle\lambda\lambda\rangle$, can take $N$ values,
\beq
\langle
\lambda^{a}_{\alpha}\lambda^{a\,,\alpha}
\rangle = -6 N\Lambda^3 \exp \left({\frac{2\pi i k}{N}}\right)\,,
\,\,\, k = 0,1,..., N-1\,, 
\label{gluco}
\eeq
labeling the $N$ distinct vacua of the theory (\ref{one}), see Fig. \ref{gc}. 
Here $\langle ...\rangle$ means averaging over the given vacuum state,
and $\Lambda$ is a dynamical scale, defined in the standard manner
(i.e. in accordance with Ref.~\cite{Hinchliffe}) in terms of the ultraviolet
parameters,
\beq
\Lambda^3 = \frac{2}{3}\, M_{\rm UV}^3\,  \,\,
\frac{8\pi^2}{Ng_0^2}\, \exp\left(-\frac{8\pi^2}{Ng_0^2}\right)
= \frac{2}{3}\, M_{\rm UV}^3\,\frac{1}{\lambda_0}\,\exp\left(-\frac{1}{\lambda_0}\right) \,,
\label{dynscgl}
\eeq
where $M_{\rm UV}$ is the ultraviolet (UV) regulator mass, while $g_0^2$
and $\lambda_0$ are the bare coupling constants. 

Note that since $\Lambda $ is expressible in terms of the
't Hooft coupling, it is explicitly $N$-independent.
Equation (\ref{dynscgl}) is exact \cite{NSVZ} in supersymmetric 
gluodynamics. If $\theta\neq 0$, the exponent in Eq.~(\ref{gluco}) is
replaced by
$$
\exp \left({\frac{2\pi i k}{N}}+\frac{i\,\theta}{N}\right)\,.
$$

All hadronic states are arranged in supermultiplets. The
simplest  is the so-called chiral supermultiplet, which includes
two (massive) spin-zero mesons (with opposite parities), and a Majorana
fermion, with the Majorana mass (alternatively,
one can treat it as a Weyl fermion). The interpolating operators
producing  the corresponding hadrons from the vacuum are $G^2$,
$G\tilde G$ and $G\lambda$.
The vector supermultiplet consists of a spin-1 massive vector particle, a $0^+$
scalar and a Dirac fermion. All particles from one supermultiplet 
have degenerate masses.
Two-point functions are degenerate too (modulo obvious kinematical
spin  factors). For instance,
\begin{eqnarray}
\langle G^2(x)\,,\, G^2(0)\rangle = \langle G\tilde G(x)\,,\, G\tilde
G(0)\rangle =
\langle G\lambda (x)\,,\, G\lambda(0)\rangle\,.
\end{eqnarray}

Unlike what happens in conventional QCD, both the meson and
fermion masses in SUSY gluodynamics are expected to scale
as $N^0$.

\section{Orbifoldization and perturbative \\
equivalence}
\label{orbifoldization}
\renewcommand{\theequation}{\thesection.\arabic{equation}}
\setcounter{equation}{0}

The technique of orbifoldization was developed in the context of string/brane
theory, and it will be discussed in the string part  of this review 
(Sect.~\ref{ovoavftsts}). Starting from supersymmetric gluodynamics
with the gauge group SU($kN$), where $k$ is an integer, one can perform a $Z_k$ orbifoldization.
  
\begin{figure}
\epsfxsize=7cm
\centerline{\epsfbox{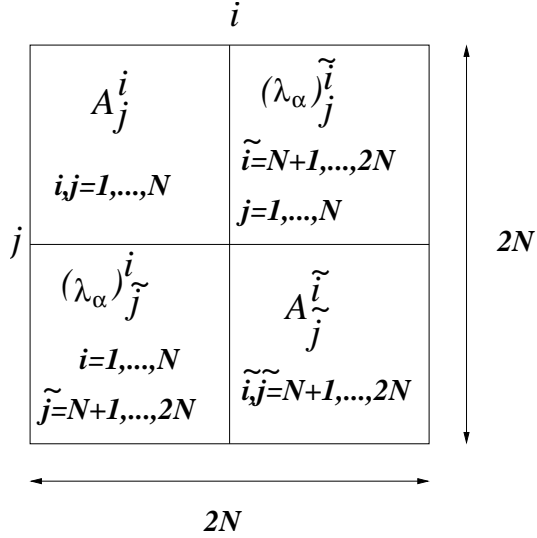}}
\caption{
Color decomposition of fields  in the $Z_2$-orbifold daughter.}
\label{z2orbi}
\end{figure}

The only thing we need to know at the moment, is that
orbifolding the parent theory (\ref{one})
one obtains a self-consistent daughter field theory by judiciously discarding  a
number of fields in the Lagrangian (\ref{one}). This is illustrated in
Fig. \ref{z2orbi} corresponding to $k=2$. The square presents the color contents 
of the fields --- gluon and gluino ---
of the daughter theory. The gauge fields, which are {\em retained} in the
daughter theory, belong to two blocks on the main diagonal.
Thus, the daughter theory has the gauge symmetry SU$(N)\times$SU$(N)$.
(Needless to say, there is no distinction at large $N$ between SU$(N)$ and U$(N)$.)
This specific theory has a realization in type-0 string theory
\cite{Armoni:1999gc}. To emphasize the fact that
the gauge group is a direct product it is convenient
to  use ``tilded" indices for one SU$(N)$, e.g. 
$(A_\mu)^{\tilde i}_{\tilde j}$,  and ``untilded" for another, e.g. $(G_{\mu\nu})^i_j
$, where $i,j, = 1,2,..., N$ while $\tilde i ,\tilde j = N+1,..., 2N$.

\vspace{2mm}

The fermion fields to be retained in the daughter
theory belong to two off-diagonal blocks
in Fig. \ref{z2orbi}. Thus, each fermion carries
one tilded and one untilded color index. In other words, they
are fundamentals with respect to the first SU($N$) and antifundamentals
with respect to the second SU($N$). Such fields are called {\em bifundamentals}.
Thus, in the daughter theory we deal with two Weyl bifundamentals,
$(\lambda _\alpha)^i_{\tilde j}$ (we will call it $\chi$)
and $(\lambda_\alpha)^{\tilde i}_j$ (we will call it $\eta$).
It is quite clear that they can be combined to form one 
four-component (Dirac) bifundamental field:
$$\eta ,\,\, \bar\chi \to \Psi^{\tilde i}_j \, .$$
The Lagrangian of the daughter theory is
\begin{eqnarray}
{\cal L} &=& -\frac{1}{4g^2} G^{i}_{j}  G^{j}_{i}  
-\frac{1}{4g^2} {G}^{\,\tilde  i}_{\tilde  j} \,{G}^{\,\tilde  j}_{\tilde i}
+\frac{\theta}{32\pi^2}\, \left( G^{i}_{j} \widetilde G^{j}_{i}  
+{G}^{\,\tilde  i}_{\tilde  j} \widetilde {G}^{\,\tilde  j}_{\tilde i}
\right)
\nonumber\\[4mm]
&+&\overline{\Psi^{\tilde  i}_{ j}}\, \, i\not\!\!D\, \, \Psi^{\tilde i}_{j}\,,
\label{three}
\end{eqnarray}
where
\beq
D_\mu = \partial_\mu \, - i\, A_\mu^a T^a \, - i\, {\tilde  A }_\mu^{\, a} \tilde 
T^a\,,
\eeq
and  the $\theta$ term is included. Note that the perturbative planar 
equivalence requires the gauge couplings of both SU($N$)'s
to be {\em the same}. Below they will be denoted by $g_D^2$. We will
then assume that the vacuum angles are the same too, as is indicated in
Eq.~(\ref{three}).

The perturbative planar equivalence that takes place
between supersymmetric gluodynamics and its orbifold daughter
(\ref{three}) requires a rescaling of
the gauge couplings in passing from the parent to
the daughter theory, namely 
\beq
g_D^2 = 2 g_P^2\,.
\label{four}
\eeq
In the general case of $Z_k$ orbifoldization the relation is
\beq
g_D^2 = k g_P^2\,.
\eeq

The orbifold daughter  is {\em not} supersymmetric.
The color-singlet supercurrent (\ref{two}) cannot be formed now
since the color indices of the gauge fields and the fermion fields
cannot be contracted. In this way the  orbifold daughter seems to be  closer
to genuine QCD than SUSY gluodynamics. In fact, for $N=3$
this is nothing but three-color/three-flavor ``QCD" with a gauged flavor group.
(Note, that flavor SU(2) is gauged anyway in the electroweak theory,
albeit the corresponding gauge coupling is small.)

The absence of supersymmetry is not the only feature
of (\ref{three}) that distinguishes this theory
from SUSY gluodynamics. The daughter theory possesses a global U(1).
The conserved fermion number current is
$
J_\mu = \overline{\Psi} \gamma_\mu \Psi\,.
$
Applying this vector current to the vacuum, one produces a vector meson
that has no counterpartners in the parent theory. One says that
the operator $J_\mu$ has no projection onto the parent theory.
One can construct  other operators with   similar properties.
There are many hadronic states in the daughter theory
that have no analogues in the parent one. For instance,
for odd $N$ one expects composite states with the 
quark content $\Psi\Psi ...\Psi\,\,  \varepsilon\,\,  \tilde \varepsilon$
and masses scaling as $N$. For even $N$, the daughter theory has no color-singlet fermions at all.

The inverse is also true. For instance, the operator of the supercurrent
produces a spin-3/2 hadron from the vacuum state of SUSY gluodynamics,
which has no counterpart in the daughter theory. Therefore,
in confronting these theories it makes sense
to compare only those sectors that have projections onto one another.
Below we will refer to such sectors as {\em common}.

Under the $Z_k$ orbifoldization the gauge symmetry of the
orbifold daughter is $ {\rm SU}(N)^k$,
and all fermion fields in the daughter theory are bifundamentals. Note
that the daughter theory is chiral for $k>2$, i.e. no mass term can be introduced
for the ``quark" fields. Such daughter theories can be only remotely
related to QCD, if at all. 

\subsection{Perturbation theory}
\label{pert-theory}

As was mentioned in Sect.~\ref{genesis}, it was discovered in 1998 \cite{KK} that the Green functions for  operators,
which exist in both theories, have  identical planar graphs\,\footnote{This
statement assumes that the daughter theory 
under consideration inherits the vacuum from the parent theory,
an important reservation given that there are more 
vacua in the parent  theory than in the daughter  one.
Perturbation theory  does not distinguish between distinct vacua
as long as they are ``untwisted", i.e. 
$Z_2$-symmetric.} provided
the rescaling (\ref{four}) is implemented. In other words, perturbatively
these Green's functions in the parent (supersymmetric) and daughter
(non-supersymmetric) theories coincide order by order in the large-$N$ limit.
Motivation for the work \cite{KK}
came from the string side. 
A detailed discussion of orbifoldization
in various gauge theories can be found in Ref. \cite{schmaltz}.

\begin{figure}
\epsfxsize=7cm
\centerline{\epsfbox{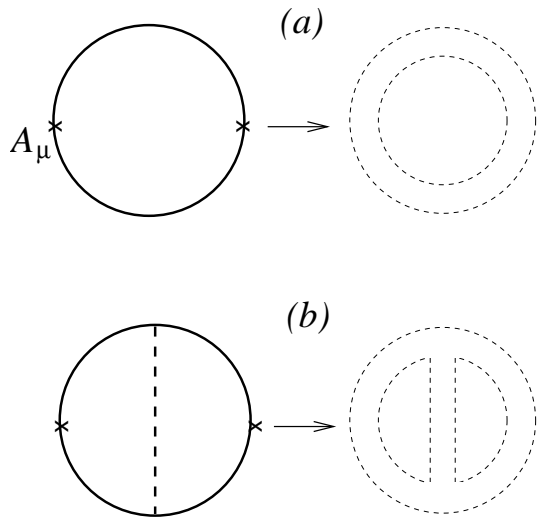}}
\caption{
Counting $N$ factors in the correlation function
of the axial currents. The 't Hooft diagrams are shown in dashed lines.}
\label{countingN}
\end{figure}

Rather than presenting a general proof (it can be found in the original
publications, see e.g. \cite{KK}) we will illustrate this statement (see 
Fig.~\ref{countingN}) by a simple 
example of the correlation function of two axial currents,
$$(2/g^2)(\bar\lambda_{\dot\alpha})^{i}_{j}(\lambda_\alpha)^{j}_{i}\,.$$
For computational purposes, it is convenient to keep
the fermion fields in the daughter theory in the same form
$(\lambda_\alpha)^{i}_{j}$, just constraining the color indices.
In the parent theory $i,j$ run from 1 to $2N$. In the daughter theory
either $i\in [1,N],\quad  j\in [N+1, 2N]$ or {\em vice versa},
$i\in [N+1,2N],\quad  j\in [1, N]$. In the first case we deal with
$\lambda^{i}_{ \tilde j}$ in the second $ \lambda^{\tilde  i}_{ j}$.

The easiest way to check the match is by analyzing relevant  graphs in the
't Hooft representation --- each gluon or fermion line carrying two indices
is represented by a double line, see Fig. \ref{countingN}. 
In this way we readily see that the weight of the graph $3a$ is 
$4N^2$ (parent), $2N^2$ (daughter), while that of $3b$ is
$8N^3 g_P^2$ (parent), $2N^3 g_D^2$ (daughter). To get the latter estimate one
notes that if the first index of $\lambda$ --- the one forming the outside loop
--- is untilded, the second index (corresponding to two inside loops
in Fig. \ref{countingN}b) is tilded, and {\em vice versa}. 
The parent-to-daughter ratio 
is 2 for both graphs $3a$ and $3b$, provided $g^2_D = 2\, g^2_P$.
Proceeding along these lines it is quite easy to see that
the parent-to-daughter ratio stays the same for any planar graph.
This proves the perturbative planar equivalence
in the case at hand. The proof can be readily extended
to any correlation function that exists in both supersymmetric gluodynamics and its orbifold daughter. In the common sector
these two theories are perturbatively equivalent at $N=\infty$.

\subsection{Consequences of perturbative planar equivalence}
\label{cotppe}

As we have just seen, a relationship can be established between two theories 
in the large-$N$
limit --- one is supersymmetric, the other is not. Although this relationship
is admittedly perturbative, a natural question to ask
is whether one can benefit from it and, if yes, how.

The {\em exact} $\beta$ function of
SU($2N$) 
SUSY gluodynamics,  established \cite{NSVZ} almost twenty years ago,
is as follows:
\beq
\beta(\alpha_P) \equiv \frac{d \alpha_P}{d\ln\mu }=
-\frac{ 1}{2\pi} \,\,\frac{6N \alpha_P^2}{1- (N\alpha_P )/\pi }\,.
\eeq
Because of  planar equivalence,  it implies
that the full $\beta$ function in the (non-supersymmetric)
daughter theory is also known. Namely, in the 't Hooft limit,
\beq
\beta(\alpha_D)
=- \frac{1}{2\pi}\,\, \frac{3N\alpha_D^2}{1-(N\alpha_D)/(2\pi )}\,  ,
\label{ten}
\eeq
where the relation (\ref{four}) between the coupling
constants in the parent and orbifold theories
is taken into account. Curiously,  the $\beta$ function (\ref{ten})
is exactly the same as in SU$(N)$ SUSY gluodynamics.

\section{Non-perturbative orbifold conjecture}
\label{npo}
\renewcommand{\theequation}{\thesection.\arabic{equation}}
\setcounter{equation}{0}

It is  non-perturbative dynamics which is our prime concern in strongly 
coupled gauge  theories. Non-perturbative
effects are at the heart of  key phenomena such as  confinement, chiral
symmetry breaking, and so on.  Strassler conjectured
\cite{strassler} that the planar Feynman graph equivalence
discussed in Sect.~\ref{orbifoldization}
extends to non-perturbative phenomena, namely
Green's functions for the color-singlet operators that appear in 
both theories are identical at
$N\to\infty$,  including non-perturbative effects.
This suggestion is referred to as the
{\em non-perturbative orbifold (NPO) conjecture}. 

Unfortunately, Strassler's suggestion, in its original form, is not valid
\cite{gorshi}, as we will see shortly.
It played an important role, however, in paving the way for further developments. Analyzing why orbifold daughters fail to retain
planar equivalence at non-perturbative level,
one can guess how one can engineer daughters free from
these drawbacks and  thus have better chances
to be non-perturbatively planar-equivalent to their supersymmetric
parents (Sect. \ref{orientifold}).

\subsection{The vacuum structure and low-energy theorems}
\label{tvsalet}

In order to go over to a non-perturbative analysis, let us first
compare the number of vacua in the SU($kN$)
supersymmetric gluodynamics to that of its $Z_k$
daughter.

The vacua in the parent theory are counted
by Witten's index \cite{wittenindex}. They are labeled by
the gluino condensate, see Eq.~(\ref{gluco}), which for
SU($kN$) takes $kN$ distinct values, so that
we have $kN$ degenerate vacua. For $k=2$,
the simplest case with which we will deal below,
there are $2N$ supersymmetric vacua in the parent theory.

At the same time, the number of vacua in the orbifold
theory is twice smaller. This is easy to see from the discrete
chiral invariance of the daughter theory. Generally speaking, the
theory (\ref{three}) is {\em not} invariant under
the chiral rotation
\beq
\chi\to \chi \, e^{i\alpha}\,,\qquad \eta \to \eta  e^{i\alpha}\,,
\label{chro}
\eeq
since under this rotation the vacuum angle $\theta$ is shifted,
$\theta\to \theta +2\, N\,\alpha$. However, if 
\beq
\alpha = k\pi /N\,,\qquad k \,\,\mbox{integer}\,,
\label{ququk}
\eeq
then $\delta\theta = 2\pi\, k$, and such a discrete rotation 
{\em is} an invariance of the
theory. The integer parameter $k$ in Eq.~(\ref{ququk})
clearly runs from $1$ to $2 N$.  However, only the phase of the bilinear
operator $\chi\eta$ distinguishes the different vacua. 
In other words, the condensate
$\langle \chi\eta\rangle $ breaks (spontaneously) $Z_{2N}$,
the discrete symmetry surviving the axial anomaly in 
the theory (\ref{three}),  down to $Z_2$. We conclude that the 
daughter theory has $N$ distinct vacua. 
 
The phase of the bifermion operator {\em per se}
is unobservable. One makes it observable by introducing a
small mass term (which we will need  anyway in order
to regularize fermion determinants, see Sect. \ref{orientifold}), 
\beq
{\cal L}_m = \left\{
\begin{array}{l}
-\frac{m}{2g_P^2}\, \lambda^a\lambda^a +\mbox{h.c.}\\[4mm]
-\frac{m}{g_D^2}\, \chi\eta +\mbox{h.c.}
\end{array}
\right.
\eeq
The vacua are no longer degenerate; the vacuum energy densities
split,
\beq
{\cal E}_{\rm vac} =\left\{
\begin{array}{l}
\frac{m}{2g_P^2}\, \langle {\rm vac}_k|\lambda^a\lambda^a |
{\rm vac}_k\rangle_0 \, e^{i\theta /(2N)}+\mbox{h.c.}\\[4mm]
\frac{m}{g_D^2}\, \langle{\rm vac}_k|\chi\eta |
{\rm vac}_k\rangle_0\, e^{i\theta /N}+\mbox{h.c.}
\end{array}
\right.\,\, +O(m^2)\,,
\label{vedorbi}
\eeq
where $|{\rm vac}_k\rangle $ denotes the $k$-th vacuum,
while the subscript 0 marks the value of the condensate at
$\theta =0$. Finally, besides the vacuum energy densities,
we will consider topological susceptibilities defined as
\beq
{\cal T} =\left.
-\frac{\partial^2{\cal E}_{\rm vac}}{\partial\theta^2}\right|_{\theta=0}
= i\int d^4 x
\left\langle \frac{1}{32 \pi^2}\,
G\widetilde G (x) ,\,\, \frac{1}{32 \pi^2}\,
G\widetilde G (0) \right\rangle_{\rm conn}\,,
\label{thureleven}
\eeq 
see Eqs.~(\ref{thetaterm}) and (\ref{three})
for the definition of $G\widetilde G$ in the parent and 
orbifold theories, respectively.

Now, we can assemble all the above elements to prove
that the NPO conjecture does not work.
Indeed, assume it does. Then 
${\cal E}_{\rm vac}^D ={\cal E}_{\rm vac}^P$.
Here we choose a pair of ``corresponding" vacua.
Remember, the parent theory has twice as many vacua.
We take the one that can be projected onto the daughter theory.
Equation (\ref{vedorbi}) implies that in the given vacuum
${\cal T}_P=\frac{1}{4}\, {\cal T}_D$.  

On the other hand, the topological susceptibilities in the given vacuum
are measurable, for instance, on lattices.
The statement that ${\cal T}_P=\frac{1}{4}\, {\cal T}_D$
is in disagreement with the definition
(\ref{thureleven}) by a factor of 2 \cite{gorshi}. Thus, the non-perturbative
sector of the orbifold theory remembers the mismatch between the 
number of vacua in the parent and daughter theories.

\subsection{What makes orbifold daughters unsuitable?}
\label{wmodu}

The underlying reason for the failure of the NPO conjecture
is most clearly seen in the string-theory language: it is the presence of {\em tachyons} in the twisted sector.
We defer the corresponding discussion to Sect.~\ref{ovoavftsts}.
Here we will present some field-theoretical arguments
illustrating the devastating consequences of the presence of the twisted sector.

First, let us explain what the twisted sector is
in the field-theoretical language. To this end we will
recast Fig.~\ref{z2orbi}, summarizing the content of the
$Z_2$ orbifold daughter,
in a slightly different way, see Table~\ref{table_N=1}.

\begin{table}
\begin{displaymath}
\begin{array}{l c@{ } c@{ } }
& \multicolumn{1}{c@{\times}}{{\rm U}_e(N)}
& \multicolumn{1}{c@{}}{{\rm U}_m(N)} \\
\hline
A _{e}^{e} & {\rm Adj} & 1  \\
A _{ m}^{m} & 1 & {\rm Adj}  \\
\lambda ^{e}_{m} &  \Yfund & \overline{\Yfund}  \\
\lambda ^{m}_{e} &   \overline{\Yfund} & \Yfund  \\
\end{array}
\end{displaymath}
\caption{The $Z_2$ orbifold theory.}
\label{table_N=1}
\end{table}

We replaced the untilded indices in Fig.~\ref{z2orbi}
by generic sub/superscripts $e$ (meaning electric) while 
the tilded indices by $m$ (meaning magnetic).
As we will see in Sect.~\ref{ovoavftsts}, the $Z_2$ orbifold theory 
has a realization in type 0A theory
\cite{Armoni:1999gc}. It lives on a brane configuration of type 0A,
which consists of  ``electric"  and ``magnetic" D-branes. This is
the origin of the sub/superscripts $e$ and $m$ in Table~\ref{table_N=1}.
The $Z_2$ orbifold theory is obviously $Z_2$-symmetric
under the exchange of all indices $e\to m$ and {\em vice versa}.

Let us divide all color-singlet operators of the $Z_2$ daughter in
two classes:  (i) those that 
are invariant (even) under the 
above $Z_2$ symmetry, and (ii) those that are non-invariant
(odd). The operators from the second class are called twisted.
For instance, 
$ G_{e}^{e}G_{e}^{e} - G_{ m}^{m}G_{ m}^{m}$ is a twisted operator, whereas
$ G_{e}^{e}G_{e}^{e} + G_{ m}^{m}G_{ m}^{m}$ is untwisted.

The perturbative relation between the orbifold theory and its
supersymmetric parent concern only the untwisted sector
\cite{K,KK}. The parent theory does not
carry information about the twisted sector of the daughter theory.
The NPO conjecture assumes that the vacuum of the daughter
theory is $Z_2$-invariant. However,  this need not be the case.

A possible sign of the $Z_2$ instability
can be obtained just in  perturbation theory. Indeed, let us assume
that at some ultraviolet (UV) scale, where perturbation theory is
applicable, the gauge couplings of the two SU($N$) factors in the
orbifold theory are slightly different. We will denote them by
$8\pi^2/(Ng^2_e) =1/\lambda_e$ and $8\pi^2/(Ng^2_m) =1/\lambda_m$.
It is not difficult to find the
renormalization group flow of $\delta \lambda $ towards the 
infrared (IR) domain.
As long as $\delta \lambda\ll \lambda_{e,m}$ we have
\beq
\frac{d (\delta \lambda)}{d\,\ln \mu}= -{3}\, {\lambda^2}\, \, (\delta \lambda)
+\mbox{higher orders}\,.
\eeq
Neglecting the weak logarithmic $\mu $ dependence of $\lambda$,
we obtain
\beq
\delta \lambda (\mu ) = \delta \lambda (\mu_0 )\, \left(\frac{\mu_0}{\mu}
\right)^{3\lambda^2}\,.
\eeq
Even if $\delta \lambda $ is small in the UV, it grows exponentially  towards the IR domain. A similar analysis of the conformal orbifold daughter of
\Nfour SUSY theory was carried out in Ref.~\cite{klebanov}, 
and a similar conclusion  about such IR instability was reached.

The advantage of the orientifold theory over the
orbifold theory is the absence of the twisted sector.
Moreover, the gauge groups are the same in the
parent and daughter theories, and so are the patterns of the
spontaneous breaking of the global symmetry
and the numbers of vacua. This is also seen from the string-theory
standpoint. The orbifold field theory originates from the type 0A
string theory, which is obtained by a $Z_2$ orbifold of type IIA. The
result is a bosonic string theory with a tachyon in the twisted
sector. In contrast, the orientifold field theory originates from
a configuration with an orientifold that removes the twisted sector
(the result is very similar to the bosonic part of the type I string).

Another way to search for instabilities in the twisted sector was
suggested in \cite{gorshi}. The logic is as follows.
Since the perturbative planar equivalence
between the SUSY parent and orbifold daughter
does not depend on the geometry of space-time,
one can compactify one or more dimensions, making
the compactified dimension small, so that  the theory
in question becomes weakly coupled. Then one can   compare 
the parent and the orbifold theories at weak coupling.

Tong considered \cite{dt} an $R^3\times S$ compactification. On $R^3\times S$, the third spatial component 
of the gluon field becomes a scalar field with a flat (vanishing)
potential at the classical level, $(A_3)^i_j\to \phi^i_j$.
(Alternatively, one can speak of the Polyakov line in the $x_3$ direction.)  
In the parent theory the flatness is perturbatively maintained to all
orders owing to supersymmetry. Non-perturbatively, the flatness is 
lifted by $R^3\times S$ 
instantons (monopoles), which generate a superpotential. 
As it  turns out, in supersymmetric vacua the non-Abelian
gauge symmetry is  broken by $\langle\phi^i_j\rangle = v_i\delta^i_j$, 
down to U(1)$^{2N}$, so that the theory is in the
Coulomb phase \cite{Davies:1999uw}.

In the daughter theory the situation is drastically different:
a potential {\em is generated at one loop}, making the point
of the broken gauge symmetry  unstable. Shifting away from this point, 
for a trial, one finds oneself in the \mbox{$Z_2$-non-invariant}
(twisted) sector, for which no planar equivalence exists,
and which proves to be energetically favored in the
orbifold theory on $R^3\times S$. In this true vacuum the energy density 
is negative rather than zero, and the full gauge symmetry
is restored. The daughter theory is in the confining phase. Obviously, there can 
be no equivalence.

\section{Orientifold field theory and \boldmath{\None} \\
gluodynamics}
\label{orientifold}
\renewcommand{\theequation}{\thesection.\arabic{equation}}
\setcounter{equation}{0}

Having concluded  that, regretfully, the planar equivalence
of the orbifold daughters does not extend to the 
non-perturbative level, we move on to another class of theories,
called {\em orientifolds}, which lately gave rise to great expectations. 
In this section we will argue that in the $N\to\infty$
limit there is a sector in the orientifold theory
exactly identical to \None SYM theory, and, therefore, exact
results on the IR behavior of this theory can be obtained.
This sector is referred to as the {\em common sector}.

The parent theory is \None SUSY gluodynamics with  gauge
group SU($N$). 
The daughter theory has the same gauge group
and  the same gauge coupling.  The gluino field
$\lambda^i_j$ is replaced by two Weyl spinors
$\eta_{[ij]}$ and $\xi^{[ij]}$,
with two antisymmetrized indices. We can combine the Weyl spinors
into one Dirac spinor, either $\Psi_{[ij]}$ or $\Psi^{[ij]}$.
Note that the number of fermion degrees of freedom in $\Psi_{[ij]}$
is $N^2-N$, as in the parent theory in the large-$N$ limit.
We call this daughter theory {\em orientifold A}.

There is another version  of the orientifold daughter --- {\em orientifold S}.
Instead of the antisymmetrization of the two-index spinors,
we can perform symmetrization, so that
$\lambda^i_j\to \, (\eta_{\{ij\}}\,,\,\,\, \xi^{\{ij\}})$.
The number of degrees of freedom in $\Psi_{\{ij\}}$ is $N^2+N$.
The field contents of the orientifold theories is shown in Table~\ref{table3}.
We will mostly focus on the antisymmetric daughter since it is of more
physical interest; see Sect.~\ref{cqciofqfsg}.

\begin{table}
\begin{center}
\begin{minipage}{2.8in}
\begin{tabular}{c||ccc }
& ${\rm SU}(N)$ & ${\rm U}_V(1)$ & ${\rm U}_A(1)$  \\
\hline \hline \\
${\eta_{\{ij\}}}$& $\symm$ & $1$ & $1$  \\
&&&\\
${\xi}^{\{ij\}}$ &$\bsymm $& $-1$ & $1$ \\
&&&\\
$A_{\mu}$ &{\rm Adj} & $0$ & $0$    \\
\end{tabular}
\end{minipage}
\begin{minipage}{2.2 in}
\begin{tabular}{c||ccc}
& ${\rm SU}(N)$ & ${\rm U}_V(1)$ & ${\rm U}_A(1)$  \\
\hline \hline \\
$\eta_{[ij]}$& $\asymm$ & $1$ & $1$  \\
&&&\\
${\xi}^{[ij]}$ &$\basymm $& $-1$ & $1$ \\
&&&\\
$A_{\mu}$ &{\rm Adj} & $0$ & $0$    \\
\end{tabular}
\end{minipage}
\end{center}
\caption{The field content of the orientifold  theories. Here, $\eta$
and $\xi$ are two Weyl fermions, while $A_{\mu}$
stands for the gauge bosons. In the left (right) parts of the
table the fermions are in the two-index symmetric (antisymmetric)
representation of the gauge group SU$(N)$. Moreover,
${\rm U}_V(1)$ is the
conserved global symmetry while the ${\rm U}_A(1)$ symmetry is lost at
the quantum level, because of the chiral anomaly. A discrete subgroup
remains unbroken.} 
\label{table3}
\end{table}

The hadronic (color-singlet) sectors of the parent and daughter theories
are different. In the parent theory composite fermions with mass
scaling as $N^0$ exist, and, moreover, they are degenerate with their
bosonic  SUSY counterparts. In the daughter theory any interpolating
color-singlet current with the fermion quantum numbers (if it exists at all) contains  a number of constituents growing  with $N$. Hence,
at $N=\infty$ the spectrum contains only bosons.

Classically the parent theory has a single global
symmetry --- an $R$ symmetry corresponding to the
chiral rotations of the gluino field. Instantons break this symmetry down to
${Z}_{2N}$. The daughter theory has, on top, a conserved anomaly-free  current
\beq
\bar \eta_{\dot\alpha } \eta_{\alpha}
-\bar \xi_{\dot\alpha } \xi_{\alpha}\,.
\label{veccur}
\eeq
In terms of the Dirac spinor this is the vector current
$\bar\Psi\gamma_\mu\Psi$.
If the corresponding charge is denoted by $Q$, then necessarily
$Q=0$ in the color-singlet bosonic sector. Thus,  the only residual global symmetry
in both theories is ${Z}_{2N}\ $ spontaneously broken
down to ${Z}_{2}\ $
by the respective bi-fermion condensates. This explains the existence
of $N$ vacua\,\footnote{At finite $N$ the parent theory
has $N$ vacua, while the orientifold daughters
have $N-2$ and $N+2$ in the antisymmetric and symmetric versions, respectively.} in both cases.

We will compare the bosonic sectors of the
parent and daughter theories. Note that 
the part of the  daughter theory's bosonic sector probed by operators
of the type (\ref{veccur}), that have no analogs in the
parent theory, is inaccessible. Such a sector of the theory does not belong
to the common sector.

\subsection{Perturbative equivalence}
\label{oripeq}

\begin{figure}
\begin{center}
\mbox{\kern-0.5cm
\epsfig{file=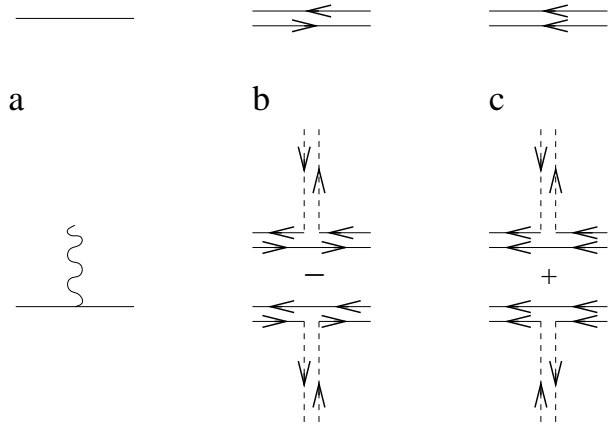,width=8.0true cm,angle=0}}
\end{center}
\caption{(a) The fermion  propagator and the fermion--fermion--gluon
vertex. (b) \None
SYM theory. (c) Orientifold daughter.}
\label{fey1}
\end{figure}

Let us start from  perturbative considerations. 
The Feynman rules of the planar theory are shown in Fig. \ref{fey1}.

The difference between the orientifold theory and \None 
gluodynamics is that the
arrows on the fermionic lines point in the same direction, since
the fermion is in the antisymmetric representation, in contrast
to the supersymmetric theory where the gaugino is in the adjoint
representation and the arrows point in  opposite directions.
This difference between the two theories does not affect planar
graphs, provided that each gaugino line is replaced by the sum
of $\eta_{[..]}$ and $\xi^{[..]}$.

\begin{figure}
\begin{center}
\mbox{\kern-0.5cm
\epsfig{file=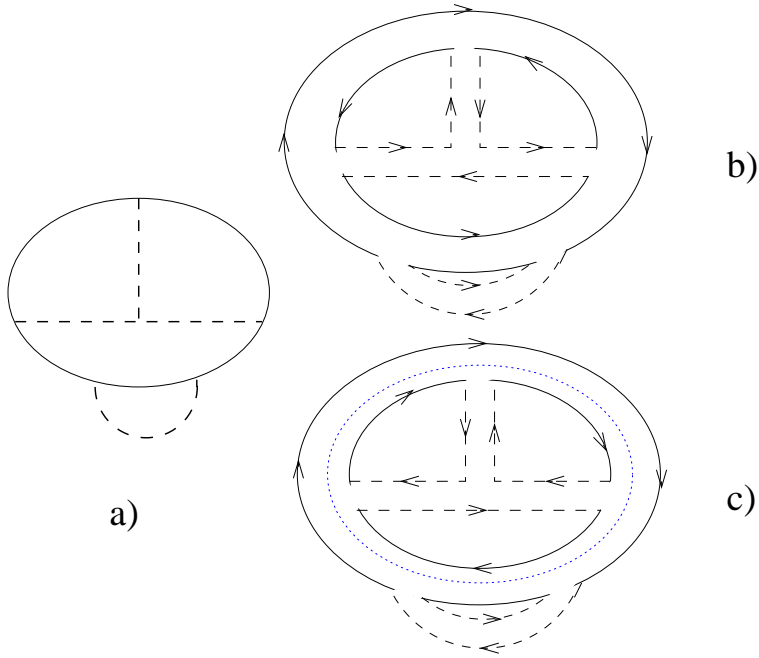,width=10.0true cm,angle=0}}
\end{center}
\caption{(a) A typical planar contribution to the vacuum energy.
(b and c) The same in the 't Hooft notation for
(b)  \None SYM theory; (c) orientifold daughter.}
\label{fey2}
\end{figure}

There is a one-to-one correspondence between the planar graphs
of the two theories. Diagrammatically this works as follows (see, for
example, Fig.~\ref{fey2}). Consider any planar diagram of the daughter theory:
by definition of planarity, it can be drawn on a sphere. The fermionic
propagators form closed, non-intersecting loops that divide the sphere into
regions. Each time we cross a fermionic line the orientation of color-index
loops (each one producing a factor $N$) changes from clock to 
counter-clockwise, and {\em vice-versa}, as  is graphically demonstrated  in
Fig.~\ref{fey2}c. Thus, the fermionic loops allow one to attribute to 
each  of the above regions a binary label (say $\pm1$),  according to whether 
the color loops go clock or counter-clockwise in the given region. 
Imagine now that one cuts out all the regions with a $-1$ label and glues them again on the sphere after having flipped them upside down. 
We will get a planar diagram of the SYM theory in which
all color loops go, by convention, clockwise.
The number associated with both diagrams will be the same since the diagrams
inside each region always contain an even number of powers of $g$, so that the relative minus signs of Fig. \ref{fey1} do not matter.

In fact, in the above argument, we cut corners, so that the careful 
reader might get somewhat puzzled.
For instance, in the parent theory gluinos are  Weyl (Majorana) fermions,
while fermions in the daughter theory are  Dirac fermions.
Therefore, we hasten to add a few explanatory remarks, which will, 
hopefully, leave the careful  reader fully satisfied.

First, let us replace the Weyl (Majorana) gluino of \None gluodynamics
by a Dirac spinor $\Psi^i_j$. Each fermion loop in the original
theory is then obtained from the Dirac loop by multiplying the latter by $1/2$.
Let us keep this factor $1/2$ in mind.

On the daughter theory side, instead of considering
the antisymmetric spinor $\Psi_{[ij]}$ {\em or}
the symmetric one, $\Psi_{\{ij\}}$, we will consider
a Dirac spinor in the {\em reducible} two-index representation
$\Psi_{ij}$, without imposing any conditions
on $i,\, \, j$. Thus, this reducible two-index representation
is a sum of $\symm + \asymm \,$. It is rather obvious that
at $N\to\infty$ any loop of $\Psi_{[ij]}$ yields the same result
as the very same loop with $\Psi_{\{ij\}}$, which implies, in turn,
that in order to get the fermion loop in, say, an antisymmetric orientifold daughter,
one can take the Dirac fermion loop in the above reducible
representation, and multiply it by $1/2$.

Given the same factor $1/2$ on the side of the parent and daughter theories,
what remains to be done is to prove that the Dirac
fermion loops for $\Psi^i_j$ and $\Psi_{ij}$ are identical at
$N\to\infty$. We will therefore focus on the color factors.

Let  the generator of SU$(N)$ in the fundamental representation
be $T^a$, while that in the antifundamental one will be $\bar T^a$,
\beq
T^a=T^a _{\Yfund}\,,\qquad \bar T^a=T^a _{\overline{\Yfund}} \,.
\label{dengen}
\eeq
Then the generator in the adjoint  representation is
\bea 
T^a _{\rm Adj} &\sim& T^a _{\Yfund \, \times \, \overline{\Yfund}} =
T^a _{\Yfund} \otimes1 + 1 \otimes T^a _{ \overline { \Yfund}}
\nonumber\\[2mm]
&\equiv &T^a \otimes1 + 1 \otimes \bar T^a 
\,,
\label{mnone}
\eea
where we made use of large-$N$, neglecting the singlet. Moreover, 
in the daughter theory the generator of the reducible $\Yfund \otimes
\Yfund$ representation can be written as
\bea
T^a_{\rm two-index}  &=& T^a _{\Yfund} \otimes 1 + 1 \otimes T^a_{\Yfund}
\equiv T^a   \otimes 1 + 1 \otimes T^a  \nonumber\\[2mm]
&\mbox{or}&\,\,\,\, \bar T^a   \otimes 1 + 1 \otimes \bar T^a \,.
\label{mn-two}
\eea
One more thing which we will need to know is the fact that
\beq
\bar T = -\tilde T= - T^*\,,
\label{bartt}
\eeq
where the tilde denotes the transposed matrix.

Let us examine the color structure of a generic planar diagram for
a gauge-invariant quantity. For example, Fig.~\ref{fey2}a
exhibits a four-loop planar graph for the vacuum energy.
The color decomposition (\ref{mnone}) and (\ref{mn-two})
is equivalent to using the 't Hooft double-line notation,
see Figs.~\ref{fey2}b,c.  In the parent theory each fermion--gluon vertex 
contains $T^a _{\rm Adj} $; in passing to the daughter theory we
replace $T^a_{\rm Adj} \to T^a_{\rm two-index} $.

Upon substitution of Eqs.~(\ref{mnone}) and (\ref{mn-two})
the graph at hand splits into two (disconnected!) parts:
the inner one (inside the dotted ellipse in Fig.~\ref{fey2}c),
and the outer one (outside the dotted ellipse in Fig.~\ref{fey2}c).
These two parts do not ``talk to each other,"  because of planarity
(large-$N$ limit). The outer parts are the same in Figs. \ref{fey2}b,c.
They are  proportional  the trace of the product of two $T$'s in the
two cases:  
$$
{\rm Tr}\,  (\bar T^a \bar T^a)\,.
$$
This is the first factor. The second one comes from the inner part of Figs. \ref{fey2}b,c.
In the parent theory the inner factor is built out of 6 $ T$'s ---
one in each fermion--gluon vertex, and three $ T$'s in the three-gluon vertex
Tr$([A_\mu A_\nu] \, \partial_\mu A_\nu )$, where $A_\mu \equiv A_\mu^a   T^a$. In the daughter theory
the inner factor is obtained from that in the parent one
by replacing all $T$'s by $\bar  T$'s. According to Eq.~(\ref{bartt}),
$\bar T = -\tilde T$ (remember that a tilde denotes the transposed matrix). 
This fact implies that the only difference
between the inner blocks in Figs. \ref{fey2}b,c is the reversal of the
direction of the
color flow on each of the 't Hooft lines. Since the inner part is a color-singlet by itself, 
the above reversal has no impact on the color factor ---
they are identical in the parent and daughter theories.

It is, perhaps, instructive to illustrate how this works
using a more conventional notation. For the
 inner part of the graph in Fig. \ref{fey2}b
we have the color factor Tr $(T^aT^bT^c)\, f^{abc}$, while in the daughter theory we have
Tr $(\bar T^a \bar T^b \bar T^c)\, f^{abc}$. Using the fact that
$$
[T^aT^b]=i f^{abc}T^c\quad\mbox{and}\quad [\bar T^a\bar T^b]=i
f^{abc}\bar T^c\, ,
$$
we immediately come to the conclusion that the above two expressions
coincide.

Thus, all perturbative results that we are aware of in \None SYM
theory apply in  the orientifold model as well. For example, the
$\beta$ function of the orientifold field theory is
\beq
\beta = -\frac{1}{2\pi}\,\frac{3N\alpha^2}{1- (N\alpha)/(2\pi)}
\left\{1+ O\left(\frac{1}{N}\right)
\right\}.
\eeq
In the large-$N$ limit it coincides with the \None SYM theory result
\cite{NSVZ}. Note that the corrections are $1/N$ rather than
$1/N^2$. For instance, the exact first coefficient
of the $\beta$ function is $-3N-4/3$ versus
$-3N$ in the parent theory. 

\subsection{Non-perturbative equivalence proof}
\label{norieq}

Now we will argue that the perturbative argument can be elevated
to the non-perturbative level in the case at hand. 
A heuristic argument in favor of the non-perturbative equivalence 
is that the coincidence of all  planar graphs 
of the two theories implies that the relevant Casimir operators of the two
representations are equivalent in the large-$N$ limit. The partition
functions of the two theories depend  on the  Casimir operators
and, therefore, must  coincide as well.   

A more formal line of reasoning is as follows.
It is essential that the fermion fields enter
bilinearly in the action, and that for any {\em given}
gauge-field configuration in the parent theory there
is {\em exactly the same} configuration in the daughter one.
Our idea is to integrate out fermion fields
for any {\em fixed} gluon-field configuration,
which yields respective determinants, and then compare them.

Consider the partition function of \None SYM theory,
\beq
{\cal Z}_0 = \int {\cal D}A\, {\cal D}\lambda \, \exp \left(
iS[A,\lambda, J] \right)\,,
\eeq
where $J$ is any source coupled to color-singlet  gluon operators.
(Appropriate  color-singlet fermion bilinears can be considered too.)

For any given gluon field, upon
integrating out the gaugino field, we obtain
\beq
{\cal Z}_0 = \int {\cal D}A \, \exp\left( iS[A, J]
\right)  \det \left (i\slsh \partial + \slsh \!A ^a T^a _{\rm Adj}  \right )
\label{Z},
\eeq
where $T^a _{\rm Adj}$ is a generator in the adjoint representation.

If one integrates out the fermion fields in the
non-supersymmetric orientifold theory, at fixed $A$,
one arrives at a similar expression, but with the generators of the
antisymmetric (or symmetric) representation instead of the adjoint,
$T^a _{\rm Adj} \rightarrow T^a _{\rm asymm}\,\,\,\mbox{or}\,\,\, 
T^a _{\rm symm}$. 

To compare the fermion determinants in the
parent and daughter theories (assuming that the gauge
field configuration $A_\mu^a (x),\,\,\, a=1,..., N^2,$ is the same and fixed),
we must cast both fermion operators in  similar forms.
To this end we will   extend both theories.
In the parent one we introduce a second adjoint Weyl fermion
$\xi^i_j$ and combine two Weyl fermions into
one Dirac adjoint fermion $\Psi^i_j$. The determinant
in this extended theory is the square of the original one.

In the daughter theory, instead of $\Psi^{[ij]}$, we will work
with the reducible representation, combining both symmetric 
and antisymmetric into the Dirac field $\Psi^{ij}$
(no (anti)sym\-metrization over the color indices).
Then the number of fermion Dirac fields, $N^2$, is
the same as in the extended parent. Again, as in the previous case,
the determinant in the extended daughter theory is the square
of the original one. To avoid zero modes, which might make determinants
ill-defined, we must impose an IR regularization. 
To this end we will introduce small fermion mass terms, 
which will render determinants well-defined. As usual,
it is assumed that the vanishing mass limit is smooth,
which must be the case in the confining regime.

Since the theory, being vector-like, is anomaly free, the determinant in 
Eq.~\eqref{Z} is a gauge-invariant object and, thus, can be expanded in Wilson-loops operators 
\beq
{\cal W}_{\cal C}[A_{\rm Adj}] = {\rm Tr}\, P\, \exp
\left(
 i \int_{\cal C} A^a_\mu T^a _{\rm Adj}\,\, dx^\mu
\right)\,.
\label{wlo}
\eeq
For other representations, the Wilson-loop operator
is defined in a similar manner. Thus, one can write
\beq
 D\equiv  \det \left (i\slsh \partial + \slsh \!A ^a T^a _{\rm Adj} - m  \right )
= \sum _{\cal C} \alpha _ {\cal C} \, {\cal W}_{\cal C}[A_{\rm Adj}]\,.
\label{det}
\eeq
Equation (\ref{mnone}) then implies that 
\beq
D = \sum _{\cal C} \alpha _ {\cal C} \, {\rm Tr}\, P\, \exp
\left(
 i \int_{\cal C} A^a_\mu \, (
T^a  \otimes 1 + 1 \otimes \bar T^a  )\, dx^\mu
\right) .
\label{det2}
\eeq
Moreover, since the commutator is such that
$$ [(T^a  \otimes
1) , \,\, (1 \otimes \bar T^a )] =0\,,
$$
the determinant \eqref{det2} can be rewritten as
\beq
D=  \sum _{\cal C} \alpha _{\cal C}\, 
{\rm Tr}\, P\, \exp \left (
i \int_{\cal C} A^a_\mu T^a \, dx^\mu \right )\,
{\rm Tr}\, P\, \exp \left (
i \int_{\cal C} A^a_\mu \bar T^a \, dx^\mu \right )\,.
\label{det3}
\eeq
As a result,  the partition function takes the form
\beq
{\cal Z}_0=
\sum _ {\cal C} \alpha _{\cal C} \,
\left\langle \,  {\cal W}_{\cal C} [A _{\Yfund}]\, \,
{\cal W} ^*_{\cal C} [A _{\Yfund}] \,  \right\rangle\,.
\label{pfttf}
\eeq
One of the two most crucial points of the proof is  the applicability of
factorization in the large-$N$ limit,
\beq
{\cal Z}_0=
\sum _ {\cal C} \alpha _{\cal C}\,
\left\langle \, {\cal W}_{\cal C} (A _{\Yfund})\,  \right\rangle\left\langle \, 
{\cal W}^*_{\cal C} (A _{\Yfund})\, 
\right\rangle = \sum _ {\cal C} \alpha _{\cal C} \left\langle \, {\cal
W}_{\cal C} (A _{\Yfund})\,  \right\rangle ^2 \, . 
\label{detN}
\eeq
In the second equality in Eq.~(\ref{detN}) we used 
the second most crucial point,
the reality of the Wilson loop,
\beq
\left\langle \, {\cal W}_{\cal C}\,  \right\rangle =\left\langle \, 
{\cal W}_{\cal C}^*\,  \right\rangle \,.
\label{promezh}
\eeq
The partition function \eqref{detN} is {\em exactly}
the same as that obtained in the (extended) orientifold theory
upon exploiting Eq.~(\ref{mn-two}), factorization,
\beq
{\cal Z}_{\rm orientifold}=
\sum _ {\cal C} \alpha _{\cal C}\,  \left\langle \, {\cal W}_{\cal C} (A _{\Yfund})\,  
\right\rangle \left\langle \,
{\cal W}_{\cal C} (A _{\Yfund})\,  \right\rangle  \, ,
\label{zorbi}
\eeq
and a third ingredient: independence of the expansion coefficients 
$ \alpha _{\cal C}$ of the fermion representation.

At this point we can take the square root of the determinants of the two extended
theories. Owing to the non-vanishing mass, the determinants do not vanish and no sign ambiguities arise.
In the parent theory we recover the (softly broken) super-Yang--Mills determinant, while in the  daughter theory, given the 
planar  equivalence of the symmetric and antisymmetric representations, we recover either one of them. We finally take the (supposedly) smooth massless 
limit and, thus, prove our central  result.

Let us ask ourselves: 
{\sl ``What is the critical difference between the orbifold and
orientifold theories, which makes the elevation of the 
perturbative planar equivalence possible to the non-perturbative level
in the latter case, while this does not work in the former?"} 
The answer is quite transparent.
In the orientifold theories  one can choose the very same
gauge field configurations as in \mbox{${\cal N}=1$} gluodynamics, one by one, and compare the fermion determinants in the given 
gauge field background. This is due to the fact that the gauge groups 
are the same in the parent and   daughter theories. In the orbifold theories
the gauge group is different from that of the parent
(SU$(N)^k$ vs. SU$(kN)$). The presence of the twisted sector makes such
strategy impossible.

\subsection{Special case: one-flavor QCD}
\label{sc1fqcd}

Let us focus our attention on the orientifold A daughter.
Why does it seem to be of more practical  importance than orientifold S?
Let us assume that $1/N$ corrections are manageable.
We will discuss them in more detail in Sect.~\ref{cqciofqfsg}.
Here, anticipating this discussion, it will be sufficient to assume
that nothing dramatic happens on the way from 
$N=\infty$ down to $N=3$, no phase transition, for instance. 
This is a pretty natural assumption implying
that the dynamical regime of the  orientifold A 
theory does not qualitatively change  in the descent from  
$N=\infty$ down to $N=3$.\footnote{If for some reasons beyond our grasp
there were a phase transition in $N$ {\em en route}, this  would be a
remarkable discovery by itself.} 

At $N=3$ the two-index antisymmetric Dirac field is {\em identical}
to the Dirac field in the fundamental representation. This is nothing but a standard
{\em quark} field. Thus, we arrive at a correspondence between
\None gluodynamics and one-flavor QCD! Although one-flavor QCD
is still not {\em the} theory of our world, it is a very close relative
into which  QCD theorists and lattice practitioners have a rich insight. 
One-flavor QCD is known to be a confining theory with a mass gap.
Therefore, we can reverse the direction of our consideration
and state that \None gluodynamics enjoys the same features ---
confinement and a mass gap.  Certainly, this was a general belief in the past.
The status of this statement can now be elevated:
we could turn a folklore statement into a motivated argument.

\subsection{Summary of the main results}

To conclude this key section we briefly summarize
its main results. We have shown that the partition functions of the 
orientifold field theory and \None SYM theory coincide  at large $N$.
All color-singlet correlation functions, which involve gluons as sources,
are the same in the parent and  daughter theories. This statement
is also valid for the fermion-bilinear sources that can be projected 
onto each other in both theories.
As a consequence, the glueball spectrum in the daughter theory is identical
to that  of SUSY gluodynamics.  In particular, there is a degeneracy between 
$0^\pm$ glueballs. This degeneracy of the bosonic 
spectrum extends much further: all mesons originating
from one and the same  supermultiplet in the parent theory
are inherited by the daughter theory, and their mass degeneracy
is part of this inheritance.
There is an {\em infinite number} of such ``unexpected" degeneracies
in the orientifold theories at $N=\infty$. At the same time,  
unlike \None gluodynamics, the orientifold daughter
theories have  no color-singlet fermions at $N=\infty$.
Finally, at $N=3$, the orientifold A theory is identical to one-flavor massless QCD.

\section{Non-supersymmetric parent--daughter \\
pairs}
\label{NSPDP}
\renewcommand{\theequation}{\thesection.\arabic{equation}}
\setcounter{equation}{0}

\noindent
A simple proliferation of the fermion fields in the form of ``flavors''
leads to non-supersymmetric parent--daughter pairs.
This was   mentioned in passing in Ref. \cite{gorshi}
in the context of $Z_2$ orbifolds, while
in the   orientifold field theories the issue of
non-supersymmetric planar-equivalent pairs was  raised in 
Ref.~\cite{adisvone} and analyzed
in \cite{AStwo}. For instance, gauge theories
with  the same number of Dirac fermions either in the 
antisymmetric two-index or symmetric two-index 
representations are planar-equivalent.

Here we would like to discuss in more
detail parent--daughter pairs which are obtained from 
\None gluodynamics and its orientifold A daughter
by introducing fermion replicas. Thus, as previously,
the parent and daughter theories share one and the same gauge group,
SU($N$), and one and the same gauge coupling.\footnote{As previously,
at large $N$ we disregard the distinction between SU($N$) and U($N$).}
The two respective fermion sectors are:

\vspace{1mm}
\noindent
(i) $k$ species of the Weyl fermions in the adjoint,
to be  denoted  as $\left(\lambda^A\right)^i_j\,$;

\begin{center}
and 
\end{center}

\noindent
(ii)  $k$ species of the Dirac fermions
$\left( \Psi^A\right)_{[ij]}$ or $\left( \Psi^A\right)^{[ij]}\,$. 

\vspace{1mm}

Here $i,\,j$ are (anti)fundamental indices
running $i,\,j=1,2,...,N$, while $A$ is the flavor index running
$A=1,2,..,k$. Note that $k\leq 5$. Otherwise we loose asymptotic freedom.
Each Dirac fermion is equivalent to two Weyl fermions,
$$
\Psi_{[ij]} \to \{ \eta_{[ij]}\,,\,\,\, \xi^{[ij]}\}\,.
$$

Of course, since both theories are non-supersymmetric, the predictive
power is significantly reduced with respect to the case  of
the SUSY parent.  Still, one can benefit from the comparison of 
both theories, in particular, the Nambu--Goldstone boson
sectors. Of some interest is also the comparison of the respective Skyrme
models. We will say a few words on that at the end of the section.

Let us start with the case (ii), $k$ species of 
$ \Psi_{[ij]}$. Since the fermion fields are Dirac and belong
to the complex
representation of the gauge group, the theory has the same (non-anomalous) chiral symmetry
as QCD with $k$ flavors, namely,
SU($k)_L\times\mbox{SU}(k)_R$.  Various arguments tell us \cite{mac,triangle,rm} that the pattern of the chiral symmetry breaking 
is the same as in QCD too, namely
\beq
\mbox{SU}(k)_L\times\mbox{SU}(k)_R \to \mbox{SU}(k)_V\,.
\label{pcsbone}
\eeq
The only distinction is that in QCD the constant $f$ scales as $\sqrt{N}\,\Lambda$
while in our case its scaling law is ${N}\,\Lambda$. Moreover, the 
coefficient $n$
in front of the Wess--Zumino--Novikov--Witten term $\Gamma$ (see e.g. Ref.~\cite{Witten:tw})
equals $N$ in QCD and $(1/2) N(N-1)$ in the case at hand (see below).

All axial (non-anomalous) currents are spontaneously broken, giving rise
to $k^2 -1$ Nambu--Goldstone ``mesons."
Some of them ---  those coupled to the  axial currents that can be elevated
from the daughter theory (ii)  to the parent theory (i) --- 
persist  in the parent theory (i), where the  fermion fields belong to the real
representation. In these channels the couplings of the
``parent" Nambu--Goldstone bosons to the corresponding axial currents
are  the same as in the daughter theory. This is
because of the planar equivalence of two theories in the common sector.

A crucial point is that not all axial currents can be elevated 
from the daughter to the parent theory.
Remember, say, that the currents with the structure $\bar\xi \xi -\bar\eta \eta$
cannot be elevated from (ii) to (i). Therefore, in comparing
these two theories, one must identify such currents and exclude them
from the parent theory.

It is not difficult to count the number of the axial currents that 
are elevated from (ii) to (i): there are
$(1/2)k(k-1)$ off-diagonal currents of the type 
$\bar\xi^A \xi^B +\bar\eta^A \eta^B$
($A\neq B$) plus  $(k-1)$ diagonal axial currents of the type
$\sum c_A (\bar\xi^A \xi^A +\bar\eta^A \eta^A)$, with $\sum c_A = 0$ 
and $C_1=1$. Altogether, we get 
$$
\frac{k(k+1)}{2}-1
$$
Nambu--Goldstone bosons. This  corresponds to the following pattern
of  chiral symmetry breaking in the parent theory:
\beq
\mbox{SU}(k) \to \mbox{SO}(k)\,,
\label{pcstwo}
\eeq
with the Nambu-Goldstone bosons in the symmetric and traceless two-index representation of $ \mbox{SO}(k)$. 

The pattern of the chiral symmetry breaking for the quarks
belonging to   real representations of the gauge group
indicated in Eq.~(\ref{pcstwo}) was advocated many times
in the literature \cite{mac,triangle,rm,Witten:tx}, but no 
complete proof was ever given.

We conclude this section by a brief comment on the topological properties
of the  chiral Lagrangian, corresponding to
the above patterns of the chiral symmetry breaking,
and how they match the underlying gauge field theory expectations.
The theory (i) is  confining
and supports flux tubes --- fundamental color charges cannot be screened.
On the other hand, we do not expect stable baryons with mass growing with $N$. Color-singlet states composed of   gluons and 
$\left(\lambda^A\right)^i_j$ form baryons with $M\sim N^0$.

At the same time, the theory (ii) does {\em not} have baryons with
$M\sim N^0$. Here the baryon masses grow with $N$.
The theory is expected to be confining too,
but two flux tubes  (each attached to a color source in the 
fundamental representation) can be screened\,\footnote{
Although two fundamental strings put together can break
via creation of a pair  $\Psi_{[ij]}+ \overline{\Psi_{[ij]}}$,
the breaking is suppressed by $1/N$ factors and does not occur at
$N=\infty$. For a recent review see \cite{ASRem}.}   
by $\left( \Psi^A\right)_{[ij]}$.

As was suggested in Refs.~\cite{Witten:tw,Witten:tx},
at large $N$ one can try to identify baryons with the 
Skyrmions supported by the corresponding chiral Lagrangians.
Since
\begin{eqnarray}
&&\pi_3\left\{\mbox{SU}(k)/\mbox{O}(k)
\right\} = Z_4\,\,\,\mbox{at}\,\,\, k=3, \quad \pi_3\left\{\mbox{SU}(k)/\mbox{O}(k)
\right\} = Z_2\,\,\,\mbox{at}\,\,\, k\geq 4;\nonumber\\[2mm]
&&\pi_3\left\{\mbox{SU}(k)
\right\} = Z\,\,\,\mbox{at all}\,\,\, k 
\end{eqnarray}
both theories, (i) and (ii),  yield Skyrmions
with $M_{\rm Skyrme}\sim N^2$, albeit the theory (ii) has a richer spectrum.
The above scaling law,
$M_{\rm Skyrme}\sim N^2$,  is due to the fact that
$f$ scales as $ N$ in the theories under consideration.

Skyrmion statistics is determined by
the (quantized) factor in front of the Wess--Zumino--Novikov--Witten term,
\beq
(-1)^{N(N-1)/2}\,.
\label{stat}
\eeq
It has half-integer spin provided that $N(N-1)/2$ is odd,
i.e. $N= 4p+2$ or $N= 4p+3$ where $p$ is an integer.
In both cases one can construct, in the microscopic theory,
interpolating baryon currents with an odd number of constituents
scaling as $N$. Why then does the Skyrmion mass scale as $N^2$?

A possible explanation is as follows.
For quarks in the  fundamental representation of
SU($N$) the color wave function is antisymmetric,
which allows all of them  to be in an $S$ wave in  coordinate space.
With antisymmetric two-index spinor fields,
the color wave function is symmetric, which would require them to 
occupy orbits with angular momentum up to $\sim N$.
Then the scaling law $M_{\rm Skyrme}\sim N^2$ seems natural.

Since $\pi_2\left\{\mbox{SU}(k)
\right\} = 0$ the chiral sector of the theory (ii) does not support flux tubes.
Albeit  disappointing, such a situation was anticipated by Witten
\cite{Witten:tx} who noted that the topology of the full space 
of the large-$N$ theory need not coincide with the topology of 
its Nambu--Goldstone sector.

In the light of this remark we can understand the complete failure
of the Skyrmion description of theory (i). In particular,
since $\pi_2\left\{\mbox{SU}(k)/\mbox{O}(k)
\right\}$ $ = Z_2$,  at $ k\geq 3$
we have flux tubes in the chiral theory, while we do not expect them
in the microscopic theory. Moreover, stable Skyrmions of 
the chiral sector should become unstable in the full
theory.

It remains to be seen whether meaningful predictions
regarding QCD baryons can be obtained from the large-$N$ limit
of the orientifold theory with two or three flavors.

\section{Orientifold large-$N$ expansion}
\label{oriln}
\renewcommand{\theequation}{\thesection.\arabic{equation}}
\setcounter{equation}{0}

\subsection{General features ($N_f>1$ fixed, $N$ large)}
\label{genf}

We will now abandon for a while the topic
of  planar equivalence, and look at the SU$(N)$
orientifold theories with $N_f$ flavors
discussed in Sect.~\ref{NSPDP}
from a more general perspective. We will focus on the antisymmetric
orientifold theories, assuming that $N_f$ does not scale with $N$
at large $N$, say, $N_f=1$, 2 or 3. We will return to the issue of planar equivalence later.

As was noted in Sect.~\ref{sc1fqcd},
if $N=3$ (i.e. if the gauge group is SU(3))
the two-index antisymmetric quark is identical to
the standard quark in the fundamental representation.
Therefore, it is quite obvious that extrapolation to large $N$,
with the subsequent $1/N$ expansion, can have distinct starting points:
(i) quarks in the {\em fundamental} representation;
(ii) quarks in the {\em two-index antisymmetric} representation;
(iii) a combination thereof.
The first option gives rise to the standard 't Hooft
$1/N$ expansion \cite{thooft1n,'tHooft:1973jz},
while the second and third lead to a new expansion \cite{adisvtwo,adisvtwop}, to
which we will refer  as the {\em orientifold large-$N$ 
expansion}.\footnote{It is curious  that Corrigan and Ramond
suggested \cite{corramond} to replace the 't Hooft model
by a model with one two-index antisymmetric quark $\Psi_{[ij]}$
and two fundamental ones $q_{1,2}^i$, as early as 1979. 
Their motivation originated from some awkwardness in the treatment
of baryons in the 't Hooft model, where all baryons, being composed of
$N$ quarks, have masses scaling as $N$ and thus
disappear from the spectrum at $N\to\infty$.
If the fermion sector contains $\Psi_{[ij]}$ and $q_{1,2}^i$,
then, even at large $N$, there are three-quark baryons of
the type $\Psi_{[ij]}\, q_{1}^i\, q_{2}^j$, which 
apparently realize a smoother extrapolation from $N=3$ to large $N$
than the one  in  't Hooft's model.}

The 't Hooft expansion is the simplest and the oldest $1/N$
expansion to have been designed specifically for QCD.
The 't Hooft limit assumes $N$ to be large, while keeping the
't Hooft coupling (\ref{thocoup})
fixed. If  the number of quark flavors
is fixed too (i.e. does not scale with $N$)
then each quark loop is suppressed by $1/N$.  Only quenched planar
diagrams (i.e. those with no quark loops)
survive in leading order in $1/N$.
Non-planar diagrams with ``handles'' are suppressed by $1/N^2$ per
handle. Each extra quark-loop insertion
in any planar graph costs $(N_f/N)$.  Thus, the corrections to the leading approximation run in powers of $(N_f/N)$ and $1/N^2$. 

The 't Hooft expansion enjoyed a  significant
success in phenomenology. It provided a qualitative explanation
for the well-known regularities of the hadronic world,
first and foremost the Zweig rule, a relative smallness of the 
meson widths, the rarity of the four-quark mesons and so on.
We hasten to add, though, that, except in  two dimensions
($D=2$), nobody  succeeded in getting  {\em quantitative} 
results for QCD, even to leading order in the $1/N$ expansion.

Although the standard large-$N$ ideology definitely
captures some basic regularities, it gives rise to certain puzzles,
as far as subtle details are concerned.
Indeed, in the  't Hooft expansion, 
the width of $q \bar{q}$ mesons scales as $1/N$,
while that of glueballs scales as $1/N^2$. In other words, 
the latter are expected to be narrower than quarkonia,
which is hardly the case in reality. No
glueballs were reliably  identified so far, in spite of decades of searches.

Moreover, the Zweig rule is not universally valid.
It is known to be badly violated for scalar and pseudoscalar mesons.
Another example of this type, thoroughly discussed in Ref.~\cite{Novikov:xj}, 
refers to the quark dependence of the vacuum energy. 
In the 't Hooft limit the vacuum energy density
is obviously independent of the quark mass, since all quark loops
die out. At the same time, an estimate based on QCD low-energy
theorems tells us \cite{Novikov:xj} that changing
the strange-quark mass from $\sim 150$ MeV to zero 
would roughly double the value of the vacuum energy density.

It is clear that the 't Hooft expansion underestimates the role of quarks.
This was noted long ago, and a remedy was suggested
\cite{TE}, a {\em topological expansion}.
The topological expansion (TE) assumes that the number of flavors
$N_f$ scales as $N$ in the large-$N$ limit, so that the ratio
$N_f/N$ is kept fixed.

The graphs that survive in the leading
order of  TE are {\em all} planar diagrams, including 
those with the quark loops. This is easily seen by  slightly  
modifying  \cite{TE} the  't Hooft  double-line notation --- adding
a flavor line  to the single color line for quarks. In the leading 
(planar) diagrams the quark loops are ``empty" inside, since 
gluons do not attach to the  flavor line.
Needless to say,  obtaining analytic results in   TE is even harder  
than in the  't Hooft case.

The orientifold large-$N$ expansion opens the way for a novel
and potentially rich large-$N$ phenomenology in which the quark loops
(i.e.  dynamical quarks) do play a non-negligible role.
An additional bonus is that in the orientifold large-$N$
expansion,  one-flavor QCD gets 
connected to  supersymmetric gluodynamics, potentially
paving the way to  a wealth of predictions.

\begin{figure}
\begin{center}
\mbox{\kern-0.5cm
\epsfig{file=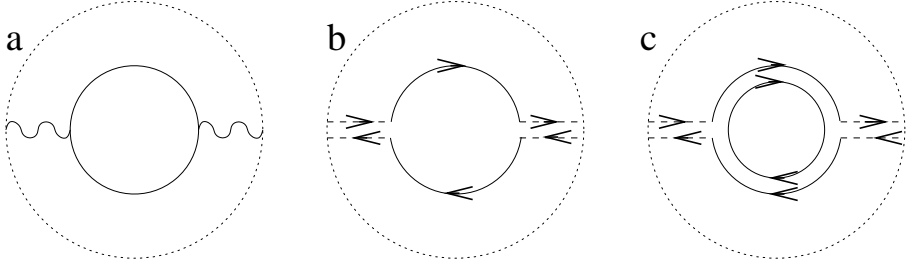,width=12.0true cm,angle=0}}
\end{center}
\caption{(a.) A typical contribution to the vacuum energy.
(b.) The planar contribution in 't Hooft large-$N$ expansion. (c.) The
orientifold large-$N$ expansion. The dotted
circle represents a sphere, so that every line hitting the dotted
circle gets connected ``on the other side."}
\label{largeN}
\end{figure}

In order to demonstrate the difference between the standard large-$N$
expansion and the orientifold large $N$ expansion we exhibit a planar
contribution to the vacuum energy in two ways in Fig.~\ref{largeN}.
Moreover, we will 
illustrate the usefulness of the orientifold large-$N$ expansion at the qualitative, semiquantitative and quantitative
levels by a few examples given below.

Probably, the most notable distinctions from the
't Hooft expansion are as follows:
(i) the decay widths of {\em both}
glueballs and quarkonia scale with $N$
in a similar manner, as $1/N^2$; this can easily be deduced
by analyzing appropriate diagrams with the quark loops
of  the type displayed in Fig.~\ref{largeN}b;
(ii) unquenching quarks in the vacuum
produces an effect that is not suppressed by $1/N$;
in particular, the vacuum energy density does depend on the quark masses
in the leading order in $1/N$.

\subsection{Qualitative results for one-flavor QCD from the  
orientifold expansion}  

In this subsection we  list some predictions
for one-flavor QCD, keeping in mind that they are expected to be
valid up to corrections of the order of $1/N =1/3$ (barring
large numerical coefficients):

(i) Confinement with a mass gap is a common feature
of one-flavor QCD and SUSY gluodynamics.

(ii) Degeneracy in the color-singlet bosonic
spectrum. Even/odd parity mesons (typically mixtures of fermionic and 
gluonic color-singlet  states) are expected to be degenerate. In particular,
\begin{equation}
{m^2 _ {\eta'} \over m^2 _ \sigma}=1 + O(1/N)
\,, \qquad\mbox{one-flavor QCD}\,,
\label{etasigma}
\end{equation}
where $\eta '$ and $\sigma$ stand for the lightest $0^-$ and $0^+$
mesons, respectively.
This follows from the exact degeneracy in \None SYM theory. Note that the
$\sigma$ meson is stable in this theory, as there are no light pions for it to decay into.

The prediction \eqref{etasigma} should be taken with care:
a rather large numerical coefficient in front of $1/N$ is not at all ruled out, since  the $\eta'$ mass is  anomaly-driven (the Witten--Veneziano (WV)
formula  \cite{Witten:1979vv,Veneziano:1979ec}), whereas the $\sigma$
mass is more ``dynamical.'' 

In more general terms the mass degeneracy is inherited
by all those ``daughter" mesons that fall into one and the
same supermultiplet in the parent theory.
The accuracy of the spectral
degeneracy is expected to  improve at higher
levels of the  Regge trajectories, as $1/N$ corrections
that induce splittings are expected to fall off (see e.g. \cite{TE}).

(iii) Bifermion condensate. \None  SYM theory has a
bifermion condensate. Similarly we predict a condensate in one-flavor
QCD. A detailed calculation is discussed in Sect.~\ref{cqciofqfsg}.

(iv) One can try to get an idea of the size of
$1/N$ corrections  from
perturbative arguments. In  one-flavor QCD
the first coefficient of the $\beta$ function is $b= 31/3$,
while in  adjoint QCD with $N_f=1$ it becomes $b= 27/3$.
One can go beyond one loop too. As was mentioned, in the very same 
approximation the $\beta$ function of the one-flavor QCD coincides
with the exact  NSVZ $\beta$ function,
\begin{equation}
\beta = -{1\over 2\pi}\,  {{9 \alpha ^2} \over {1- 3\alpha / 2\pi}}\, .
\end{equation}
Thus, for the (relative) value of the two-loop $\beta$-function
coefficient, we predict $+3 \alpha / 2\pi$, to be compared
with  the exact value in one-flavor QCD,
$$
+ \frac{134}{31}\frac{ \alpha }{2\pi} \approx 4.32 \,\frac{\alpha }
{ 2\pi}\,.
$$

We see that the orientifold large-$N$ expansion somewhat  overemphasizes
the quark-loop contributions, and, thus, makes the theory less 
asymptotically free than in reality--- the  opposite of what happens in 't Hooft's  expansion. Parametrically, the error is  $1/N$ rather than  $1/N^2$. 
This is because there are $N^2-1$ gluons and $N^2-N$ fermions in the 
orientifold  field theory.

If one-flavor QCD  is related, through the planar
equivalence,  to SYM theory, while the 't Hooft $1/N$
expansion connects it with   pure Yang--Mills theory,
it is tempting to combine both to ask a seemingly
heretical  question:
``Are there any traces of supersymmetry in {\em  pure Yang--Mills theory}?"
We try to answer this question in Appendix A,
which presents some evidence for  a ``residual SUSY" in pure Yang--Mills theory.

\subsection{An estimate of the conformal window in QCD from the 
orientifold large-$N$ expansion}

One can try to use the orientifold large-$N$ expansion 
for an estimate of the position of the {\em lower} edge
of the conformal window. The logic behind this analysis is as follows.
Since this expansion overestimates the role of the quark loops,
in leading order in $1/N$ one can expect to obtain a lower boundary on the onset of the conformal regime, which, hopefully,
will be close to the actual lower edge of the conformal window. 
The 't Hooft large-$N$ limit \cite{'tHooft:1973jz}
is not suitable for this purpose at all, since in this limit
the quark effects disappear altogether. 

We  start from a brief summary of the
issue of the conformal window in QCD. 
The standard definition of the coefficients
of the $\beta$ function  from the Particle Data Group (PDG) 
is\,\footnote{With respect to the  PDG, we dropped a factor 2 in the middle equality, $2 \beta (\alpha)_{\rm PDG} \to \beta (\alpha)$,
which is neither conventional nor convenient.} 
\beq
\mu\,\frac{\partial\alpha}{\partial \mu}\equiv \beta (\alpha)
=-\frac{\beta_0}{2\pi } \alpha^2 -\frac{\beta_1}{4\pi^2 } \alpha^3 +...
\label{gmlf2}
\eeq
(see Ref. \cite{Hinchliffe}).
The first two  coefficients of the
Gell-Mann--Low function in QCD have the form \cite{Hinchliffe}
\beq
\beta_0 = 11 -\frac{2}{3} N_f\, , \,\,\, \beta_1 = 51 -\frac{19}{3} 
N_f\,  ,
\label{betaQCD}
\eeq
where $N_f$ is the number of  quark flavors, all taken to be massless.
At small $\alpha_s$ the $\beta$ function is negative (asymptotic
freedom!) since the $O(\alpha_s^2)$ term dominates. With the scale 
$\mu$ decreasing the running gauge coupling constant grows,
and eventually the second term becomes important. The second term 
takes over the first one at $\alpha_s/\pi \sim 1$, when all terms in 
the $\alpha_s$ expansion are equally important, i.e. in the strong 
coupling regime. We have no firm knowledge regarding
what happens at $\alpha_s/\pi \sim 1$.
 
Assume, however, that for some reason the first 
coefficient $\beta_0$ is abnormally small, and this smallness does 
not propagate to higher orders. Then the second term catches up with 
the first one when $\alpha_s/\pi \ll 1$: we are at weak coupling, 
and higher-order terms are inessential. Inspection of 
Eq.~(\ref{betaQCD}) shows that this happens when $N_f$ is close
to $33/2$, say 16 or 15 ($N_f$ has to be less than $33/2$ to ensure 
asymptotic freedom). For these values of $N_f$ the second coefficient
$\beta_1$ turns out to be negative!
 
This means that the $\beta$ 
function develops a zero at weak coupling, at
\beq
\frac{\alpha_{*}}{2\pi} = \frac{\beta_0}{-\beta_1} \ll 1\, .
\label{yaus}
\eeq 
For instance, if  $N_f = 15$ or 16 the critical values are at 
\beq
\lambda_*\equiv\frac{N\alpha_*}{2\pi }=\frac{3}{44}\,,\quad
N_f = 15;\qquad  \lambda_* = \frac{3}{151}\,,\quad
N_f = 16 ,
\eeq 
where in the expression above $N=3$.
At such values of $\alpha$, the Gell-Mann--Low function 
vanishes, i.e. $\beta(\alpha_*)=0$.
This zero is nothing but the infrared fixed point of the theory. 
At large distances $\alpha_s \to \alpha_{s}^{*}$,
and $\beta (\alpha_{s}^{*}) = 0$, implying that the trace of the 
energy-momentum vanishes. Then the theory is in the conformal 
regime. There are no localized particle-like states in the spectrum of
this theory; rather we deal with massless unconfined interacting 
``quarks and gluons."  All correlation functions at large distances
exhibit a power-like behavior. In particular, the potential between
two heavy static quarks at large distances $R$ will behave as
$\sim \alpha_{s}^{*}/R$,  a pure Coulomb behavior. The 
situation is not drastically different from conventional QED.
As long as $\alpha_{s}^{*}$
is small, the interaction of the massless quarks and gluons
in the theory is weak at all distances, short and large, and is 
amenable to the standard perturbative treatment (renormalization 
group, etc.). Chiral symmetry 
is not broken spontaneously,  quark condensates do not develop, and
QCD becomes a fully calculable theory.
The fact that, at $N_f$ close to 16, QCD becomes
conformal and weakly coupled in the infrared limit has been known
for about 30 years \cite{MB,BZ}. If the conformal regime takes place
 at $N_f= 16 $
and 15,  let us ask ourselves: How far can we descend in $N_f$
without ruining the conformal nature of the theory in the infrared?
Certainly, when $N_f$ is not close to 16, the gauge coupling 
$\alpha_{s}^{*}$ is not small. The theory is {\em strongly coupled},
like conventional QCD, and, simultaneously, it is
conformal in the infrared. Presumably, chiral symmetry  
is unbroken,  $\langle \bar qq\rangle = 0$. The vacuum structure is 
totally different, and so are all properties of the theory.

The range of $N_f$ where this phenomenon occurs is called
the {\em conformal window}. The right edge of the conformal 
window is at $N_{f<}= 16 $. According to our  conjecture,
the orientifold large-$N$ expansion allows one to get an idea of
the left edge of the conformal window, $N_{f>}$, which 
happens to lie not too far from actual QCD 
(i.e. not too far from $N_f=3$)!

For antisymmetric quark ``flavors"
\begin{eqnarray}
\beta_0 &=&\frac{11}{3}N-\frac{2}{3}NN_f+\frac{4}{3}N_f\,,\\[3mm]
\beta_1&=&\frac{17}{3}N^2-\frac{5}{3}N(N-2)N_f-\frac{(N-2)^2(N+1)}{N}N_f\,.
\end{eqnarray}
We will consider the limit $N\to\infty$. At $N_f=6$, asymptotic freedom is lost.
At $N_f=5$ we have a conformal infrared fixed point with
\beq
\lambda_*\equiv\frac{N\alpha_*}{2\pi }=\frac{1}{23}\,.
\eeq
The above assertion seems to be solid, since 
$\lambda_*$ is small enough, cf. Eq.~(\ref{yaus}).
At $N_f<5$ the position of the would-be infrared fixed point
shifts to rather large values of $\lambda_*$ (namely, $\lambda_*=1/5$),
and is thus unreliable.

Interpolating down to the actual value of
the number of quark colors, $N=3$, in the most naive manner,
we would conclude that the left edge of the conformal window is
around $N_f=5$. 

How does this compare with other theoretical
and/or numerical data?
First, in the instanton liquid model it was found 
\cite{CW1} that  the chiral condensate disappears at $N_f=5$,
$\langle \bar qq\rangle = 0$. Although nothing is said about the 
onset of the conformal regime in this work, the vanishing
of $\langle \bar qq\rangle $ may be a signal. 
It is believed that color confinement  in QCD implies 
chiral symmetry breaking \cite{Casher}. Another evidence in favor 
of $N_f=5$ arises from the condition 
$\gamma(N_f)=1$ \cite{Grunberg:2000ap}.

More definite is the conclusion of  lattice investigations
\cite{CW2}. It is claimed that at $N_f=7$ not only does the quark
condensate disappear, but the decay law of 
correlation functions at large distances changes from  exponential
to power-like, i.e.  the theory switches from the confining regime 
to the conformal one.
 
Of course, at $N_f=3$ we know from experiment that QCD is in the
confining rather than conformal regime. From the above argument, it
looks likely that going up  from $N_f=3$ to $N_f=5$ or so ---  quite a modest 
variation in the number of massless quarks --- 
one  drastically changes the vacuum structure of the theory,
with the corresponding abrupt change of the  picture of the hadron world.
Let us emphasize again that this phenomenon would be totally
missed by 't Hooft's  expansion.

An interesting question is what happens slightly below the left edge 
of the conformal window, i.e. say at  $N_{f>}-1$. If $N_f$ were a continuous 
parameter, and the phase transition in $N_f$ were of the second 
order, slightly below $N_{f>}$ the string tension $\sigma$ would be 
parametrically small in its natural scale given by $\Lambda^2$. In reality
$N_f$ changes discretely. Still it may well happen that
at $N_f = N_{f>}-1$ the ratio $\sigma/\Lambda^2$
is numerically small. This would mean that the string and its 
excitations, whose scale is set by $\sigma$, are abnormally light.
Under the circumstances one might hope to build
an effective low-energy approach  analogous
to the  chiral Lagrangians of actual QCD. In the limit 
$\sigma/\Lambda^2 \to 0$ the string  becomes, in a sense, classic;
quantum  corrections are unimportant. We do not rule out that the string
representation of QCD --- the Holy Grail of two generations of
theorists --- is easier to construct in this limit. One could even
dream of an expansion in $N_{f>}-3$. At this moment,
this is  pure speculation, though.

\subsection{Orientifold field theories in two and three dimensions}
\label{oftittd}

The large-$N$ equivalence of gauge theories with adjoint
matter and anti-symmetric (or symmetric)
matter is not restricted to four dimensions.
The equivalence holds in any number of dimensions.

Let us consider first the two-dimensional case. We assert that planar 
two-dimensional Yang--Mills theory with fermions in the adjoint representation (this
theory is {\em not} supersymmetric) is equivalent to planar
two-dimensional QCD with fermions in the antisymmetric representation. 

When the fermions are massless, the equivalence of the two theories can
be obtained (besides our derivation, see Sect.~\ref{norieq})
from the universality theorem of Kutasov and Schwimmer
\cite{Kutasov:1994xq}. The theorem states that  physics of massless 
two-dimensional QCD
depends on the total chiral anomaly, but not
on the specific representation of the fermions. The anomaly in two dimensions
is dictated by the vacuum polarization diagram, ${\rm Tr}\, T^a T^b =T(R)
\delta ^{ab}$. For the adjoint representation, $T=N$, whereas for
the antisymmetric representation $T=N-2$. At large $N$ the two
anomalies coincide and, hence, so do  physics of the two theories.

The large-$N$ equivalence extends further; for instance, it takes place for
$m\neq 0$, where $m$ is the fermion mass term. In this
case we have two dimensionful parameters, $m$ and $g^2$,
and one dimensionless, $N$. The most interesting physical question one can ask is 
the rate of convergence of the $1/N$ expansion.

Of course, this question is of great interest in four dimensions too.
It seems that it will be much easier and faster to answer it first in two
(or, perhaps, three) dimensions. Indeed, two-dimensional Yang-Mills theory with fermions in  the adjoint representation can be readily studied
on lattices, by methods of discrete light-cone
quantization (DLCQ) and, to some extent, analytically.
In fact, it was studied quite extensively 
in this way in the 1990's, see 
Refs.~\cite{Demeterfi:1993rs}--\cite{Armoni:1998kv}.
Now these analyses can be repeated with two-index antisymmetric (or symmetric) fermions,
with the aim of studying  the rate of convergence of the
orientifold expansion.

Also very promising is three-dimensional \None Yang--Mills
theory. In this case the parent theory (with  adjoint fermions)
is supersymmetric, provided one deals with 
three-dimensional Majorana fermion fields.\footnote{Introducing
adjoint Dirac fermion fields would render the theory at hand
non-supersymmetric. Even so, 
studying the parent--daughter planar equivalence in this case
is likely to be rewarding. This case is similar to the
``flavor proliferation" treated in Sect.~\ref{NSPDP}.}
In three dimensions this theory admits a Chern--Simons term,
which generates a mass for the gauge fields, and a counterpart for the fermions.
A remarkable feature of this theory is the
occurrence of a phase transition at a certain critical value of the
Chern--Simons coefficient, and the possibility of an
analytic study of this phase transition \cite{Witten:1999ds}.
It is not ruled out that consequences of  the large-$N$ equivalence 
between the three-dimensional  SYM
theory and the orientifold theory could be used in 
analytical  studies of confinement in  a $T\neq 0$ regime 
of four-dimensional  one-flavor QCD ($T$ standing here for the temperature).

\subsection{$1/N$ corrections}
\label{onencor}

The statement that large-$N$ Yang--Mills theories
with adjoint and two-index fermions
are equivalent both perturbatively and non-perturbatively
will become a critical advance, only provided that
we will be able to get an idea of $1/N$ corrections.
Indeed, our final goal, after all,
is understanding nature; and in nature $N=3$.
No matter how good toy models are for various explorations,
they acquire a real value only as long as they 
properly reflect  the pattern of actual QCD.

From this standpoint, the orientifold large-$N$ expansion
is far from being optimal, since, as was mentioned,
corrections to the planar limit scale as $1/N$ rather than
$1/N^2$; they can thus be large, leading to drastic deviations
from the predictions obtained on the basis of planar equivalence. 
In fact, as we will see below (Sect.~\ref{cqciofqfsg}), in some problems
one can encounter corrections of the type
$1-(2/N)$. Further progress will depend on whether we 
will be able to handle them.

We would like to make a  remark  concerning this aspect.
Our first observation will be useful for lattice-QCD practitioners.
On the lattices it does not make much difference whether one simulates
fermions in the antisymmetric, symmetric or reducible
two-index representation (i.e. no (anti)sym\-metrization at all).
Let us focus on the latter case.
The number of flavors need not be equal to 1;
it can be, say, 2 or 4. The important thing is that 
these numbers  match.
Say, considering two Majorana adjoint fermions in the
parent theory requires one Dirac $\Psi^{ij}$ (no (anti)symmetrization)
in the daughter one. This is the sum of two irreducible representations
A and S. 

Assume that we have certain information regarding
the mass of a meson with the quantum numbers
$0^-$ in the parent theory. Then, on the lattice, one can measure
the mass of its counterpart in the orientifold theory described above.
Our assertion is: in 
such a study the deviation  from the adjoint theory will be
$O(1/N^2)$ rather than $O(1/N)\, $!

Another idea of how to minimize the $1/N$ corrections is to add extra
fermions in the fundamental representation to the fermion in the antisymmetric
representation. At least perturbatively, it is obvious that an adjoint
fermion (the gluino) is better approximated at finite $N$ by an
antisymmetric fermion and two additional fundamental fermions. For
example, the one-loop $\beta$-function coefficient of the latter theory is $3N$,
exactly as the one-loop $\beta$ function of \None SUSY Yang--Mills theory. 
The similarity of the two theories at finite $N$ and the potential
 phenomenological applications will be further discussed in Sect.~\ref{v1nc}. 

In addition, some insight about $1/N$ corrections can be obtained from
string theory, where this problem translates into the understanding of
string loops corrections.

Work on other methods of minimizing (or suppressing)
$1/N$ corrections in the transition from the parent theories
to orientifold daughters is in progress.

\section{Calculating the quark condensate in one-flavor QCD
from supersymmetric gluodynamics}
\label{cqciofqfsg}
\renewcommand{\theequation}{\thesection.\arabic{equation}}
\setcounter{equation}{0}

{\em Estimates presented in this section are arranged in a somewhat
different manner than in the
original work \cite{adisvtwop}. The final result is the same.}

\vspace{3mm}

The gluino condensate in SU($N$) SYM theory
is presented in Eq.~(\ref{gluco}). Let us reproduce it here again, for 
convenience,
\beq
\langle \lambda^{a}_{\alpha}\lambda^{a\,,\alpha}
\rangle = -6 N\Lambda^3\,,
\label{dopgluco}
\eeq 
where we set $\theta=0$, $k=0$, and the parameter 
$\Lambda$ in Eq.~(\ref{dopgluco})  refers to the Pauli-Villars scheme.
As is explained in detail in Appendix B, $\Lambda_{\rm Pauli-Villars}$  
identically coincides in supersymmetric theories
with the scale parameter defined in the dimensional-reduction scheme, 
DRED +$\overline{\rm MS}$
procedure. In perturbative calculations one usually
uses dimensional regularization rather than reduction
(supplemented by $\overline{\rm MS}$).
The difference between these two $\Lambda$'s is insignificant
--- it amounts to a modest factor $\exp(1/12)$, see  Appendix B --- and we will disregard it hereafter.

It is not difficult to show\,\footnote{
One can establish
the mapping of $\left\langle 
\lambda^{a\,,\alpha} \lambda^a_{\alpha}  \right \rangle $
onto $ \langle \bar\Psi_{[ij]} \Psi^{[ij]} \rangle$ as follows.
The simplest way is the comparison of the
corresponding  mass terms. The Dirac fermion $\Psi$  of the
orientifold theory can be replaced by two Weyl spinors, $\xi_{[ij]}$
and $\eta^{[ij]}$, so that the fermion mass term becomes:
$$
m\bar\Psi\Psi = m\xi\eta + {\rm h.c.}\; ,
$$
while in softly broken SYM the mass term has the form
$$
\frac{m}{2}\, \lambda\lambda + {\rm h.c.}
$$
Thus,
$$
\frac{1}{2} \langle \lambda\lambda\rangle \leftrightarrow 
\langle \xi\eta \rangle\,,\quad\mbox{or}\quad \langle \lambda\lambda\rangle 
\leftrightarrow \langle \bar\Psi\Psi \rangle
\quad\mbox{at}\quad \theta =0\,.
$$
The same identification is obtained from a comparison
of the two-point functions in the scalar and/or pseudoscalar channels in
both theories.}
that the correspondence between 
the bifermion operators is as follows:
\beq
\langle \lambda^{a}_{\alpha}\lambda^{a\,,\alpha}
\rangle  \leftrightarrow \langle \bar\Psi\,\Psi
\rangle \,.
\label{projectPD}
\eeq
The left-hand side is in the parent theory,
the right-hand side is in the orientifold theory A,
and they project onto each other with the unit coefficient.
It is worth emphasizing  that Eq.~(\ref{projectPD})
assumes that $\langle \lambda^{a}_{\alpha}\lambda^{a\,,\alpha}
\rangle$ is real and negative in the vacuum under consideration
(which amounts to a particular choice of vacuum).
It also assumes that the gluino kinetic term is normalized
non-canonically, as in Eq.~(\ref{cnormprim}).

Thus, the planar equivalence gives us a prediction for 
the quark condensate in the one-flavor orientifold theory
at $N=\infty$, for free. 
Our purpose here is to go further, and to estimate
the quark condensate at $N=3$, i.e. in
one-flavor QCD. To have an idea of the value
of $1/N$ corrections, as a first step,  one can examine the first
two coefficients in the $\beta$ function and the anomalous dimension
of the matter field in three distinct  gauge theories, presented in Table~\ref{TTT}. We confront pure Yang-Mills  theory (YM),  
QCD with one Dirac fermion in the fundamental representation 
(1f-QCD), orientifold theory A with one flavor 
(Orientifold A), and, finally, SYM theory --- all with the gauge group SU($N$).

\begin{table}
\begin{center}
\begin{tabular}{|c|c|c|c|c|}
\hline
$\frac{\mbox{ Theory} \to  }{\mbox{  Coefficients} \downarrow} $ \rule{0cm}{0.7cm} &YM& 1f-QCD &  Orienti A  & SYM \\[3mm]
\hline\hline
\rule{0cm}{0.7cm}$\beta_0$  & $\frac{11}{3}\, N$ & $\frac{11}{3}\, N-\frac{2}{3}$&$3N+\frac{4}{3}$ &$3N$\\[2mm]
\hline   
\rule{0cm}{0.7cm}$\beta_1$ &$\frac{17}{3}\, N^2$ & $\frac{17}{3}N^2-\frac{13}{6}N+\frac{1}{2N}$&
$3N^2 +\frac{19}{3}N-\frac{4}{N}$ &$3N^2$\\[2mm]
\hline
\rule{0cm}{0.7cm}$\gamma$ & $\star$   & $\frac{3(N^2-1)}{2N}$&$\frac{3(N-2)(N+1)}{N}$ &$3N$\\[2mm]
\hline
\end{tabular}
\end{center}
\caption{Comparison of the anomalous dimensions and the first
two coefficients of the $\beta$ function. The notation is explained in the text.}
\label{TTT}
\end{table}

We use the standard definition of the coefficients
of the $\beta$ function from PDG, see Eq.~(\ref{gmlf2}).
The coefficients can be found in \cite{Tarasov:1980kx}, where formulae 
up to three loops are given. The  anomalous dimension $\gamma$
of the fermion bilinear operators $\bar\Psi\Psi$ is normalized in such a way that
\beq
\left(\bar\Psi\Psi\right)_Q =
\kappa^{\gamma/\beta_0}\,\left(\bar\Psi\Psi\right)_\mu\,,\qquad \kappa
\equiv \frac{\alpha (\mu)}{\alpha (Q)}\,, \label{mRGI}
\eeq
and $\mu$ and $Q$ denote the normalization points. For our present 
purposes  we can limit ourselves to the two-loop
$\beta$ functions and the one-loop anomalous dimensions. We can 
easily check that the various coefficients of the
orientifold theory A go smoothly from those of YM ($N=2$) through
those of 1f-QCD ($N=3$) to those of SYM theory 
at $N \rightarrow \infty$. Note, however, that some corrections
are as large as $\sim 2/N$. This is an alarming signal.

The gluino condensate $\langle \lambda^{a}_{\alpha}\lambda^{a\,,\alpha}
\rangle$ is renormalization-group-invariant (RGI) at any $N$, and so is $\Lambda$. This is not the case for the quark condensate in the non-SUSY
daughter (i.e. the orientifold theory) at finite $N$.
Unlike SUSY theories, where it is customary to normalize
the gluino kinetic term as 
\beq
\frac{1}{g^2}\,\bar\lambda\, i \slsh\! D \lambda\,,
\label{cnormprim}
\eeq
the standard normalization 
of the fermion kinetic term  in non-SUSY theories is canonic,
\beq
\bar\Psi\, i \slsh\! D \Psi\,.
\label{cnorm}
\eeq
Hence, in fact, the correspondence between operators is
\beq
\langle \lambda^{a}_{\alpha}\lambda^{a\,,\alpha}
\rangle  \leftrightarrow \langle g^2 \bar\Psi\,\Psi
\rangle \,.
\label{projectPD1}
\eeq
In the canonic normalization (\ref{cnorm}),  the RGI combination
is
\beq
\left(g^2\right)^{\gamma/\beta_0}\, \bar\Psi\Psi \equiv\left(g^2\right)^{1-\delta (N)}\, \bar\Psi\Psi\,,
\label{rgic}
\eeq
where
\beq
\frac{\gamma}{\beta_0} =\frac{\left(1-\frac{2}{N}\right)\left(1+\frac{1}{N}\right)}{1+\frac{4}{9N}}
\label{odin}
\eeq
and
\beq
\delta (N) \equiv 1-\frac{\gamma}{\beta_0} = O(1/N)\,,\qquad \delta (3) = \frac{19}{31}\,.
\label{oprdelta}
\eeq
Combining Eqs.~(\ref{dopgluco}),  (\ref{projectPD1}) and  (\ref{rgic})
we conclude that
\beq
\left[g^2 (\mu) \right]^{\delta (N)}
\left\langle \left(g^2\right)^{1-\delta (N)}\,\bar\Psi\Psi
\right\rangle = - 6 (N-2) \, \Lambda^3_{\overline{\rm MS}}\,\,  {\cal K} (\mu, N)\,,
\label{prepred}
\eeq
where $\mu $ is some {\em fixed} normalization point;
the correction factor
\beq
{\cal K} (\mu, N=\infty ) =1\,, 
\eeq
simultaneously with $\delta (N=\infty ) =0$. At finite $N$
the correction factor ${\cal K}  (\mu, N) -1 = O(1/N)$ 
and ${\cal K}$ depends on $\mu$ in the same way 
as $\left[g^2 (\mu) \right]^{\delta (N)}$. 
The combination $ \left(g^2\right)^{1-\delta (N)}\,\bar\Psi\Psi$
on the left-hand side is singled out because of its RG invariance.
Equation (\ref{prepred}) is our master formula.

\vspace{1mm}

The factor $N-2$ on the right-hand side of Eq.~(\ref{prepred}), a descendant 
of $N$ in Eq.~(\ref{dopgluco}), makes 
$\left\langle \left(g^2\right)^{1-\delta (N)}\,\bar\Psi\Psi \right\rangle$  vanish at $N=2$. This requirement is obvious, given that
at $N=2$ the antisymmetric fermion loses color. In fact,
we replaced $N$ in Eq.~(\ref{dopgluco}) by $N-2$ by hand, 
assembling all other $1/N$ corrections in $ {\cal K}$,
in the hope that all other $1/N$ corrections collected in ${\cal K}$
are not so large. There is no obvious reason for them to be large.
Moreover, we can try to further minimize them
by a judicious choice of $\mu$.

It is intuitively clear that $1/N$ corrections  in ${\cal K}$
will be minimal, provided that $\mu$ presents a scale ``appropriate to
the process", which, in the case at hand, is the formation of the quark
condensate. Thus, $\mu$ must be chosen as low as possible,
but still in the interval where the notion of $g^2 (\mu)$
makes sense. Our educated guess is $\lambda (\mu) =(1/2)$
corresponding (at $N=3$) to
\beq
\left[g^2 (\mu) \right]^{\delta (3)} \approx 4.9\,.
\eeq
As a result, we arrive at the conclusion that in one-flavor QCD
\beq
\left\langle \left(g^2\right)^{1-\delta (3)}\,\bar\Psi\Psi
\right\rangle = - 1.2 \,  \Lambda^3_{\overline{\rm MS}}\,\,  {\cal K} ( N)\,.
\label{arrco}
\eeq
Empiric determinations of the quark condensate with which we will
confront our theoretical prediction are usually quoted for the
normalization point 2 GeV.  To convert the RGI combination on the left-hand side of
Eq.~(\ref{arrco}) to the quark condensate at 2 GeV, we must divide by
$[g^2 (2\,\,{\rm GeV})]^{1-\delta (3)} \approx 1.4$.
Moreover, as has already been mentioned, we expect non-planar  corrections
in ${\cal K}$ to be in the ballpark $\pm 1/N$. If so, three values
for ${\cal K}$,
\beq
{\cal K} = \{ 2/3,\,\, 1,\,\, 4/3\}\, ,
\label{set}
\eeq
give a representative set. 
Assembling all these factors together we end up with
the following prediction for one-flavor QCD:
\beq
\left\langle \, \bar\Psi_{[ij]} \Psi^{[ij]}
\right\rangle_{2\,\,{\rm GeV}} = -\, (0.6\,\,{\rm to}\,\, 1.1) \,\,  \Lambda^3_{\overline{\rm MS}}\,.
\label{predi}
\eeq

Next, our task is to compare it with empiric determinations,
which, unfortunately, are not very precise.
The problem is that one-flavor QCD is different   both from actual QCD, with three
massless quarks, and from quenched QCD, 
in which lattice measurements have recently been carried out~\cite{lattice}.
In   quenched QCD there are no quark loops in the running
of $\alpha_s$; thus, it runs steeper than in one-flavor QCD.
On the other hand, in three-flavor QCD  the running
of $\alpha_s$ is milder than in one-flavor QCD.

To estimate the input value of  $\lambda_{\overline{\rm MS}}  $
we resort to the following procedure.
First, starting from $\alpha_s (M_\tau )=0.31$ (which is close to the
world average) we determine  $\Lambda^{(3)}_{\overline{\rm MS}}$.
Then, with this $\Lambda$ used as an input, we evolve  the coupling constant back
to 2 GeV according to the {\em one-flavor} formula. In this way we obtain
\beq
\lambda (2\,\,\rm GeV) = 0.115\,.
\label{perots}
\eeq
A check exhibiting the scatter
of  the value of $\lambda (2\,\,\rm GeV)$
is provided by lattice measurements.
Using the results of Ref.~\cite{Luscher:1993gh} referring to {\em pure} Yang--Mills
theory one can extract $\alpha_s (2\,\,\rm GeV)=0.189$.
Then, as previously, we find  $\Lambda^{(0)}_{\overline{\rm MS}}$,
and evolve back to 2 GeV according to the {\em one-flavor} formula.
The result is
\beq
\lambda (2\,\,\rm GeV) = 0.097\,.
\label{vtorots}
\eeq
The estimate (\ref{vtorots}) is smaller than (\ref{perots}) by
approximately one standard deviation $\sigma$. This is natural, since the lattice
determinations of $\alpha_s$ lie on the low side, within 
one $\sigma$ of the world average. In passing from Eq.~(\ref{arrco}) to Eq.~(\ref{predi})
we used the average value $\lambda_{\overline{\rm MS}} \, (2\,\,\rm GeV) =0.1$.

One can summarize the lattice (quenched) determinations of the quark condensate, and the chiral theory determinations extrapolated to one flavor, available in the literature,  as follows:
\beq
\left\langle \, \bar\Psi_{[ij]} \Psi^{[ij]}
\right\rangle_{2\,\,{\rm GeV,\,\, ``empiric"}} = -\, (0.4\,\,{\rm to}\,\, 0.9) \,\,  \Lambda^3_{\overline{\rm MS}}\,.
\label{empi}
\eeq
We put empiric in quotation marks, given all the uncertainties
discussed above.

Even keeping in mind all the uncertainties involved in our numerical
estimates, both from the side of supersymmetry/planar equivalence
$1/N$ corrections, and from the ``empiric" side, a comparison of Eqs.~(\ref{predi}) and (\ref{empi}) reveals 
an encouraging overlap.
We  claim, with satisfaction, that we are on the right track!
 
\section{The orientifold Large-$N$ expansion and \\
 three-flavor QCD}
 \label{v1nc}
 \renewcommand{\theequation}{\thesection.\arabic{equation}}
\setcounter{equation}{0}

So far we established an analytical relation between the orientifold
theory (A or S) and \None SYM for $N \rightarrow \infty$. This relation was
used to make predictions for one-flavor QCD. It seems that
our method is restricted to it, since the \None theory contains
one gluino. Could we use our new 
technique to calculate quantities in three-flavor QCD, while using
\None SYM as a parent theory ?
 
To this end we can try to deform the orientifold A theory
in such a way that the leading planar approximation
will remain intact, while $1/N$ corrections hopefully will change
in the direction of making the $N=3$  deformed model
closer to  the parent theory.  Amusingly enough,
at $N=3$, the daughter theory will be QCD with three massless flavors! 

The model to be outlined in this section
was in fact formulated in 1979 \cite{corramond} (see footnote in Sect.~\ref{genf});
we  refer to it as orienti/2f model.
Its fermion sector consists of a single 
two-index antisymmetric Dirac field
$\Psi_{[ij]}$ plus two fundamental Dirac fields
$q_1^i$ and $q_2^i$ ($i,j =1,2,...,N$). At $N=3$ it reduces
to {\em three-flavor QCD}, in contradistinction with
orientifold A theory whose $N=3$
limit is    one-flavor QCD. On the other hand, at $N\to \infty$
fundamental quark loops can be neglected. Each of the  two 
determinants --- one for $q_1$ and another for $q_2$ ---
becomes unity, up to $1/N$ corrections. As a result,
the orienti/2f theory is planar-equivalent
to supersymmetric gluodynamics, much in the same way
as the orientifold theory. It is important to note
that only those operators of the daughter theory that do not
explicitly contain 
$q_{1,2}$ can be matched onto operators of the parent one. The latter can appear only in loops.

Why may one  hope that adding $q_{1,2}$ one suppresses 
$1/N$ corrections?  A hint comes from 
perturbation theory. Indeed, in 
orienti/2f the first coefficient of the $\beta$
function exactly coincides with that
of SUSY gluodynamics, while the deviation
of the second coefficient is $\sim 1/3$ of the deviation
inherent to orientifold A (cf. Tables~\ref{TTT}
and \ref{TaT} and Appendix C). On the non-perturbative side,
orientifold A has no color-singlet baryons with
mass scaling as $N^0$, while orienti/2f does have
such baryons (remember, in SUSY gluodynamics 
there are baryons of the type $G\lambda$ whose mass
$\sim N^0$).

The orienti/2f theory has not been analyzed in earnest yet.
This is an obvious task for the future. Here we limit ourselves
to an introductory remark. The vacuum structure in the
orienti/2f theory is quite different from that in SUSY gluodynamics.
For generic values of $N$ the orienti/2f theory possesses the
global symmetries\,\footnote{The symmetry is enlarged to $  {\rm U}(1)_V\times
{\rm U}(1)_A  \times   {\rm SU}(3)_R\times {\rm SU}(3)_L$ for $N=3$.}
\beq
 \{ {\rm U}(1)_V\times
{\rm U}(1)_A\} \times
\{ {\rm U}(1)_V\times {\rm U}(1)_A\}
\times \{  {\rm SU}(2)_R\times {\rm SU}(2)_L\}  \,,
\eeq
where the first two factors correspond to
$\Psi$ while the last four to $q_{1,2}$.
All vector symmetries are realized linearly. As for the axial
symmetries, SU(2)$_A$ is spontaneously
broken by $\langle \bar q_{1}q_{1} +\bar q_{2}q_{2}
\rangle \neq 0$. The two ${\rm U}(1)_A$'s are separately anomalous,
but one can construct an anomaly-free combination,
which is spontaneously broken by
$\langle \bar \Psi\Psi
\rangle \neq 0$ as well as by $\langle \bar q_{1}q_{1} +\bar q_{2}q_{2}
\rangle \neq 0$.
The anomaly-free ${\rm U}(1)_A$ is generated
by the current
\beq
A^\mu = \bar\Psi\gamma^\mu\gamma^5\Psi -\frac{N-2}{2}
\left(
\bar q_{1}\gamma^\mu\gamma^5 q_{1} +
\bar q_{2}\gamma^\mu\gamma^5 q_{2}
\right).
\label{afc}
\eeq
The anomalous ${\rm U}(1)_A$ current is ambiguous, since we 
can always add to it a  multiple of the conserved current. The above symmetries give
rise to a vacuum manifold of the type $S_3\times S_1$,
corresponding to four Nambu-Goldstone bosons. The fact that the anomalous
${\rm U}(1)_A$ current has an anomaly-free discrete
subgroup has no impact on the vacuum structure.

Thus, the parent theory has a mass gap while
the daughter one does not. It is not difficult to see, however,
that the massless modes in the daughter theory,
in the common sector, manifest themselves only at 
a subleading in $1/N$
level. For instance, the Nambu--Goldstone bosons
associated with the spontaneous breaking of 
$ {\rm SU}(2)_A$ are pair-produced.
The Nambu--Goldstone boson
associated with the spontaneous breaking of 
the anomaly-free ${\rm U}(1)_A$ can be produced by an appropriate
operator, but the residue will necessarily be suppressed by $1/N$. 

The most visible  effect of the
$q_{1,2}$ quarks in loops is the change
in the gauge coupling $\beta$ function mentioned
at the beginning of this section. If this is numerically the most 
important effect, one can readily arrive at an estimate of the quark condensate in the three-flavor theory. In fact, $\langle \bar \Psi \Psi
\rangle$ would be predicted to be the same as in Eq.~(\ref{prepred}), up to  modifications in $\delta$
due to the changes in $\beta_0$ and $\beta_1$. Good agreement is obtained
once more with phenomenological estimates and with lattice data,
which are not very precise, though. In this case there is no
need for extrapolation in the  quark flavors.

\section{Domain walls and the cosmological constant in 
the orientifold field theories}
\label{secDW}
\renewcommand{\theequation}{\thesection.\arabic{equation}}
\setcounter{equation}{0}

Domain walls are BPS objects  in \None supersymmetric
gluodynamics \cite{Dvali:1996xe}. As we know from Sect.~\ref{parent},
if the gauge group is SU($N$), there are $N$ distinct discrete  
vacua labeled by the order parameter, the gluino condensate,
\beq
\langle \lambda \lambda \rangle_k = - N \Lambda ^3 
\exp\left( i {2\pi k\over N}
\right)\,,\qquad k=0,1,2,..., N-1
\label{condensate}
\eeq 
(the vacuum angle is set to zero, and 
in this section we include  the factor 6, cf. Eqs.~(\ref{gluco}) and (\ref{dopgluco}), in the definition of $\Lambda$).
The ``elementary" domain wall $W_{\{k,k+1\}}$ interpolates
between the $k$-th and $(k+1)$-th vacua.
Moreover, at $N\to\infty$ two parallel domain walls
$W_{\{k,k+1\}}$ and $W_{\{k+1,k+2\}}$
are also BPS --- there is neither attraction nor repulsion between 
them \cite{Dvali:1996xe}.

It is known that the BPS domain walls  in \None  gluodynamics
present a close parallel  to D branes in string theory \cite{Witten:1997ep}. 
In particular,
a fundamental flux tube can end on the BPS domain wall, much as F1 end
 on D-branes in string theory \cite{Witten:1997ep}
(see also \cite{Kogan:1997dt}--\cite{Loewy:2001pq}). 

The orientifold theory has $N$ discrete degenerate vacua.\footnote{At $N\to\infty$. In fact, the number of vacua in
orientifold A is $N-2$, while it is $N+2$ in orientifold S.} Hence,
one can expect domain walls. Indeed, the daughter theory inherits 
domain walls from its supersymmetric parent.
Two parallel  walls of the type
$W_{\{k,k+1\}}$ and $W_{\{k+1,k+2\}}$
are in an indefinite equilibrium. In this sense they are
``BPS,'' although the standard definition of
``BPS-ness,'' through central charges and supercharges,
is certainly not applicable in the non-supersymmetric theory.
In this section we elaborate on  physics of the ``BPS" domain walls
 in the orientifold theory.
 
We will show that these walls carry charges similar to the
Neveu-Schwarz--Neveu-Schwarz (NS-NS) and Ramond--Ramond
(R-R) charges. In addition, we will argue that an open-closed
string channel duality holds for the analogous field theory
annulus amplitude. Moreover, by exploiting the similarity
between string theory and field theory we will provide a reason 
why the vacuum energy density (the
``cosmological constant")  in the orientifold gauge theory vanishes
at order $N^2$ despite the fact that the hadronic spectrum of the theory  
contains only bosons --- for color-singlet states there is no Bose-Fermi 
degeneracy typical of supersymmetric theories.

\vspace{1mm}

We will assume that our gauge theory has a string theory dual in the
spirit of Ref.~\cite{Maldacena:1997re} (yet to be found, though). It is 
presumably of the type 0B on a curved background, similarly to the
orientifold field theory analogue
of \Nfour SYM theory, which is type 0B on ${\rm AdS}_5 \times RP ^5$
\cite{Angelantonj:1999qg}. In this picture the closed strings correspond
to the infrared (IR) degrees of freedom:  the glueballs and ``quarkonia.''
Indeed, the type 0 (closed) strings are purely bosonic, in agreement 
with our expectation from the confining orientifold field theory.
Moreover, the bosonic IR spectrum of the gauge theory is even/odd parity-degenerate, in accordance with degeneracies between the NS-NS and R-R towers of the type-0 string.

\subsection{Parallel Domain Walls versus D Branes}
\label{PDWDB}

\noindent 
As was mentioned, the gauge theory fundamental flux tubes
can end on a BPS domain wall. Let
us assume that, for \None gluodynamics/orientifold theory, 
the domain walls have a realization in terms of D$p$-branes ($p>1$) of the
corresponding type IIB/0B string theory. 
Their world volume is $012 + (p-2)$ directions transverse to the 
four-dimensional  space-time  $0123$. Specific AdS/CFT
realizations of domain walls in \None theories are given in Refs.
\cite{Polchinski:2000uf}--\cite{Maldacena:2000yy},
mostly in terms of wrapped D5-branes. We deliberately do not specify which particular branes are used to model the BPS walls, since we  do not
perform actual AdS/CFT calculations.  

D-branes carry the NS-NS charge, 
as well as the R-R charge \cite{Polchinski:1995mt}. 
Moreover, interactions induced by these charges exactly cancel,
ensuring that the parallel D-branes neither attract nor repel
each other. Let us see how this is realized in \None SYM theory.  
We start from two parallel BPS walls  $W_{\{k,k+1\}}$ and $W_{\{k+1,k+2\}}$.
Each of them is BPS, with the tension~\cite{Dvali:1996xe}
\beq
T_{\{k,k+1\}} = T_{\{k+1,k+2\}} = \frac{N}{8\pi^2}\, | \langle \mbox{tr} \lambda\lambda
\rangle | \,2\, \sin\frac{\pi}{N}\,.
\label{tension}
\eeq
The tension of the configuration $W_{\{k,k+2\}}$ is
\beq
T_{\{k,k+2\}} = \frac{N}{8\pi^2}\, | \langle \mbox{tr} \lambda\lambda
\rangle | \, 2 \, \sin\frac{2\pi}{N}\,.
\eeq
This means that at leading order in $N$ (i.e.$\,\, N^1$)
two parallel walls  $W_{\{k,k+1\}}$ and $W_{\{k+1,k+2\}}$
do not interact. There is no interaction at the level
$N^0$ either. An attraction emerges at  the level $N^{-1}$.
That the inter wall interaction potential is $O(N^{-1})$
can be shown on general kinematic grounds.
A relevant discussion can be found in Ref. \cite{adam}, see Eq. (37);
see also Ref.~\cite{RSV01}. This and other aspects of dynamics of
the inter wall separation  are challenging questions, which are 
currently under investigation. Here we summarize some 
findings \cite{AStwo}. 

Our task is to understand the dynamics
of this phenomenon from the field theory side. Assume that two walls under
consideration are separated by a distance $Z \gg m^{-1}$,
where $m$ is the mass of the lightest composite meson. 
What is the origin of  the force between these walls?

The interaction is due to the meson exchange in the bulk.
Consider  the lightest mesons, scalar and pseudoscalar.
The scalar meson $\sigma$, the ``dilaton,'' is coupled \cite{Migdal:jp}
to the trace of the energy-momentum
tensor $\theta^\mu_\mu$:
\beq
\frac{\sigma}{f}\,\theta^\mu_\mu \,  =
\frac{3N}{16\pi^2 f}\,\, \sigma\, \mbox{tr}\,  G^2 \,.
\label{anomc}
\eeq
It is very easy to show that the   coupling constant $f$ scales as
\beq
f\sim N\, \Lambda\,.
\label{ccs}
\eeq
Integrating over the transverse direction, and using the
fact that $\theta^\mu_\mu$ translates into mass, we find that the
``dilaton''--wall coupling (per unit area) is 
\beq
T\,\sigma/f\,,
\label{odin2}
\eeq
where $T$ is the wall tension, see Eq. (\ref{tension}). Since the wall
tension scales as $N$, the $\sigma$--wall coupling  scales as $N^0$.

The scalar meson is degenerate in mass with the pseudoscalar one, 
the ``axion'' or ``$\eta '$'' which, thus, must be included too. 
For brevity  we will denote this field by $\eta$;
it will have a realization in terms of the RR 0-form of type IIB/0B.
The coupling of this ``axion" to the wall is related to the change 
of the phase of the gluino condensate across the wall. Therefore, 
a natural estimate of the axion-wall coupling
is\,\footnote{Equation (\ref{dva2})
guarantees, automatically, that the  axion--wall coupling
is saturated inside the wall. Outside the wall, in the vacuum,
 $\alpha =$ const.,
while $\langle  \mbox{Tr}\,  G^2\rangle = \langle  \mbox{Tr}\,  G\tilde{G}\rangle=0 $. }
\beq
\frac{\eta}{f}\,\int dz\,\,  \frac{N}{8\pi^2}\,\,  \mbox{tr}\,  G\tilde{G}
\to 
\frac{\eta}{f}\,\int dz\,\, \frac{N}{8\pi^2}\, |\,\mbox{tr}\,\lambda\lambda\, |\,\,
\frac{\partial\alpha}{\partial z}\,,
\label{dva2}
\eeq
where $z$ is the coordinate transverse to the wall,
and $\alpha$ is the phase of the order parameter.
At large $N$ the absolute value of the order parameter
stays intact across the wall, while $\int dz\, ({\partial\alpha}/{\partial z})
=2\pi /N$. 
Thus, the  axion--wall coupling  scales as $N^0$. Note that the sign of the
coupling depends on whether we cross the wall from left to right or
from right to left. This is why the  exchange 
of the dilaton between two parallel BPS walls leads
to the wall attraction, while that of the axion leads to repulsion.

Since the  dilaton and axion  couplings to the wall
are proportional to $  N^0$, we  get no force
between the walls  at the level $N^1$ just for free.
The underlying reason is that the BPS domain-wall tension scales 
not as $N^2$, as could be naively expected to be the case
for  solitons, but rather as $N^1$,  the D-brane type of behavior.

We want to make a step further, however.  We will show momentarily
that $\sigma$ and $\eta$ contributions (at the level $N^0$)
are exactly equal in absolute value but are opposite in sign.
Needless to say,   this requires the degeneracy of their masses and their couplings to the walls, except for the relative sign.

Taking the right-hand side of Eq.~(\ref{dva2})
at its face value we get the  axion--wall coupling in the form
\beq
T\,\eta/f + O(1/N)\,,
\label{tri}
\eeq
i.e. indeed the same as Eq.~(\ref{odin2}).

At finite $N$ the wall thickness, which scales as \cite{Gabadadze:1999pp}
$(N\Lambda)^{-1}$, is finite too.   On the wall world volume
one has a $(2+1)$-dimensional supersymmetry,
which places scalars and pseudoscalars into distinct
(non-degenerate) supermultiplets \cite{AV}. At $N=\infty$
the wall thickness vanishes, and it is natural that 
in this limit we deal with the lowest component of $(3+1)$-dimensional
chiral superfield which has the form $\sigma + i\eta$.

Summarizing, the  wall ``R-R charge'' is due to the axion-like nature 
of $\eta$.   The ``NS-NS charge'' is due to the wall tension. 
Both charges are indeed equal  and scale as $N^0\, \Lambda^2$. 
This guarantees that at order $N^0$ there is no force. 
The $\sigma$ and $\eta$ coupling to the wall split at order $N^{-1}$.

The very same arguments can be repeated {\em verbatim}
in the orientifold theories. In the limit $N\to\infty$
the degeneracy of the $\sigma$ and $\eta$ masses holds, and so does
the degeneracy of  the wall ``NS-NS and R-R charges.'' The reason
why  the couplings to the wall (associated
with the trace of the energy-momentum tensor for the  dilaton  and
the axial charge for the  axion) are the same is that these 
charges and couplings are inherited
from the parent supersymmetric theory at  $N\to\infty$, see \cite{adisvone}.
It will be useful, for what follows, to note that
 $\sigma$ exchange alone (before cancellation) generates 
an interaction potential between the walls
(per unit area)
\beq
\frac{V}{A} \sim N^0\,\Lambda^3 \, e^{-mz}\,.
\label{intpotu}
\eeq

Let us now see how this picture is implemented in string theory.
From the open string standpoint the result of a vanishing force  is
natural. This is the Casimir force between the walls. Since the 
UV degrees of freedom are Bose--Fermi-degenerate, the vacuum energy
and, hence, the Casimir force is zero.

\begin{figure}
\begin{center}
\mbox{\kern-0.5cm
\epsfig{file=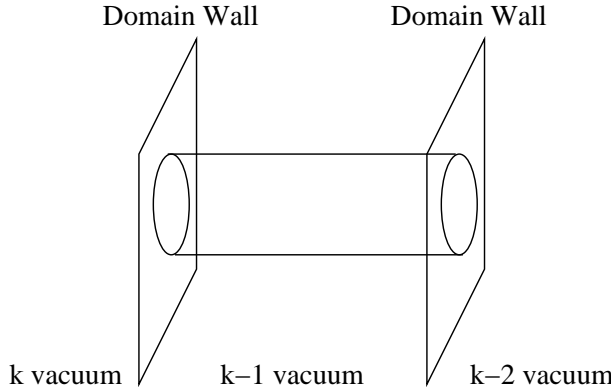,width=8.0true cm,angle=0}}
\end{center}
\caption{The annulus diagram for the orientifold field theory. The
D branes are domain walls. Closed strings are bosonic glueballs and
open strings are UV degrees of freedom.}
\label{dw}
\end{figure}

In the gauge theory the closed
strings are field-theoretic glueballs~ \cite{Witten:1998zw,Csaki:1998qr}. 
Let us consider first the large separation limit. At the lowest level we 
have a massive scalar and a pseudoscalar (we assume a mass gap).
These two exactly degenerate states correspond to the  dilaton 
and  axion  (the RR 0-form of type IIB/0B). They are expected to become
massive when the theory is defined on a curved background~\cite{Witten:1998zw}. The type IIB/0B action contains the couplings
\beq
e^{-\Phi}\, {\rm tr}\, G^2 + C\, {\rm tr}\, G \tilde G,
\label{one2}
\eeq
where $\Phi$ denotes the dilaton and $C$ the R-R 0-form. In addition
we have a coupling of the graviton and the R-R 4-form
\beq
\eta ^{\mu \rho} h ^{\nu \lambda }\, {\rm tr}\, G_{\mu \nu} G_{\rho
\lambda} +
C ^{\mu \nu \rho \lambda }\, {\rm tr}\, G_{\mu \nu} G_{\rho
\lambda}\,.
\label{two2}
\eeq
In the gauge theory  the ``graviton'' (tensor meson) and the 4-form are heavier
glueballs, since they carry higher spins than the dilaton and the 0-form.
Similarly, the whole tower of degenerate bosonic hadrons of the 
orientifold field theory should correspond to the NS-NS and R-R fields 
of type II/0 string theory.  This gives
us a new picture of why the force between the domain walls vanishes
in terms of the glueball exchanges: even-parity glueballs lead to
an attractive force between the walls, whereas odd-parity glueballs
lead to a repulsion. The sum of the two is exactly zero
at the leading $N^0$ order. The $1/N$ force between the walls is related to
a possible non-vanishing force between parallel D-branes in a curved
space at $O( g^2_{\rm st})$.

We can also exploit the above picture to estimate the potential between
a given wall and an anti-wall. Since this configuration is not BPS
a non-vanishing force is expected at  leading $N^0$ order. 
At large separations, the force is controlled by an exchange of the lowest massive closed strings, 
the dilaton and the zero-form. Now their contributions add up and we
get an attractive potential, as indicated in 
Eq.~(\ref{intpotu}).

\subsection{Vanishing of the Vacuum Energy in \None SYM and Orientifold Theories}
\label{VVESOT}

\noindent
One of  surprising results is that the  $N^2$ part of the vacuum 
energy density vanishes in the  orientifold field theory. While this result 
makes sense from the UV point of view, where we do have Bose--Fermi degeneracy, it looks  rather  mysterious from the IR standpoint, since 
at the level of the composite color-singlet states
we have only bosonic degrees of freedom and no Bose--Fermi degeneracy.
Since at large $N$ we have free bosons, it is legitimate
to sum the bosonic contributions  to the vacuum energy density 
${\cal E}$ as follows:
\beq
{\cal E}= \sum _n \sum _{\vec k} {1\over 2} \sqrt {{\vec k}^2 + M_n^2}, 
\label{ve}
\eeq
where $M_n$ are the meson masses. An apparent paradox arises
since the sum runs over positive contributions. How can positive
contributions sum to zero?

In fact, there is no paradox and no contradiction with the statement
of vanishing of ${\cal E}$ at $O(N^2)$.
We will present below arguments that will, hopefully, make transparent
the issue of  the vanishing of the vacuum energy density (at the $N^2$
level) in the orientifold theory. 
Usually it is believed that one needs full supersymmetry
to guarantee that ${\cal E}=0$. It turns out that a milder requirement
--- the degeneracy between scalar and pseudoscalar glueballs/mesons ---
does the same job. Of course, in supersymmetric theories this degeneracy
is automatic. The orientifold field theory is the first example
where it takes place (to leading order in $1/N$) without 
full supersymmetry because it is inherited through the planar equivalence.

Two comments are in order here.
First, the sum \eqref{ve} is not well-defined since the expected
Regge trajectories are not bounded from above and, therefore, a
regularization is needed. Second, in the above sum \eqref{ve} 
the $N$ dependence of each individual mode is $N^0$. The expected
$N^2$ dependence of the vacuum energy is hidden in the sum over
all hadronic modes and in the regularization. 

Thus, even though \eqref{ve} represents the
vacuum energy density of the theory, it cannot be calculated from this equation. 
Below, we present an 
alternative  definition of the vacuum energy density, with a UV regulator built in, which will illustrate   the above surprising  
nullification. 

Let us consider the contribution to ${\cal E}$
from the open string sector. At large $N$ it is dominated
by the annulus diagram where each boundary consists of $N$ D-branes,
and a summation over the various D-branes is assumed. 
The M\"obius and Klein-bottle as well as higher-genus amplitudes
are suppressed at large $N$.  From this standpoint it is not surprising that the
${\cal E}$ vanishes, as we have $N^2$ bosons (NS open
strings) and $N^2$ fermions (Ramond fermionic open strings).

The annulus diagram has another interpretation, however. It represents
the force between the D-branes. The force is mediated by
bosonic closed strings. In a SUSY setup (the type II string),
D-branes are the BPS objects --- hence,  the  zero force. 
As has been discussed above, the balance, at large separations, is achieved
thanks to a cancellation between the dilaton, the graviton
and the massless R-R forms.

It is interesting that the force between parallel self-dual D-branes vanishes  also in type-0 string theory~\cite{Klebanov:1998yy,Angelantonj:1999qg}
\beq
{\cal A}= N^2 (V_8-S_8) \equiv 0\,.
\eeq
This is due to the underlying supersymmetry on
the world-sheet. The mechanism is exactly as in the type II case:
the R-R modes cancel the contributions of the NS-NS modes.  Note
that since we are interested only in the planar gauge theory, 
we can restrict ourselves to $g_{st}=0$ on the string theory side.
Therefore, higher-genus amplitudes are irrelevant to our discussion.
At this level the relevant type-0 amplitudes, as well as the bosonic
spectrum, are identical to the type
II ones, in a not too surprising similarity with the situation in the
large-$N$ dual gauge theories (the type-0 string becomes, in a sense, 
supersymmetric at tree level). 
In particular, the induced dilaton tadpole and cosmological constant
are irrelevant and, thus,  the background inherited from
the supersymmetric theory remains intact.

The vanishing of the annulus diagram leads to an explanation of the mysterious
vanishing of the cosmological constant in the orientifold field
theory:  if one views the hadrons (in the spirit of the AdS/CFT
correspondence) as closed strings, the
degenerate bosonic spectrum is the reason behind the vanishing result
for both wall--wall interaction and ${\cal E}$.

We hasten to add that although the mechanism
is similar, there is a difference between the two cases: the wall--wall
interaction involves the force between ``D2''-branes (wrapped D5-branes), whereas the vanishing cosmological constant involves the force between ``D3''-branes. The two sorts of branes are not necessarily the same --- 
everything  depends on the specific realization. However, from the bulk point of
view, the mechanism is identical. The only requirement is the degeneracy
of the NS-NS and the R-R towers and their couplings to the branes. 

The difference between string theory and field theory is that in
string theory the force between D-branes, from the closed string
standpoint, is related to the contribution to the cosmological
constant from the open string sector. However, closed strings and
open strings are independent degrees of freedom and the contribution
of the closed strings to the cosmological constant should therefore be
added. In the gauge theory string
picture, the closed strings are simply hadrons made
out of the constituent open strings --- the gluons and quarks.
Therefore, the value of the cosmological constant can be determined by
either UV or IR degrees of freedom: it is the
same quantity.

Let us now switch to the field theory language. 
At first, we will acquaint the reader with
some general relations, relevant to 
${\cal E}$, which are valid in any gauge theory with no mass scale
other than the dynamically generated $\Lambda$.
The vacuum energy density ${\cal E}$ is defined through
the trace of the energy-momentum tensor,
\begin{eqnarray}
{\cal E} & =& \frac{1}{4}\,
\left\langle \theta ^\mu _\mu
\right\rangle  = \frac{1}{4}\, \int DA\, D\Psi\,\,  \theta ^\mu _\mu\,\, \exp (iS)\,,
\nonumber\\[3mm]
\theta ^\mu _\mu & =& -\frac{3N}{32\pi^2} \, G^2\,,
\label{traceT}
\end{eqnarray}
where
\beq
G^2 \equiv G_{\mu\nu}^a\,  G^{\mu\nu ,a}\,, \qquad G\tilde G\equiv 
G_{\mu\nu}^a\,  \tilde{G}^{\mu\nu ,a}\,.
\eeq
The second line in Eq. (\ref{traceT}) is exact in \None gluodynamics and is valid
up to $1/N$ corrections in the orientifold theory.

Now, we use an old trick \cite{Novikov:xj} to express the trace of the
energy-momentum tensor in terms of  a two-point function. The idea is to vary
both sides of Eq.~\eqref{traceT} with respect to $1/g^2$
and exploit the fact that the only dimensional parameter
of the theory, $\Lambda$, exponentially depends on $1/g^2$. In this
way one  obtains~\cite{Novikov:xj} 
\begin{eqnarray}
{\cal E} & =& -i\int\, d^4x\,\left\langle {\rm vac}\left|
T\left\{ \frac{1}{4}\,\theta ^\mu _\mu (x)\,,\,\, \frac{1}{4}\,\theta ^\nu _\nu (0)
\right\}\right|{\rm vac}\right\rangle_{{\rm conn}}
\nonumber\\[4mm]
& \equiv & -i\int\, d^4x\,\left\langle \frac{3N}{128\pi^2}\, G^2 (x)\,,
\,\, \frac{3N}{128\pi^2}\, G^2 (0)
\right\rangle_{{\rm conn}}\,.
\label{twopoint2}
\end{eqnarray}
The above expression \eqref{twopoint2} is formal --- it 
is not well defined in the UV.
This is the same divergence that plagues any
calculation of ${\cal E}$ (remember, Eq. (\ref{twopoint2})
is general,  not specific to supersymmetry).

In SUSY, we are prompted to a natural regularization. Indeed, let us consider
the two-point function of the lowest components
of two chiral superfields $D^2 W^2$,
\beq
\langle{\rm tr} D^2 W^2 (x)\,,\,\,{\rm tr} D^2 W^2(0)\rangle\,.
\label{doppyat}
\eeq
The supersymmetric Ward identity tells us that this two-point function
vanishes identically. There are two remarkable facts encoded in Eq.~(\ref{doppyat}).
First, since $D^2 W^2 \propto \left (G^2 + i G\tilde G\right)$,
the vanishing of (\ref{doppyat}) is not due to the boson--fermion
cancellation but, rather, to the cancellation between
even/odd parity mesons (glueballs).

Second, Eq. (\ref{doppyat}) generalizes Eq.~(\ref{twopoint2}),
so that the expression for the vacuum energy density takes the form
\begin{eqnarray}
{\cal E} =  -i\left(  \frac{3N}{128\pi^2}\right)^2 \int\, d^4x\,
\left( \left\langle
 G^2 (x)\,,
\,\, \, G^2 (0)
\right\rangle - \left\langle
G\tilde G (x)\,,
\,\, \, G\tilde G (0)
\right\rangle \right),
\label{ereg}
\end{eqnarray}
where the connected correlators are understood on the
right-hand side. To see that Eq. (\ref{ereg}) is the same as 
(\ref{twopoint2}), notice that
\beq
0=  -i\int\, d^4x\,\left\langle \frac{3N}{128\pi^2}\, G\tilde G (x)\,,
\,\, \frac{3N}{128\pi^2}\, G\tilde G (0)
\right\rangle\,.
\label{formzero}
\eeq
The above zero is explained by the fact
that $G\tilde G$ is proportional to the divergence 
of the axial current $A_\mu$
both in \None gluodynamics and in the orientifold theory.
Adding a formal zero we in fact achieved the ultraviolet regularization so that (\ref{ereg}) obeys an unsubtracted dispersion relation.

It is absolutely clear that this regularization
works perfectly both in SUSY theories and
in the orientifold theories (at the $N^2$ level).
Indeed, the part of the two-point function that involves $G^2$ is saturated  by
even-parity glueballs, while the part
that involves $G\tilde G$
is saturated by  odd-parity ones. We then get
\beq
{\cal E} = \sum _{\rm even\,\, parity} {\lambda _n ^2 \over M_n ^2}
-\sum _{\rm odd\,\, parity} {\lambda _n ^2 \over M_n ^2}\,,\qquad \lambda_n^2\sim N^2
\quad \mbox{for all $n$}\,, 
\label{T-glueballs}
\eeq
where $\lambda _n$ are the couplings to $T_\mu^\mu$ and $\partial^\mu a_\mu$, respectively, and
$M_n$ are the glueball masses.  Clearly, if  the masses and the
couplings of the glueballs are even/odd-parity degenerate, as 
is the case in \None gluodynamics {\em and} in the
large-$N$ orientifold field theories, ${\cal E}$ vanishes. 

In summary, in the UV calculation the Fermi--Bose degeneracy
was responsible for the vanishing of the cosmological constant both in 
supersymmetric gluodynamics and in the orientifold theory (where 
the cancellation was at the $N^2$ level). In dealing with ${\cal E}$,
a certain regularization procedure is needed. In SUSY it is implicit.
In passing from the UV to the IR language, we make it explicit
through Eq. (\ref{ereg}).

Formula \eqref{T-glueballs} is in remarkable agreement with 
our string theory picture. Only bosonic glueballs are involved, and   the 
even/odd parity glueballs contribute with  opposite signs.

\section{A chiral ring for non-supersymmetric theories}
\label{cring}
\renewcommand{\theequation}{\thesection.\arabic{equation}}
\setcounter{equation}{0}

In four-dimensional supersymmetric  theories there is a notion of the
{\em chiral ring}. An assembly of chiral operators (i.e. those annihilated
by supersymmetries $\bar Q$)
forms a ring (this is a mathematical term). The product of two chiral operators is also 
a chiral operator. Chiral operators are usually considered
modulo operators of the form $\{\bar Q, ...\}$. The equivalence classes
can be multiplied and form a ring. Chiral operators
are the lowest components of chiral superfields.

A crucial property of the
chiral operators is that their correlation functions are  space-time
independent \cite{Novikov:ee}, e.g.
\beq
\langle O_1(x) O_2(y)\rangle = {\rm const.},
\label{onedopone}
\eeq
where the constant on the right-hand side depends neither on $x$ nor on $y$.
Consider for example the correlation function
\beq
\langle T\left\{ \lambda^a_\alpha (x)\lambda^{a\alpha}(x)\,
,\, \lambda^b_\beta (0)\lambda^{b\beta}(0)\right\}\rangle\,.
\label{tpf}
\eeq
The proof of the constancy is quite straightforward \cite{Novikov:ee} and is 
based on three elements: (i) the supercharge 
${\bar Q}^{\dot\beta}$ acting on the vacuum state annihilates it; 
(ii) ${\bar Q}^{\dot\beta}$ anticommutes with $\lambda\lambda$; 
(iii) the derivative $\partial_{\alpha\dot\beta}(\lambda\lambda)$ is 
representable as the anticommutator of ${\bar Q}^{\dot\beta}$ and 
$\lambda^\beta G_{\beta\alpha}$. 
One differentiates Eq. (\ref{tpf}), substitutes 
$\partial_{\alpha\dot\beta}(\lambda\lambda)$ by $\{
{\bar Q}^{\dot\beta}, \lambda^\beta G_{\beta\alpha}\}$, and obtains 
zero.$\,$\footnote{The operator $\lambda\lambda$
is called $Q$-closed, while the operator
$\partial_{\alpha\dot\beta}(\lambda\lambda)$ is $Q$-exact.}  Thus, 
supersymmetry requires the $x$ derivative of (\ref{tpf}) to vanish.  
It does not require the vanishing  of the correlation function {\em 
per se}.  A constant is alright. 

It is quite clear that the necessary and sufficient condition for the
constancy of the correlation function (\ref{tpf}) is the degeneracy of
the glueball  spectra in the scalar and pseudoscalar channels.
If the spectra were not degenerate, the dispersion representation
would require a non-trivial $x$ dependence.  Since the above 
degeneracy is inherited by the orientifold theories at $N\to\infty$,
one expects that in this limit they do carry a remnant of this 
phenomenon. Our task is to identify this aspect.

Equations (\ref{onedopone})  imply
that in SUSY theories the expectation values of the chiral operators
factorize,  
\beq
\langle O_1 O_2 ... O_n \rangle = \langle O_1 \rangle \langle O_2
\rangle...\langle O_n \rangle.
\eeq 
This factorization is {\em exact}. Since in 
\None gluodynamics   the chiral ring is spanned by the operator
$S={\rm Tr}\,\, W_\alpha ^2$, see \cite{Cachazo:2002ry} for a
comprehensive discussion,  for definiteness we can 
limit ourselves to the operators $ \lambda^a_\alpha \lambda^{a\alpha}$.
Their projection onto orientifold theory
is $\bar \Psi _L \Psi _R \equiv \eta_{\{ij\}}\xi^{\{ij\}}$
or $\eta_{[ij]}\xi^{[ij]}$
in the notation of Sect.~\ref{orientifold}. 
The question to ask is what is the consequence of the exact
factorization
\beq
\langle {\rm Tr } \lambda \lambda {\rm Tr }  \lambda \lambda  \rangle
=  \langle {\rm Tr } \lambda \lambda \rangle^2
\label{exaf}
\eeq 
in the parent theory for the daughter operators $\bar \Psi _L \Psi _R$. 

As in any large-$N$ theory, the nonfactorizable part
of $\bar \Psi _L \Psi _R \bar \Psi _L \Psi _R$
is suppressed by $1/N$.
More exactly, in the general case one can expect
\beq
\langle \bar \Psi _L \Psi _R \bar \Psi _L \Psi _R  \rangle
=  \langle \bar \Psi _L \Psi _R \rangle^2
\left( 1 + O(1/N^2)\right)\,,
\eeq 
where   $O(1/N^2)$ corrections on the right-hand side is due to
non-factorizable contributions; while the factorizable part is
$\langle \bar \Psi _L \Psi _R \rangle^2 = O( N^4)$.
The orientifold theory is special.
The exact factorization (\ref{exaf}) in conjunction with the planar equivalence implies that in the orientifold theories
the non-factorizable part is {\em additionally suppressed}, namely,
\beq
\langle \bar \Psi _L \Psi _R \bar \Psi _L \Psi _R  \rangle
=  \langle \bar \Psi _L \Psi _R \rangle^2
\left( 1 + O(1/N^3)\right)\,.
\label{rels3}
\eeq 
The relative suppression $O(1/N^3)$ in Eq.~(\ref{rels3}) rather than 
$O(1/N^2)$ is the remnant of the chiral ring. This result replaces, in the orientifold theories, the statement of the existence of the chiral ring in SYM theories.

In more general terms  one can write
\beq
\langle \bar \Psi _L \Psi _R (x_1)\,,\, \bar \Psi _L \Psi _R (x_2) \rangle  = \langle
\bar \Psi _L \Psi _R \rangle ^2\left[ 1 + O(N^{-3})\, f (x_1-x_2)\right]. 
\eeq 
This is the meaning of a chiral ring in non-supersymmetric
theories (orienti A/S). It would be interesting to explore the phenomenological consequences of our finding.

\section{Parent theories related to  \Ntwo SUSY}
\label{ptn2}
\renewcommand{\theequation}{\thesection.\arabic{equation}}
\setcounter{equation}{0}

\subsection{Planar equivalence between
distinct supersymmetric theories and an estimate
of the accuracy of 1/N expansion}
\label{pebd}

While the original proof of the (non-perturbative)
planar equivalence between a supersymmetric parent
theory and its orientifold daughter \cite{adisvone}
connects ${\cal N}=1$ gluodynamics with an 
${\cal N}=0$ QCD-like theory, it can be generalized, practically {\em verbatim}, to include other parent--daughter pairs, in 
particular, supersymmetric pairs.\footnote{{\em Orbifold} pairs with
${\cal N}=1$ supersymmetric  parent {\em and} daughter theories  were
analyzed in \cite{Erlich:1998gb}.}

Here we will consider the mass-deformed Seiberg--Witten theory
as a parent, and its particular ${\cal N}=1$ orientifold daughter.
The presence of the mass term is important in order to make
relevant superdeterminants well-defined.

The Lagrangian of the parent/daughter theories are
\bea
{\cal L} = \left\{
\frac{1}{4g^2_0}  \int\!{\rm d}^2\theta \,\mbox{Tr}\, W^2 + \, 
\mbox{h.c.}\right\} + {\cal L}_{\rm matter}
\label{p21}
\eea
where the matter part for the parent theory is
\beq
{\cal L}_{\rm matter}=\frac{1}{4} \int \!{\rm  d}^2\theta {\rm 
d}^2\bar\theta\,\bar \Phi e^V \Phi + \left\{
\frac{m_0}{4}\int\! {\rm d}^2\theta\, \Phi^2
+ {\rm h.c.}\right\}\,.
\eeq
while for the daughter theory it is
\beq
{\cal L}_{\rm matter}=\frac{1}{4} \int \!{\rm  d}^2\theta {\rm 
d}^2\bar\theta\,\left( \bar \xi e^V \xi +
\bar \eta e^V \eta \right) + \left\{
\frac{m_0}{2}\int\! {\rm d}^2\theta\, \xi\eta
+ {\rm h.c.}\right\}\,.
\eeq
Here $\Phi$ is the chiral superfield in the adjoint representation,
while $\xi$ and $\eta$ are two-index antisymmetric chiral superfields of the
type $\xi^{[ij]}$ and $\eta_{[ij]}$.
The gauge group is SU($N$), the bare gauge coupling is denoted by
$g_0^2$, while the bare-mass parameter is $m_0$. Remember that
in our conventions $\int \, \theta^2\, d^2\theta =2$.

In the limit $m_0\to 0$ the parent  theory has ${\cal N}=2$,
while its orientifold daughter has ${\cal N}=1$.
We will keep $m_0$ fixed (i.e. $N$-independent);
this is important for the proof of the non-perturbative planar equivalence.
The ratio $m/\Lambda$ can be arbitrary but {\em non-vanishing}. 

The fact that, at $m=0$ (or, more generally, $m\to 0$
as $N\to \infty$), the proof does not hold is
seen from the fact that in this limit the parent theory develops a
moduli space, which is much larger than that of the daughter theory.
With $m/\Lambda$ fixed both theories have $N$ discrete vacua.
Physically interesting is the case of non-vanishing but small
$m/\Lambda$.

The advantage of having the supersymmetric orientifold daughter
is that we can now calculate the gluon condensates exactly and
independently in the two theories, and compare the degree of coincidence at 
large $N$. At $N=\infty$ they must coincide; the question we address is
$1/N$ corrections.

We will start from the introduction of appropriate scale parameters. In
general, the two-loop $\beta$-function equation leads to the following
convention \cite{Hinchliffe}:
\beq
\Lambda^{\beta_0} = M_{\rm UV}^{\beta_0}\,  \left(\frac{16\pi^2}{\beta_0 \, g_{0}^2}
\right)^{\beta_1/\beta_0}\,\, \exp
\left(-\frac{8\pi^2}{ g_{0}^2}
\right)\,.
\label{stcon}
\eeq
In our specific case we thus have\,\footnote{In fact, the expressions for 
$\Lambda_P$, the renormalization-group-invariant mass in the parent theory
and -- what is most important --  for the gluino condensates in 
 Eqs.~(\ref{odinn}) -- (\ref{dvan}), are exact.
This is due to a   holomorphic  structure of the NSVZ $\beta$ function
 and  the gluino condensates in   supersymmetric theories.} 
\bea
\Lambda_P 
&=& 
M_{\rm UV} \, \exp
\left(-\frac{8\pi^2}{ 2N\, g_{P0}^2}\right)
\,,
\nonumber\\[4mm]
\Lambda_D 
&=& 
M_{\rm UV}\,  \left[ \frac{16\pi^2}{(2N+2) \, g_{D0}^2}
\right]^{\frac{2}{N}\frac{N-1}{N+1}}\,\, \exp
\left[-\frac{8\pi^2}{(2N+2)\,  g_{D0}^2}
\right] .
\label{rgis}
\eea
In deriving the above expressions 
we used the data collected in Table~\ref{tablesec13}.

\begin{table}
\begin{center}
\begin{tabular}{|c|c|c|c|}
\hline
$\frac{\mbox{ Data} \to  }{\mbox{ Theory} \downarrow} $ \rule{0cm}{0.7cm} &$\beta_0$ &$\beta_1$ & $\gamma$ \\[3mm]
\hline\hline
\rule{0cm}{0.7cm}${\rm Parent}$  & $2N$ &$0$ & $2N $
 \\[2mm]
\hline
\rule{0cm}{0.7cm}${\rm Daughter}$ &$2N+2$ &$\frac{8(N^2-1)}{N}$& $\frac{2 (N-2)(N+1)}{N}$  \\[2mm]
\hline
\end{tabular}
\end{center}
\caption{Coefficients $\beta$ and $\gamma$ 
pertinent to the discussion in Sect.~\ref{pebd}. }
\label{tablesec13}
\end{table}

Next, we would like to define renormalization-group invariant (RGI) masses in
the two theories.  Applying Eq. \eqref{mRGI} and  the data from 
Table~\ref{tablesec13} we get
\beq
\frac{m_P(\mu )}{N g_P ^2 (\mu ) } ={\rm RGI} \,,\qquad
\frac{m_D(\mu )}{\left[ N g_D ^2 (\mu )\right]^{\frac{N-2}{N}}}
={\rm RGI}\,.
\eeq
It is clear, that physically it 
makes sense to compare the parent and daughter theories provided
that they have one and the same dynamical scale, $\Lambda_{  P}
=\Lambda_{  D}\equiv \Lambda$, and the same RGI mass,\footnote{In principle,
there is a legitimate alternative: instead of identifying
the RGI masses one could impose the condition of coincidence
of the running masses at some relevant scale $\mu$. The running masses are
measurable. Distinction  between these two options arises at the $1/N$ level and
is insignificant in the limit of small $m/\Lambda$. }
\beq
 \frac{m_P}{N g^2 _P} = \frac{m_D} {(N g_D ^2)^{\frac{N-2}{N}}}
\equiv m\,.
\label{idrgim}
\eeq

On general grounds, from generalized $R$ parity, one can fix the  dependence
of the gluino condensate on the dynamical scale and the RGI mass \cite{SV}.
Accordingly, the value of the gluino condensate in the parent theory is
\beq
\langle  \lambda^{a}_\alpha \lambda^{a\,\alpha}\rangle _P
= C _P \times \Lambda ^2_P \times {m_P \over N g^2_P}\,,
\label{odinn}
\eeq
and
 \beq
\langle  \lambda^{a}_\alpha \lambda^{a\,\alpha}\rangle _D
= C _D \times \Lambda ^{2N+2 \over N} _D \times \left[ {m_D  \over
(N g^2 _D)^{({N-2 \over N})}}\right]^{{N-2 \over N}}
\label{dva}
\eeq
in the daughter theory. Two constants $C_P$ and $C_D$ can be fixed by taking
the limit $m_0 \rightarrow M_{\rm UV}$ in the two theories. Then the
theories reduce to ${\cal N}=1$  gluodynamics, and the {\it exact} value of the
condensate is known. One readily finds
\beq
C_P = - 4 N (8\pi^2)     ,\quad  C_D = - 4 N (8\pi^2)^{(1-2/N)^2} 
\left(1+\frac{1}{N}\right)^{\frac{4(N-1)}{N^2}} ,
\label{dvan}
\eeq  
so that
\beq
\frac{\langle  \lambda^{a}_\alpha \lambda^{a\,\alpha}\rangle_{  D}}{\langle  \lambda^{a}_\alpha \lambda^{a\,\alpha}\rangle_{  P}}
={C_D \over C_P} \left(\frac{\Lambda}{m}\right)^{2/N}=
1+\frac{2}{N}\, \ln\frac{\Lambda}{64 \pi^4 m} + O\left ({1\over N^2} \right )\,.
\label{ir}
\eeq
The comparison of the gluino condensates is moderately interesting,
since in both theories gluinos are in the adjoint representation, and  orientifoldization
has only an indirect impact on the daughter-theory gluino. Of more
interest is the ratio  of the matter condensates, since it is the matter field
that is subject to orientifoldization.
The matter condensates can be expressed in terms of those in Eq.~(\ref{ir})
by virtue of the Konishi anomaly \cite{Konishi:1983hf}:
\bea
 \bar{D}^2\, (\bar\Phi e^{V} \Phi)& =& 
4 \, m_0 \Phi^2 +
\frac{N}{2\pi^2}\,{\rm Tr}\, W^2\, ,
\nonumber\\[4mm]
 \bar{D}^2\, (\xi e^V \xi +
\bar \eta e^V \eta)& =&
4 \, m_0 \xi\eta  +
\frac{N-2}{2\pi^2}\,{\rm Tr}\, W^2\, ,
\label{ka1}
\eea 
which then implies
\beq
\frac{m_{D0}\, \langle\xi\eta\rangle}{m_{P0}\, \langle\Phi^2\rangle}=\frac{N-2}{N}\, \,
\frac{\langle  \lambda^{a}_\alpha \lambda^{a\,\alpha}\rangle_{ D}}{\langle  \lambda^{a}_\alpha \lambda^{a\,\alpha}\rangle_{ P}}
=\frac{N-2}{N}\, {C_D \over C_P} \, \left(\frac{\Lambda}{m}\right)^{2/N}\to 1\,,
\label{irp}
\eeq
at $N\to\infty$.
This formula, as well as Eq.~(\ref{ir}), show
in an absolutely transparent form that, as the ratio
$m/\Lambda$ decreases, a critical  value of $N$ needed for the onset
of the planar equivalence  increases logarithmically,
\beq
N_*\sim \ln \frac{\Lambda}{m}\,.
\eeq
On the other hand, Eq.~(\ref{irp}) exhibits the vanishing of the matter
condensate in the daughter theory at $N=2$.
This is natural too, since at $N=2$ the daughter-theory matter
loses its color. 

\subsection{Dijkgraaf--Vafa deformations}
\label{dijvafa}

So far we only considered the relation between mass deformed \Ntwo
SYM and a massive orientifold \None theory. 
Recently there was a lot of interest in more general deformations of 
the \Ntwo SYM theory, of the type
\beq
{\cal W}_{\rm tree}=\sum _k g_k\, {\rm Tr}\, \Phi ^k.
\label{repertl}
\eeq
Dijkgraaf and Vafa \cite{DV}, following an earlier work by Cachazo,
Intriligator and Vafa \cite{Cachazo:2001jy}, showed that the effective superpotential
of such a theory can be obtained from a matrix model. It was shown
later that
the same result could be derived using standard field theory techniques
\cite{G,Cachazo:2002ry} (generalized  Konishi anomalies), hence confirming the Dijkgraaf--Vafa conjecture. 

We would like  to extend the analysis of Sect.~\ref{pebd} with a purely
quadratic superpotential  to a more general setup
where we have an \None SYM theory with matter in the antisymmetric
representation and a generic tree-level superpotential
\beq
{\cal W}_{\rm tree}=\sum _k g_k\, {\rm Tr}\, (\xi\eta) ^k\,.
\eeq
We wish to compare it with \None SYM theory with matter in the adjoint
representation and a tree-level superpotential
\beq
{\cal W}_{\rm tree}=\sum _k g_k\, {\rm Tr}\, (\Phi) ^{2k}.
\label{repertlp}
\eeq
The work is in progress, and it is premature to give a detailed report
of relevant results. We would like to note only that at
large $N$ the effective superpotentials of the above two theories  
coincide. Therefore, the planar equivalence is instrumental here too,
albeit only for holomorphic data.

\subsection{Non-holomorphic information}
\label{cwcnhi}

In Sect.~\ref{dijvafa} we mentioned that softly broken \Ntwo
SYM and \None SYM theories, with  massive antisymmetric 
matter field, must effectively coincide at large $N$ in the holomorphic sector.

The planar equivalence between SUSY gluodynamics and
non-SUSY orientifold field theories extends further.
Our results were not restricted to
holomorphic data in the parent \None theory. For example, we argued
that the glueball spectra of the two theories coincide at large $N$.

A natural question is whether for \Ntwo related parents and their daughters
the  correspondence
is restricted to holomorphic data or can be extended, e.g. to $D$-terms.
It is conceivable, for instance, that the prepotential of the softly broken \Ntwo
theory can tell us something about the $D$ term of the orientifold 
\None daughter theory, thus providing important information on its spectrum.

A ``nearly holomorphic'' result is the string tension formula
of Douglas and Shenker for $k$-strings in softly broken \Ntwo SYM  theory
\cite{Douglas:1995nw}
\beq
\sigma _ k \sim m \Lambda \sin \pi {k \over N}.
\label{st}
\eeq
An extended large-$N$ correspondence would predict
a similar sine formula for the $k$-string tension
in \None orientifold field theory, cf.~\cite{ASRem}.

\section{A view from the string theory side}
\label{ovoavftsts}
\renewcommand{\theequation}{\thesection.\arabic{equation}}
\setcounter{equation}{0}

The orientifold field theory has a nice and natural 
description in type-0 string theory. We will 
describe this connection in detail. We start from a short review about
orbifolds and orientifolds in type II string theory. We introduce the
type-0 string and its various orientifolds. Then we explain
how the orientifold field theory is realized via a type-0 brane
configuration. We will use this realization to conjecture
a supergravity description of the large-$N$ theory. The lift to M-theory
of the type 0A brane configuration is also briefly discussed.

\subsection{Orbifolds, orientifolds and quiver theories}
\label{ooqt}

One of the most interesting aspects of string theory is that (in contrast
to the point-particle theory) a consistent theory can be 
formulated on singular spaces, called ``orbifolds.'' The textbooks by
Polchinski \cite{Polchinski} contain an excellent introduction to orbifolds.

Let us focus on closed string theories, such as type IIA/B.
The $Z_2$ orbifold, obtained by the identification $X^m\rightarrow -X^m$,
is the simplest example. Closed strings can propagate in the ``bulk,''
but another consistent possibility is to consider also strings
that are stuck at the fixed point
$X^m=0$. This is the twisted sector of the closed string theory.

One can introduce D-branes and place them on the orbifold singularity
\cite{Douglas:1996sw}. The massless open string spectrum on $N$ D-branes placed
on a $Z_k$ singularity is identified with a U$(N)^k$ gauge theory with
chiral bifundamental matter. These field theories are also known in the
literature as ``quiver theories.''

Following Maldacena \cite{Maldacena:1997re}, Kachru and Silverstein
considered \cite{kachru} D3-branes on $R^6/Z_k$ orbifold singularity. They
conjectured that the U$(N)^k$ quiver theory that lives on the
branes is dual to type IIB string theory on $AdS_5 \times S^5/Z_k$.
As we already mentioned in the introduction, some of these theories
are non-supersymmetric and, hence, Kachru and Silverstein predicted
a large class of large-$N$ non-supersymmetric finite theories.    

The AdS/CFT correspondence relates open strings to closed strings.
Clo\-sed string modes are identified with operators in field theory. 
In general, untwisted (``bulk'') closed strings are identified with the
operators that are invariant under the orbifold action. For example,
the dilaton is identified with the operator $G_1^2 + G_2 ^2 + ... +
G_k ^2$. Twisted modes are identified with non-invariant operators
(therefore, they are called ``twisted operators''). Examples of the
correspondence in orbifold spaces are given in \cite{Oz:1998hr}
and \cite{Gukov:1998kk}.  An important comment is that when the
orbifold breaks supersymmetry, there are always tachyons in the twisted
sector. From  duality one then expects instabilities at strong
coupling on the gauge theory side \cite{klebanov}.

Similarly to orbifolds, one can consider orientifolds. The latter are
obtained by gauging the transformation 
$$
\Omega:\,X^m (z,\bar z) \rightarrow
-X^m (\bar z,z)\,,
$$ 
a combination of world-sheet parity and a space-time
reflection. One can think about orientifolds as mirrors that sit at 
the fixed point $X^m=0$. The closed strings and their mirrors propagate
in the bulk. The strings become unoriented. In contrast to orbifolds, there is 
{\em no} twisted sector in the case of orientifolds. 

\begin{figure}
\begin{center}
\mbox{\kern-0.5cm
\epsfig{file=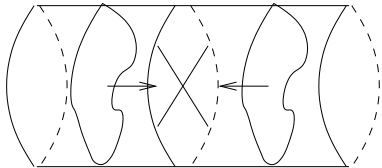,width=5.0true cm,angle=0}}
\end{center}
\caption{A D-brane, a closed string and their mirrors in the presence of an orientifold plane.}
\label{mobius}
\end{figure}

There are two options:  one can place either (i) a D-brane on the orientifold fixed point, 
or (ii) a D-brane and its mirror with respect to the orientifold, see Fig.~\ref{mobius}. 
Usually we have a U$(N)$ gauge factor on a stack of $N$ D-branes ($N^2$
combinations of massless open strings). If the D-branes
sit on top of an orientifold, we will obtain either  SO($N$)  or  Sp($N$)  gauge
groups. The reason is that only $(N^2 \pm N)/2$ combinations of open
strings survive the projection. The second situation in which a closed string
and its mirror are propagating between a D-brane and an orientifold
plane is illustrated in Fig.~\ref{mobius}. The same process
can be described in terms of circulating open strings.

The generalization of the AdS/CFT correspondence to  \Nfour  theories with 
SO($N$)  or  Sp($N$) gauge
groups is obtained by considering $N$ D3-branes on top of an O3-plane. The
resulting space is AdS$_5 \times RP^5$. Since there are no twisted
sectors for orientifolds, the AdS/CFT correspondence dictionary between 
the operators and fields for SO/Sp theories is obtained by selecting $Z_2$ invariant fields and their corresponding operators from the U($N$) dictionary. 
 
\subsection{Type 0 string theory}
\label{tzst}

Type 0A/B string theories are non-supersymmetric closed string theories
with world-sheet supersymmetry. The world-sheet definition of these 
theories is exactly the same as that  of the type II string theories. The difference between the theories is the non-chiral (diagonal) Gliozzi--Scherk--Olive (GSO) projection \cite{GSOp}. This projection omits the space-time fermions 
(the NS-R sector)
but keeps the following bosonic sectors \cite{Klebanov:1998yy}:
\bea
& & {\rm type\,\, 0A:}\,\,\,(NS-,NS-)\oplus (NS+,NS+)\oplus
(R+,R-)\oplus (R-,R+)\,,  \nonumber  \\[3mm]
& & {\rm type\,\, 0B:}\,\,\,(NS-,NS-)\oplus (NS+,NS+)\oplus
(R+,R+)\oplus (R-,R-) \nonumber 
\eea
in the spectrum. 

In particular, there is a {\em tachyon} in the
(NS-,NS-) sector. Because of the doubling (with respect to the type II) of
the R-R sector, there will be two kinds of D-branes. These two kinds
are often called ``electric" and ``magnetic" branes, or fractional
branes. The type-0 string can also be thought of as an orbifold
of the type-II string. In particular, type 0B is obtained by  
orbifolding type IIB by the space-time fermion number $(-1)^{F_s}$.
Then, we can think about   untwisted   and
twisted fields and, accordingly, about untwisted and twisted D-branes.
In the following we will be interested in the brane configurations that
consist  of coincident pairs of $N$ electric and $N$ magnetic
D-branes, or a set of $N$ untwisted branes in the orbifold
language. The untwisted branes are analogous to the type-II branes.

\subsection{Type-0 orientifolds}
\label{tzo}

The type-0B string in ten dimensions admits three kinds of
orientifolds~\cite{Blumenhagen:1999uy}. We will be
interested mostly in Sagnotti's {\em non-tachyonic} 
orientifold~\cite{Sagnotti:1995ga,Sagnotti:1996qj}. The
projection is 
$
\Omega' = \Omega (-1)^{f_R}\,.
$
It acts on the oscillators as follows:
\bea
& & \Omega ' \ \alpha ^\mu _n \ \Omega ' = \tilde \alpha ^\mu _n\,,
\nonumber \\[3mm]
& & \Omega ' \ \Psi  ^\mu _r  \ \Omega '   = \tilde \Psi ^\mu_r \,,
\nonumber  \\[3mm]
& & \Omega ' \ \tilde \Psi ^\mu _r \ \Omega '   = \Psi ^\mu _r\,,
\nonumber \\[3mm]
& & \Omega ' \ \state{0}_{\rm NS-NS} = - \state{0}_{\rm NS-NS}\,. \nonumber 
\eea
In particular, the tachyon (the NS-NS vacuum) is not invariant under
$\Omega'$ and, hence, it is projected out. In addition, $\Omega'$ projects
out one set of the R-R fields. In order to cancel the R-R tadpole one has to
introduce 32 D-branes. The massless open string spectrum on the branes forms
the ten-dimensional  U(32)  gauge theory with an antisymmetric Dirac
fermion. In this rather unusual orientifold, the projection
acts only on the fermionic open strings and, therefore, there are $N^2$ gauge bosons
as in the oriented theory and only $2\times {1\over 2} (N^2-N)$
massless fermions. In a sense, the
closed string part of the resulting theory looks as the bosonic part of the type-I 
string. It is a tachyon-free bosonic string theory with an open string sector.

\subsection{Brane configurations}
\label{bc}

The orientifold field theories (either orienti A or orienti S) 
can be embedded in a brane configuration
of type-0A string theory. Let us briefly review the corresponding construction.

\begin{figure}
\begin{center}
\mbox{\kern-0.5cm
\epsfig{file=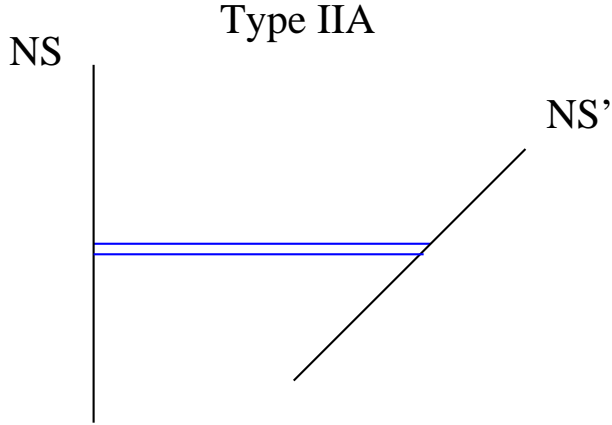,width=8.0true cm,angle=0}}
\end{center}
\caption{The type-IIA brane configuration. The field theory on the
branes is \None SYM theory.}
\label{conf0}
\end{figure}

Let us first describe how \None SYM theory is realized in 
type-IIA string theory \cite{Hori:1997ab,Witten:1997ep,Brandhuber:1997iy}
(see also \cite{Giveon:1998sr}). One can consider a 
brane configuration that consists of
$N$ D4-branes suspended between orthogonal NS5-branes. The
world-volume of the D4-branes is 01234, those of the NS5-branes are 012356
and 0123478. This brane configuration is depicted in Fig.~\ref{conf0}.

\begin{figure}
\begin{center}
\mbox{\kern-0.5cm
\epsfig{file=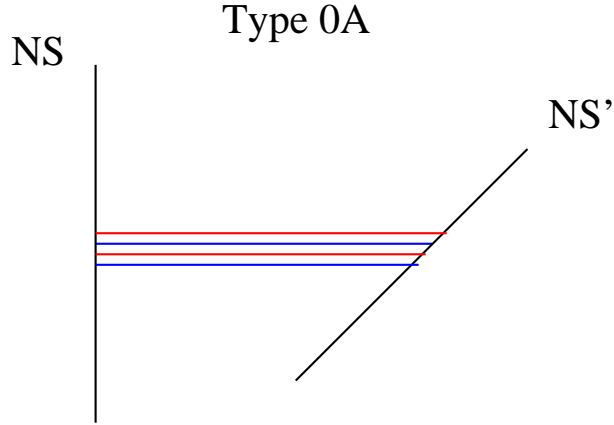,width=8.0true cm,angle=0}}
\end{center}
\caption{A type-0A brane configuration. The field theory on the
branes is a gauge theory with gauge group  U($N$)$\times$U($N$)  and a bifundamental  Dirac fermion (the $Z_2$  orbifold field theory).}
\label{conf1}
\end{figure}

Moreover, one can consider an analogous brane configuration in type-0A
string theory~\cite{Armoni:1999gc}, see Fig.~\ref{conf1}. The 
type-IIA D4-branes have to be replaced by the untwisted D4-branes of type 0A
($N$ electric  plus $N$ magnetic branes).  The field theory on
the brane is a gauge theory with a U($N$)$\times$U($N$) gauge group and 
a bifundamental  fermion (the $Z_2$  orbifold field theory discussed in 
detail in Sect.~\ref{orbifoldization}). 
Note that, since   type-0A string theory contains a closed string tachyon
whose mass is $-2/ \alpha'$, it is not obvious at all that
there is a decoupling limit. Moreover, the tachyon that lives in
the twisted sector couples to the twisted operator ${\rm Tr}\, G_{ee}^2
- {\rm Tr}\, G_{mm} ^2$ on the brane. If we adopt an AdS/CFT
philosophy, then the above-mentioned field theory should suffer from
instabilities \cite{klebanov}.  And it does! These problems were discussed 
in the field-theory part of this review, see Sect.~\ref{npo}.

In order to have a better-behaved model, one can use the orientifold
construction  from the previous section. We can add an O'4-plane on top of the 
D4-branes. The resulting field theory is our orientifold field theory
(A or S, depending on the orientifold charge).

\begin{figure}
\begin{center}
\mbox{\kern-0.5cm
\epsfig{file=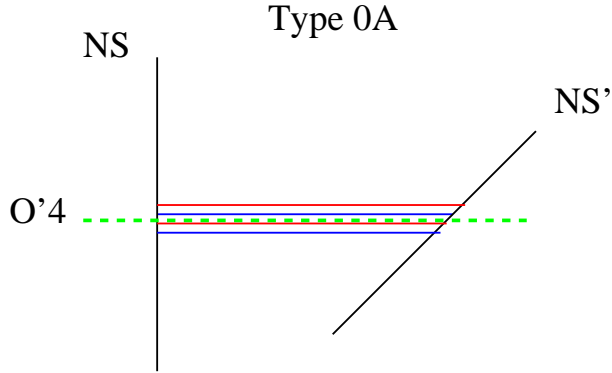,width=8.0true cm,angle=0}}
\end{center}
\caption{A type-0A brane configuration with an orientifold plane. 
The field theory on the
branes is a U$(N)$ gauge theory with a Dirac fermion in
the antisymmetric (symmetric)  two-index representation.}
\label{conf2}
\end{figure}

This is the four-dimensional U$(N)$ gauge theory with a Dirac fermion in
the antisymmetric (symmetric) 
two-index representation. Note that the bulk tachyon does not
couple to the field theory and, therefore, the decoupling limit is no
longer problematic. In addition (and as a consequence),  we do not expect any instabilities in field theory. This is the underlying reason   why 
non-supersymmetric orientifold field theories are ``better'' than
orbifold field theories. 
The brane configuration that realizes the orientifold field theory 
is depicted in Fig.~\ref{conf2}.

Since we discuss a non-supersymmetric brane configuration, a natural
question is the stability issue. In general, non-BPS configurations are
unstable. In the present case the situation is better, however. The untwisted
D4-branes of type-0 string theory behave, at tree level ($g_{\rm st}=0$),   
like the BPS branes of type-IIB string theory. The annulus diagram 
vanishes, implying that there are no forces between the branes. 

From the analysis of the above brane configuration  we learn that the orientifold field
theory is  better than the orbifold field theory due to the absence
of the twisted sector in the former. The brane realization also teaches us
that a useful way of thinking about the orientifold field theory
is as if we had a U($N$)$\times$U($N$) field theory, but with the two gauge
factors identified. We exploited this way of thinking of the orientifold
field theory in our proof \cite{adisvone} of planar equivalence  between \None SYM theory and orientifold theories.

\subsection{AdS/CFT} 
\label{adscft}

In this section we would like to provide a dual-gravity description
of the large-$N$ orientifold field theory. The large-$N$ gauge
theories were conjectured to have a dual-string description.
The prime example is the SU$(N)$ \Nfour SYM theory, which is dual
to the type-IIB string theory on AdS$_5 \times S^5$.

Consider the orientifold
field theory, which is the daughter of  \Nfour SYM theory. The gauge 
group is  SU$(N)$ in both the parent and daughter theories. 
The  orientifold daughter has six scalars in the adjoint representation 
of SU$(N)$ and four Dirac
fermions in the antisymmetric two-index representation. 

This theory lives
on a collection of self-dual (untwisted) D3-branes and an O'3-plane
of type-0B string theory. It is natural, by considering the
near-horizon geometry, to suggest \cite{Angelantonj:1999qg}
that its string dual is type-0B  string theory on AdS$_5 \times RP^5$. 
In the gravity approximation, the solution, as well as the dynamics,
are almost identical to those of the parent supersymmetric theory.
This is because the massless closed string modes of type 0B are
the bosonic modes of type IIB. The effective gravity
action of the non-tachyonic type 0B is also identical to the bosonic
part of the type-IIB action. Hence, the dual string theory
predicts that the large-$N$ orientifold field theory will
coincide with the supersymmetric theory in the bosonic sector.
This is in agreement with field-theory expectations.  

Let us consider now the field-theory analogue of \None SYM theory. 
In this case the problem is more involved, since there is no
known supergravity dual to just \None SYM theory. There is, however,
a supergravity dual to a theory that flows to \None SYM theory in the
IR limit.  This is the Maldacena-Nu\~{n}ez solution
\cite{Maldacena:2000yy}.
 We can consider an analogous solution in non-tachyonic
 type-0 string theory.
 Since the gravity action is identical to type II, it is guaranteed that such
a solution exists also in the non-supersymmetric case. Thus,
all  predictions of supergravity can be copied in the 
large-$N$ orientifold field theory, as long as only
the bosonic sector is considered. In particular, all correlators will be the same and
the bosonic glueball spectra should agree. Note that in type-0 string
theory there are no fermionic closed strings,  in agreement with the
absence of fermionic glueballs in the orientifold field theory. 
Moreover, we will have $N$ degenerate vacua and a gluino condensate. Again, all the above is in full  agreement with the field-theory results.

Can we use the dual string theory
description to calculate the $1/N$ corrections that make the supersymmetric
parent theory and  non-supersymmetric daughter  different?
This is a real challenge. On the string theory
side, we expect that these effects will be due to a dilaton tadpole
term that exists in the  type-0 effective action but does not exist in the 
type-II action \cite{Angelantonj:1999qg}. Understanding
the finite-$N$ theory could lead to predictions for one-flavor QCD
from gravity! Were this feasible, it would represent a major breakthrough.

\subsection{The bifermion condensate from M-theory}
\label{bcfmt}

In this section we discuss the strong-coupling regime of the orientifold field
theory at large-$N$ starting from the relation of type-0A string theory to M-theory.

Similarly to type-IIA string theory, type-0A can be obtained from
M-theory by a compactification of the eleventh dimension on a
Scherk--Schwarz circle \cite{Bergman:1999km}. By using this conjecture
we can lift the brane configuration displayed in Fig.~\ref{conf2} to 
M-theory, analogously to the lift of the analogous type-IIA brane configuration~\cite{Hori:1997ab,Witten:1997ep,Brandhuber:1997iy}.
Note that the presence of the orientifold plane can be neglected in the 
large-$N$ limit, since its R-R charge is negligible with respect to the R-R charge of the branes ($N$). Similarly to the type-IIA situation, we will obtain
a smooth M5-brane and the resulting curve (the shape of the M5-brane)
will be the same as the curve of \None SYM 
theory \cite{Brandhuber:1997iy},
\beq
S^N =1 .
\eeq
The meaning of this curve is that there are $N$ vacua with the
bifermion condensate as the order parameter, in agreement with our 
field-theory results.

\section{Conclusions}
\label{conclu}
\renewcommand{\theequation}{\thesection.\arabic{equation}}
\setcounter{equation}{0}

This review is devoted to a recent development, which will hopefully
produce a strong impact on QCD and QCD-like theories:
planar equivalence between distinct gauge theories at strong coupling,
proved both at the perturbative and at the non-perturbative level.
The ``main" parent gauge theory is \None gluodynamics.
Although it is not completely solved, a rather  extensive understanding of this theory emerged in the 1980s and '90s. The orientifold field theories, being planar-equivalent to \None gluodynamics, inherit its most salient features:
an infinite number of spectral degeneracies 
in the bosonic sector at $N=\infty$,
discrete well-defined vacua labeled by an order parameter, the quark condensate, confinement with a mass gap, etc. The orientifold theory 
A is special. At $N=3$ it  is not just another cousin of QCD; it identically coincides with one-flavor QCD. This gives a hope, for the first time ever,
to obtain a direct relation between 
physical observables in SYM theories and those in one-flavor QCD.
To this end working tools must be found, which will allow us to treat
$1/N$ corrections (say,  the leading post-planar-equivalence corrections).
Finding such tools would amount to a major breakthrough in QCD.
At the moment, progress in this direction is modest.
Performing a rather crude estimate of 
$1/N$ corrections, we were able to evaluate the quark condensate
in one-flavor QCD starting from the known exact expression for
the gluino condensate in \None gluodynamics. Our result is in 
agreement with the existing empiric
evaluations, although the uncertainties on both sides --- theoretical and
experimental --- are rather large.

Our presentation sometimes follows and sometimes deviates
from the historical path. The very notion of the 
planar equivalence between distinct gauge theories
(with varying degrees of supersymmetry)
was born in the depths of string theory. We discuss the genesis of the 
idea in the very beginning of our review, and then essentially leave
string theory to Sect.~\ref{ovoavftsts}. This section collects together, in
concise form, string-theory interpretations of various  results on  the 
planar equivalence discussed in the bulk of the review.
This is done on purpose. We hope that the above interpretations
will turn out to be inspirational in solving the issue of $1/N$ corrections
--- today's problem \# 1 as we see it.

Meanwhile we initiate 
and carry out discussions of other issues, which are conceptually
related to the statement of planar equivalence.
These include ``BPS" domain walls and ``chiral rings"
in orientifold theories (both turn out to be well-defined
in the absence of SUSY!), the vanishing of the vacuum energy density
(cosmological term) at $O(N^2)$, 
parent--daughter pairs with no supersymmetry (proliferation of flavors),
and so on. Although each of the above issues can (and must) be elaborated in more detail  in the future, we feel that outlining novel directions
is as important now as the pursuit of the actual solutions. In this aspect our review is unconventional: besides summarizing results already published in the original literature, we pose and discuss new questions, e.g. in 
Sects.~\ref{v1nc}, \ref{cring}, \ref{ptn2}, and Appendix C.
 
\section*{Acknowledgments}
\addcontentsline{toc}{section}{Acknowledgments}

We are grateful to O. Aharony, C. Angelantonj, J. Barbon, O. Bergman, 
I.~Brunner, K. Chetyrkin, A. Czarnecki, P. H. Damgaard, P.~de~Forcrand, 
P.~Di~Vecchia, Y. Frishman, G. Gabadadze, L. Giusti, A. Gorsky,  A. Hanany, A. Hoang, N. Itzhaki, A. Kataev, D. Kazakov, V. Khoze, I. Kogan, 
W.~Lerche,  A. Loewy, B. Lucini, M. L\"{u}scher,  J. Maldacena, Y. Oz, M.~Peskin, 
J.~Polchinski, R. Rabadan,  E. Rabinovici, P. Ramond, A. Ritz, F. Sannino, A. Schwimmer, Y. Shadmi, A. Smilga, J. Sonnenschein, L. Susskind, M. Teper, D. Tong, and A. Vainshtein, for discussions.  The initial stages of this
work were carried out while M.S. was visiting CERN Theory Division, to
which he is grateful for the kind hospitality. The work  of M.S. was
supported in part by DOE grant DE-FG02-94ER408.

\section*{Appendix A:  SUSY relics in pure Yang-Mills theory}
\addcontentsline{toc}{section}{Appendix A:  SUSY relics in pure Yang-Mills
theory}
\label{appone}
\renewcommand{\theequation}{A.\arabic{equation}}
\setcounter{equation}{0}

In Sect.~\ref{oriln} we suggested a new ``orientifold"
large-$N$ expansion that
connects  one-flavor QCD to \None gluodynamics. The  't Hooft
$1/N$ expansion  connects one-flavor QCD  to 
pure  SU(3)  Yang-Mills theory. 
Therefore, the pure  SU(3) gauge  theory is also
approximated,  in a sense,  by a supersymmetric theory.

Although it is reasonable to  suspect that in this case
the  approximations errors accumulate,  
it still makes sense to ask whether there are any relics of SUSY in
the  SU(3)  Yang-Mills theory.

In fact, it has been known for a long time \cite{Novikov:xj}
that such relics do exist in pure gauge theory, although at that time
they were not interpreted in terms of supersymmetry.
Indeed, it was shown  \cite{Novikov:xj}
that one can use an approximate holomorphic formula\, 
\begin{eqnarray}
\langle {\rm Tr}\, G^2 + i\,{\rm Tr}\tilde G G \rangle &=&
M^4_{\rm UV}\, e^{-\tau / 3}\,,\\[4mm]
\tau &=&\frac{8\pi^2}{g^2} +i\,  \theta\,.
\label{hf}
\end{eqnarray}
(For this equation to be holomorphic in $\tau$,
renormalization-group-invariant and,
in addition, to  have the correct $\theta$ dependence,
we must use $\beta_0=12$ rather than the actual value $\beta_0=11$ in SU(3) 
Yang--Mills theory. The decomposition $11=12-1$ has a deep physical
meaning: $-1$ presents a unitary contribution in the one-loop $\beta$
function, while $12$ is a {\em bona fide} antiscreening, see e.g. \cite{Netal}).
The holomorphic dependence is a relic of supersymmetry.
Equation (\ref{hf}) implies, in particular,
that the topological susceptibility $\chi$
is expressible in terms of the expectation value of
${\rm Tr}\,G^2$:
\begin{eqnarray}
12 \chi & \equiv & -i\int\, d^4x\,\left\langle \frac{\sqrt 3}{16\pi^2}\,
G^a \tilde{G} ^a (x)\,,
\,\, \frac{\sqrt 3}{16\pi^2}\, G ^a \tilde{G} ^a(0)
\right\rangle_{{\rm conn}} \nonumber \\[4mm] 
 & = & \left\langle\frac{1}{8\pi^2 } G^a G^a 
\right\rangle \, .
\label{hf1}
\end{eqnarray}
The numerical value of the left-hand side is known, either from the
WV formula \cite{Witten:1979vv,Veneziano:1979ec}
or from lattice measurements \cite{Alles:1996nm}, to be
$\approx 1.3\times 10^{-2}\,\,{\rm GeV}^4$. The gluon condensate
on the right-hand side is that of  pure Yang--Mills theory and
was estimated to be approximately twice larger  (see
Ref.~\cite{Novikov:xj}, Sect. 14)
than the gluon condensate in actual QCD, $$\left \langle {1\over 4\pi^2} G^a
G^a \right \rangle _{\rm QCD} \approx 0.012 \, {\rm GeV}^4,$$ see
Ref. \cite{SVZ}. If so, the numerical value of the right-hand
side of (\ref{hf1}) is $\approx 1.2\times 10^{-2}\,\,{\rm GeV}^4$. Note that the  phenomenological estimate of the gluon condensate and the factor 2
enhancement mentioned above are valid up to $\sim 30\%$. 

\section*{Appendix B: Relation between \boldmath{$\Lambda_{\rm PV}$} and \boldmath{$\Lambda$'s}
used in perturbative calculations}
\addcontentsline{toc}{section}{Appendix B. Relation between \boldmath{$\Lambda_{\rm PV}$} and \boldmath{$\Lambda$'s}
used in perturbative calculations}
\label{app-lambdas}

\renewcommand{\theequation}{B.\arabic{equation}}
\setcounter{equation}{0}
\renewcommand{\thesubsection}{B.\arabic{subsection}}
\setcounter{subsection}{0}

The standard regularization in performing loop
calculations in perturbation theory in non-Abelian gauge theories is
the dimensional regularization, supplemented by
${\overline{\rm MS}}$ renormalization. 
Determinations of the scale parameter $\Lambda$ in QCD
that can be found in the literature refer to this scheme.
At the same time,
in  calculating nonperturbative effects 
(e.g. instantons) one routinely uses the Pauli-Villars (PV)
regularization scheme. This is because the instanton field is (anti)self-dual, and this notion
cannot be continued to $4-\varepsilon$ dimensions.
The PV regularization/renorma\-lization scheme
in this context was suggested by 't Hooft \cite{th};
it was further advanced in supersymmetric instanton calculus in 
Ref.~\cite{Novikov:ic}; see also the review paper \cite{Shifman:1999mv}.

This previously created no problem, since there were no analytical predictions
for non-perturbative quantities in QCD in terms of $\Lambda$
that could be compared with measured quantities.
With the advent of the orientifold $1/N$ expansion, this becomes a problem.
Indeed, in the parent SUSY theory non-perturbative
quantities are known in terms of $\Lambda_{\rm PV}$.
We then copy them in the large-$N$ limit of the orientifold daughter theory,
extrapolate down to $N=3$, and thus get a prediction in one-flavor QCD
in terms of $\Lambda_{\rm PV}$. In order to calibrate the prediction
one needs to know the relation between $\Lambda_{\rm PV}$
and $\Lambda_{\overline{\rm MS}}$.

In fact, the situation is slightly more 
complicated.
Dimensional regularization {\em per se} cannot be used in SUSY theories,
since it breaks the balance between the number
of the fermionic and bosonic degrees of freedom.
For instance, in SUSY gluodynamics, the standard dimensional regularization would  effectively imply
$d-2$ bosonic degrees of freedom and $2$ fermionic.
The problem is solved by elevating dimensional regularization to
{\em dimensional reduction} (DRED),
with the subsequent application of the $\overline{\rm MS}$ procedure.
The supersymmetry is maintained because even in $d= 4-\varepsilon$
dimensions, the numbers of the  fermionic and bosonic degrees of freedom
match. In the above example of SUSY gluodynamics
it is 2 versus 2.

Our task here is to derive the relation between $\Lambda_{\rm PV}$
and $\Lambda_{\overline{\rm MS}}$. At the one-loop level,
this is a rather simple exercise, a direct generalization
of 't Hooft's argument \cite{th}. An exact relation between two renormalized 
gauge couplings (one in PV, another in DRED +  ${\overline{\rm MS}}$)  at one loop
will allow us to obtain  the ratio $\Lambda_{\rm PV}/\Lambda_{\overline{\rm MS}}$ 
up to $O(\alpha )$ corrections. This is sufficient  for
practical purposes.

\subsection{Brief history}
\label{briefhistory}

Before delving into this simple calculation it would be in order to
say a few words on history and literature, since it is rather confusing.
The first analysis of $\alpha_{\rm PV}$ versus $\alpha_{\overline{\rm MS}}$
at one loop was carried out by 
't Hooft, in connection with instantons; see Sect. XIII of his
pioneering paper \cite{th}. Unfortunately, the key expression
(13.7) contained an error\,.\footnote{Unfortunately, this error propagated in part in some reviews, e.g.
Ref.~\cite{abc}.} The error was noted and corrected by Hasenfratz and Hasenfratz
\cite{Hasenfratz:1981tw}, see also Ref.~\cite{Shore:1978eq}.
What is confusing is the fact that a
later reprint of 't Hooft's paper (see \cite{msth}) reproduces the 
corrected derivation, and the updated
result (13.9), without mentioning that the required  corrections had been 
made.\footnote{ Equation (13.7) of the reprinted article still contains a typo, $-1$ on the right-hand side
should be replaced by $-1/2$. This misprint has no impact on 
the subsequent expressions of the reprinted article.}
A subtle point in the derivation, which in fact led to the above controversy,
is present only in non-supersymmetric theories, such as pure Yang--Mills
theory or QCD. It is absent in supersymmetric theories.
This circumstance will immensely facilitate our analysis.
 
\subsection{Relating $\alpha_{\rm PV}$ and $\alpha_{\overline{\rm MS}}$ at one loop}      
\label{relatingalphas}

We will show that in supersymmetric gauge theories, at one loop,
\beq
\alpha_{\rm PV}^{\rm renorm} = \alpha_{\overline{\rm MS}}^{\rm renorm}\,,
\label{appalpha}
\eeq
where the left-hand side refers to the Pauli-Villars procedure, while the
right-hand side to DRED+${\overline{\rm MS}}$. The important part in this
equality is that it takes place for non-logarithmic terms. Equation
(\ref{appalpha}) implies, in turn, that 
\beq
\Lambda_{\rm PV}= \Lambda_{\overline{\rm MS}}\left( 1+O(\alpha_0 )\right)\,.
\eeq

To keep our derivation as simple as possible, we will limit ourselves to
SQED. Generalization to non-Abelian theories is absolutely straightforward.

The  Lagrangian of SQED is
\bea
{\cal L } &=& \left( \frac{1}{8 e^2_0}\int d^2\theta \, W^2 + \mbox{h.c.}\right) +
\frac{1}{4}\int d^4\theta \left(\bar{S}e^V S + \bar{T}e^{-V} T \right)
\nonumber\\[4mm]
 &+& \left( \frac{m_0}{2}\int d^2\theta \, S\,T +  \mbox{h.c.} \right)\,,
\label{sqed}
\eea 
where $S$ and $T$ are chiral superfields of the electric charge $+1$ 
and $-1$,
respectively. The subscript 0 marks the bare 
coupling constant and mass. 
The Lagrangian (\ref{sqed}) describes photon, electron 
and two  selectrons. We will impose the Wess-Zumino gauge
and will use the background field formalism. (For a detailed introduction 
see the review paper \cite{Novikov:gd}).

The renormalized gauge coupling $1/e^2$ is defined as the coefficient
in front of the operator $-\frac{1}{4}\, G_{\mu\nu}G^{\mu\nu}$ in the effective
Lagrangian, presented as a sum in various Lorentz- and gauge-invariant
operators (which are classified according to the number of derivatives).

The background field formalism assumes that a background field
$(A)_{\rm bck}$ is introduced; the matter fields in the loop
propagate in the given background. For instance, the contribution of one
charged scalar field is given by
\beq
\left\{ {\rm Det} (-D_\mu^2 - m^2)
\right\}^{-1}\,, 
\label{scadet}
\eeq
where $D_\mu$ is the covariant derivative with respect to the
background field,
\beq
D_\mu = \partial_\mu +i  (A_\mu)_{\rm bck}\,.
\label{codi}
\eeq
The mass term $m$ is assumed to be real; at one loop the distinction 
between $m$ and $m_0$ is unimportant.

The contribution of one charged Dirac fermion
can be written as 
\beq
\left\{ {\rm Det}\left[  \left(-D_\mu^2 - m^2 \right)I-\frac{i}{2}\,\sigma^{\mu\nu}(G_{\mu\nu})_{\rm bck}\right]\right\}^{1/2}\,,
\label{ferdet}
\eeq
where $I$ is the $4\times 4$ unit matrix in the space of the 
spinorial Lorentz indices, and $\sigma^{\mu\nu} =\frac{1}{2}\,(\gamma^\mu
\gamma^\nu -\gamma^\nu \gamma^\mu)$.

The second term in the square brackets is the magnetic interaction
of the electron with the background field.
The scalar field (selectron), having no spin, has no magnetic interaction;
in this case the background field is coupled to the electric charge
(through the covariant derivative). 

Let us switch off, for a moment, the magnetic interaction of the electron.
Then the electron contribution to the effective Lagrangian
reduces to 
\beq
\left\{ {\rm Det}  \left(-D_\mu^2 - m^2 \right)I\right\}^{1/2}
= \left\{ {\rm Det}  \left(-D_\mu^2 - m^2 \right)\right\}^{2}\,,
\eeq
and is exactly canceled by the contribution of two selectrons (the square 
of  Eq.~(\ref{scadet})). This cancellation is due to the
balance of the fermion and boson degrees of freedom
inherent in supersymmetry. It is important
that DRED does not destroy this balance --- the cancellation will
hold in $4-\varepsilon$ dimensions too. 

Thus, the charge coupling drops out,\footnote{The charge contribution is the one
that is most labor-consuming. In terms of the instanton
calculation of
the gauge coupling renormalization, the charge contribution is
in one-to-one correspondence with the non-zero modes, while
the magnetic contribution corresponds to zero modes.
For more details see the review paper \cite{abc}.} and it is only the magnetic interaction
of the electron that determines the gauge-coupling renormalization at one loop.
One obtains it by expanding Eq.~(\ref{ferdet}) in $\sigma^{\mu\nu}(G_{\mu\nu})_{\rm bck}$. It is quite obvious that
the linear term vanishes; the first non-vanishing term
is quadratic in $\sigma^{\mu\nu}(G_{\mu\nu})_{\rm bck}$, and
it is proportional to (after the Wick rotation of the integration momentum)
\beq
\left. \mbox{Tr}\left\{  \sigma^{\mu\nu}(G_{\mu\nu})_{\rm bck}\,\,
\sigma^{\alpha\beta}(G_{\alpha\beta})_{\rm bck}\right\} \,\int
\frac{d^4 p}{(2\pi)^4}\,\frac{1}{(p^2+m^2)^2}\,\,\right|_{\rm reg}\,,
\label{befre}
\eeq
where the subscript reg means that the expression must be regularized.
In DRED we replace
\beq
\frac{d^4 p}{(2\pi)^4} \to \frac{d^{4-\varepsilon} p}{(2\pi)^{4-\varepsilon}}\,,
\eeq
while in the Pauli-Villars regularization we replace the integrand
\beq
\frac{1}{(p^2+m^2)^2}\to\frac{1}{(p^2+m^2)^2}- 
\frac{1}{(p^2+M^2_{\rm UV})^2}\,.
\eeq
As a result, expression (\ref{befre}) reduces to
\bea
-\frac{1}{\pi^2}\, (G_{\mu\nu})_{\rm bck}\, (G^{\mu\nu})_{\rm bck}\times
\left\{\begin{array}{l}
\frac{1}{\varepsilon}-\ln m-\frac{\gamma}{2}+\frac{\ln (4\pi )}{2}
\,,\quad\mbox{DRED}\\[4mm]
\ln\frac{M_{\rm UV}}{m}\,,\qquad\qquad\qquad\quad\, \mbox{PV}
\end{array}
\right.
\eea
where $\gamma$ is Euler's constant.

The ${\overline{\rm MS}}$ renormalization prescribes one to
replace $\frac{1}{\varepsilon}-\frac{\gamma}{2}+\frac{\ln (4\pi )}{2}$
by $\ln \mu$, while the PV renormalization replaces 
$M_{\rm UV}$ by the same $\ln \mu$. Thus, in these two schemes,
at one loop, the renormalized gauge couplings are related to the bare coupling in one and the same way,
i.e.
\beq
\alpha_{\overline{\rm MS}}^{-1}- \alpha_{\rm PV}^{-1}= 0(\alpha )
\,,
\label{ohalpha}
\eeq
implying, in turn, that
\beq
\Lambda_{\rm PV} = \Lambda_{\overline{\rm MS}}\,.
\eeq

\section*{Appendix C: Getting two-loop $\beta$ functions for free}
\addcontentsline{toc}{section}{Appendix C: Getting two-loop $\beta$ functions for free}
\label{forfree}
\renewcommand{\theequation}{C.\arabic{equation}}
\setcounter{equation}{0}
\renewcommand{\thesubsection}{C.\arabic{subsection}}
\setcounter{subsection}{0}
 
{\it The above title is a Pickwickian joke --- remember 
``The Posthumous Papers of the Pickwick Club", by Charles Dickens.
The calculation of the two-loop $\beta$ function outlined below
does reduce to a few simple algebraic manipulations, and,
thus, is indeed ``for free", except that the price had been previously paid.
}
\\ \\

In our above considerations we have repeatedly
used the values of the second coefficients in the $\beta$ functions
referring to various theories, which are derivatives of supersymmetric gluodynamics. 
These coefficients can be borrowed from the literature.
They were calculated in the 1970's and 80's \cite{7080}
through a direct
computation of relevant two-loop graphs, exploiting
the technique of dimensional regularization \cite{dimreg} ---
the only regularization scheme  sufficiently developed
to allow us to get practical results in the Yang--Mills theory
beyond one loop. 
Calculation of two-loop diagrams in the
Yang--Mills theory (let alone three loops) is a rather cumbersome
task. It requires some experience, and usually, goes beyond the level accessible to field-theory students.

Here we will show how the observations we made 
in the body of the review allow us
to reduce the calculation of the two-loop coefficients of the $\beta$
functions, in a class of theories related to supersymmetric gluodynamics,
to simple algebraic manipulations. The theories that can be treated in this way
must have  the same $T_{\rm eff}$ as in 
supersymmetric gluodynamics, i.e. $N/2$ (the definition of
$T_{\rm eff}$ is given below). 
Besides practical simplicity, our analysis will allow us to address
another issue, which is very hard (if  at all possible) to study in a direct
computation of   two-loop graphs in
dimensional regularization. The   dimensional-regularization-based
analysis makes no distinction between UV and IR contributions.
In supersymmetric gluodynamics
we know that  the {\em first} coefficient of the
$\beta$ function is  Wilsonian.
The second and all higher coefficients in the $\beta$ function
{\em vanish} in the Wilsonian action; they occur when passing
to the canonic action as a result of an anomaly
(sometimes referred to as the {\em holomorphic
anomaly} \cite{holom-anom}). A detailed discussion 
of this topic can be found in the original publication
\cite{wvnv} and later elaborations \cite{Arkani-Hamed:1997mj}.
In non-supersymmetric Yang--Mills theories,
the problem of what is Wilsonian and what is not
has not been addressed so far, even at the two-loop level, let alone higher
loops.  We turn to this question at the end of this appendix.

\vspace{2mm}

In order to be able to follow the presentation below, the reader must be familiar with the background field method, as well as
instanton calculus. We recommend the
following reviews to study instanton calculus:

\vspace{2mm}

(a) {\em ABC of Instantons}, by V.~A.~Novikov, M.~A.~Shifman,
A.~I.~Vainshtein and V.~I.~Zakharov, 
Sov.\ Phys.\ Usp.\  {\bf 24}, 195 (1982); 
an updated and expanded version can be found
in M. Shifman, {\em ITEP Lectures on Particle Physics and
Field Theory} (World Scientific, Singapore, 1999),
Vol. 1, pp. 201-300; especially Sects. 9.2, 9.3, and 9.4.

(b) {\em Instantons Versus Supersymmetry: Fifteen Years Later,}
by M.~Shifman and A.~Vainshtein,
in M. Shifman, {\em ITEP Lectures on Particle Physics and
Field Theory} (World Scientific, Singapore, 1999)
Vol. 2, pp. 485-647; especially Sect. 4.1
[hep-th/9902018].

\vspace{2mm}

As for the former, the background-field method ascends to Julian Schwin\-ger's ideas. Extensive 
references can be found in the review article \cite{Novikov:gd} as well 
as in a recent publication \cite{Bornsen:2002hh}. 
The former paper adapts various versions of the background-field technique to non-Abelian gauge-field theories. The latter 
presents the state-of-the-art of the background field technology.
Building on  the ideas of Ref.~\cite{pyarnu}, 
the authors perform an economic and elegant calculation of the 
$\beta$ function at {\em three} loops.

\vspace{1mm}

\subsection{The class of theories we will be dealing with}
\label{wtwwdw}

The theories to be discussed below are SU($N$) Yang--Mills theories
with a judiciously chosen fermion sector. The composition
of the fermion sector must be such that
\beq
T_{\rm eff} \equiv \sum_{\rm repr}\, T(R) = \frac{N}{2}\,,
\label{mex7}
\eeq
where the sum runs over all fermions involved in the theory.
For instance, in supersymmetric gluodynamics 
the only fermions are gluinos (the Weyl spinors)
in the adjoint representation of SU($N$) and, hence, the
condition (\ref{mex7}) is satisfied (the factor 1/2 is due to the
``Weyl-ness" of the gluinos).

Another theory in which (\ref{mex7}) is satisfied is
$N_f =N_c\, $ QCD. For brevity, we refer to this theory as
$N^2\,$-QCD. The number of colors is $N$, and so
is the number of the quark flavors (the Dirac spinors in the
fundamental representation of SU($N$)).
This theory is obviously non-supersymmetric, but
$T_{\rm eff} = N\times (1/2)$, where 1/2 comes from each flavor.

Finally, the third example is  Yang--Mills theory with one
Dirac spinor in the antisymmetric two-index representation
(``orienti") plus two fundamental quarks (``2f").
We refer to this theory as orienti/2f-QCD. At $N=3$
it becomes three-flavor QCD. In this theory  
\beq
T_{\rm eff} = \frac{N-2}{2} + 2\times \frac 12 =\frac N2\,,
\label{mex8}
\eeq    
i.e. the effective Casimir coefficient coincides with that
in supersymmetric gluodynamics.

What remains to be added?
The standard definition of the coefficients
of the $\beta$ function  from the Particle Data Group (PDG) 
is\,\footnote{Compared to PDG, we dropped a factor 2
 in the middle equality, $2 \beta (\alpha)_{\rm PDG} 
\to \beta (\alpha)$; this factor of 2 is neither conventional 
nor convenient.} 
\beq
\mu\,\frac{\partial\alpha}{\partial \mu}\equiv \beta (\alpha)
=-\frac{\beta_0}{2\pi } \alpha^2 -\frac{\beta_1}{4\pi^2 } \alpha^3 +...
\label{gmlf22}
\eeq
The values of the first two coefficients borrowed from the literature are displayed in 
Table~\ref{TaT}.

\begin{table}[h]
\begin{center}
\begin{tabular}{|c|c|c|c|}
\hline
$\frac{\mbox{ Theory} \to  }{\mbox{  Coefficients} \downarrow} $ \rule{0cm}{0.7cm} &$N^2$-QCD& Orienti/2f &SYM  \\[3mm]
\hline\hline
\rule{0cm}{0.7cm}$\beta_0$  & $3\, N$ & $3\, N$ & $3\, N$ \\[2mm]
\hline   
\rule{0cm}{0.7cm}$\beta_1$ &$\frac{7N^2 +1}{2}$ & $3\, N^2 +2\, N-\frac{3}{N}$ & 3$N^2$ \\[2mm]
\hline
\end{tabular}
\end{center}
\caption{The first
two coefficients of the $\beta$ function in $N^2$-QCD and   
orienti/2f-QCD (SUSY gluodynamics is shown for comparison).}
\label{TaT}
\end{table}

Equation (\ref{gmlf22}) corresponds to the following effective action:
\beq
{\cal L}_{\rm eff} =
-\frac{1}{4}\,\left\{\frac{1}{g_0^2} - \frac{\beta_0}{8\pi^2}\,\ln\frac{M_{\rm UV}}{\mu}
-\frac{\beta_1}{64\pi^4}\, g_0^2\,\ln\frac{M_{\rm UV}}{\mu} +...
\right\} \, G_{\mu\nu}^a G^{\mu\nu ,\, a}\,.
\label{t-lea}
\eeq

\subsection{\boldmath{$T_{\rm eff}$}}
\label{tef}

In SUSY gluodynamics, the gluino 
contribution is characterized by 
\beq
T(G)_{\rm eff} = \frac{N}{2}\,,
\label{gripone}
\eeq
where $1/2$ reflects the fact that 
gluinos are the Majorana rather than Dirac spinors. 
In the daughter theories we define
\beq
T_{\rm eff} = \sum_R T(R)\,,
\label{defte}
\eeq
where the sum runs over all fermion representations
present in the given theory, and
it is assumed, for simplicity, that the daughter fermions are Dirac. 
Otherwise, we should have introduced a 1/2 factor
for each Weyl fermion in the sum (\ref{defte}).

The class of daughter theories to be considered below is defined
by the condition
$T_{\rm eff} =N/2$. It is quite obvious that 
in any theory belonging to this class the first coefficient
of the $\beta$ function is the same as in 
SUSY gluodynamics, by construction. The second coefficient
is almost the same, but not quite.

\begin{figure}
\epsfxsize=8cm
\centerline{\epsfbox{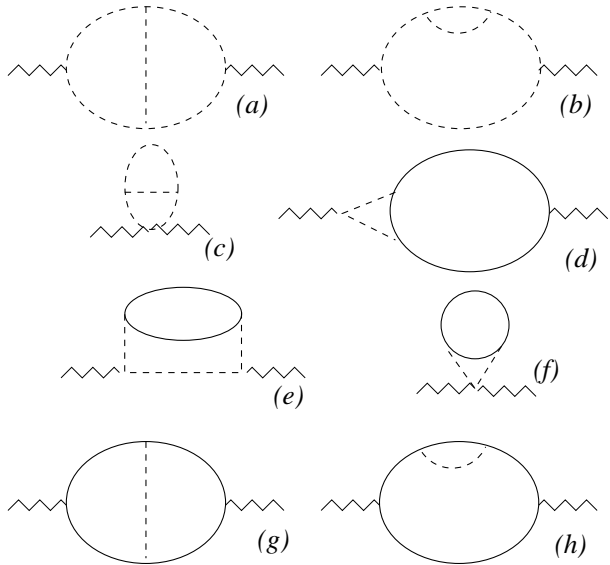}}
\caption{
Two-loop contributions
to the $\beta$ function. The wavy line is the background (soft) gluon field, 
the dashed line stands for the quantum gluon field, the solid line for the
fermion fields. The fermion contribution is represented
by diagrams $d$ through $h$.
}
\label{mstwo}
\end{figure}

Why do we say ``almost"? Because the vast majority of graphs relevant
to the calculation of the $\beta$ function at two loops
are either completely insensitive to fermions (Fig.~\ref{mstwo}a--c)
or, if sensitive,  
depend only on $T_{\rm eff}$ (Fig.~\ref{mstwo}d--f), and once 
$T_{\rm eff}$
is set to be the same as in SUSY gluodynamics, these graphs 
produce the same result. In fact, the only two diagrams that 
differentiate
SUSY gluodynamics on the one hand, and daughter theories with
$T_{\rm eff}=N/2$ on the other, are those depicted in 
Figs.~\ref{mstwo}g and h. Note that these two graphs are the simplest
in the set.

\subsection{What is known about the last two graphs in\\
 Fig.~\ref{mstwo}?}
\label{wika}

\begin{figure}
\epsfxsize=6cm
\centerline{\epsfbox{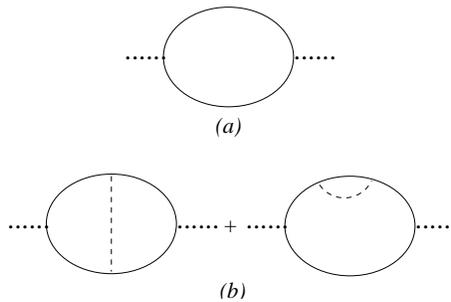}}
\caption{
$\gamma^* \to \bar\psi\,\psi\to \gamma^*$ in one and two loops.
A virtual photon is denoted by the dotted line.
}
\label{msptwo}
\end{figure}

Let us compare these two graphs (Figs.~\ref{mstwo}g  and  h), 
with those determining the inclusive cross section of $e^+e^-$
annihilation in hadrons. More exactly, we will truncate the
$e^+e^-$ vertex, which is irrelevant to our purposes, and will focus
on the part describing the conversion of the virtual photon
into quarks (Fig.~\ref{msptwo}).

These graphs are treated in textbooks (e.g. \cite{Peskin}), 
an easy exercise feasible in more than one way.
The well-known answer is as follows:
the ratio of the diagrams in Fig.~\ref{msptwo}b to that in 
Fig.~\ref{msptwo}a is
\beq
3\, \frac{g^2}{16\pi^2}\,C(R)
\label{alphapi}
\eeq
(for quarks in SU(3) QCD, we have $C(R) =4/3$,
and Eq.~(\ref{alphapi})   is nothing but the famous $\alpha_s/\pi$
correction!). The last two graphs in Fig.~\ref{mstwo} differ from the
last two graphs in  Fig.~\ref{msptwo} only by color factors.

The color factor relevant to Fig.~\ref{mstwo}h is  
\beq
\mbox{Tr}\,\, (T^aT^c T^b T^c)\to 
\left[  C(R)
-\frac{N}{2}\right]
T(R)\,\delta^{ab}\,.
\label{mex88}
\eeq
We can drop the term $-NT(R)/2$, since this term is one and
the same in SUSY gluodynamics and daughter theories
satisfying the condition (\ref{mex7}) ---
remember, we consider only such theories
here.  The color factor relevant to Fig.~\ref{mstwo}g
is 
\vspace{1mm}
\beq
\mbox{Tr}\,\, (T^aT^c T^c T^b )\to 
 C(R)
T(R)\,\delta^{ab}\,.
\label{mexny8}
\eeq

\vspace{1mm}

\noindent
Taking account of the previous remark, we conclude that
the overall color factor in the last two diagrams
of Fig.~\ref{mstwo} is $ C(R)T(R)$, and
the ratio
\vspace{1mm}
\beq
\epsfxsize=2.5cm
\begin{array}{ll}\\[-4mm]
\epsfbox{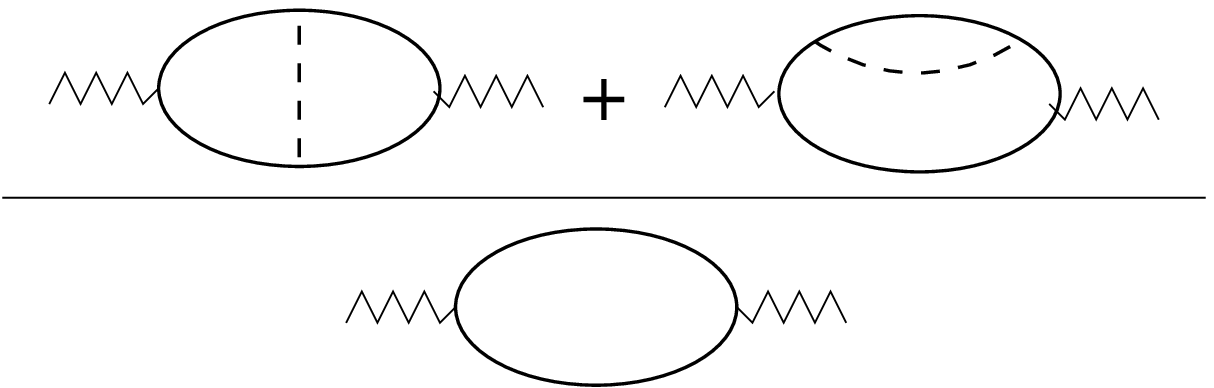}\end{array}
=3\, \frac{g^2}{16\pi^2}\,C(R)\,,
\label{picture-eq}
\eeq

\vspace{1mm}

\noindent
i.e. the same as in Eq.~(\ref{alphapi}).

\vspace{1mm}

To summarize,
the change of the two-loop coefficient in the $\beta$ function
in passing from SUSY gluodynamics to any daughter theory
with $T_{\rm eff} =N/2$ is governed by the combination
\beq
\sum_R C(R)T(R)\,.
\eeq
In SUSY gluodynamics the above combination reduces
to $N^2/2$. If it were equal to $N^2/2$
in the daughter theory under consideration, this would have the same
two-loop coefficient as the SYM theory, namely
$3N^2$. With our judicious choice of the fermion
sector the value of $\sum_R T(R)$ is fine-tuned, Eq.~(\ref{mex7}),
but that of $\sum_R C(R)T(R)$ is not. The change in $\sum_R C(R)T(R)$
is very easy to take into account, however. Combining Eq.~(\ref{t-lea})
with (\ref{picture-eq}) we obtain\,\footnote{For the sign convention, see
Eq.~(\ref{dede}).}
\beq
\Delta \beta_1 =-2\,\Delta (CT)\,.
\label{deltact}
\eeq
In deriving the above formula 
we also used the fact that $$\beta_0 = (11/3)N -(4/3)\sum T(R)\,.$$

\subsection{\boldmath{$N^2\,$}-QCD }
\label{n2qcd}

In QCD with $N$ fundamental
quark flavors,
$T_{\rm eff}$ is  $N/2$, so that this theory belongs
to our class. At the same time
\beq
\sum C_{\rm fund}\, T_{\rm fund} = \frac{N}{2}\,\left(
1-\frac{1}{N^2}\right)\, \frac{N}{2} = -\frac{1}{4} +\frac{N^2}{4}
\label{n2qcdct} 
\eeq
versus $N^2/2$ in SUSY gluodynamics.
Hence, $\Delta (CT) = -(N^2+1)/4$, and
\beq
\Delta \beta_1 = (\beta_1)_{N^2{\rm QCD}} -
(\beta_1)_{{\rm SYM}} = \frac{N^2+1}{2}\,,
\eeq
in full accord with the value
presented in Table~\ref{TaT}.

\subsection{Orienti/2f-QCD}
\label{f2qcd}

The analogue of Eq.~(\ref{n2qcdct}) is
\begin{eqnarray}
\sum_{A,\,\,\rm fund} \,\,   C(R) T(R)
&=&  
\frac{(N-2)^2 \, (N+1)}{2N}+ 
\frac{N}{2}\,\left(1-\frac{1}{N^2}\right)
\nonumber\\[4mm]
&=&
\frac{N^2}{2} - N +\frac{3}{2N}\,,
\label{mex14}
\end{eqnarray}
to be compared with $N^2/2$ in SUSY gluodynamics.
Equation~(\ref{mex14})
implies that $\Delta (CT) = -N + 3 \, (2N)^{-1}$.
Note that the leading term proportional to
$N^2$ cancels in $\Delta (CT)$. This cancellation was certainly 
expected: as we know from the bulk of the review, in the planar limit,
the $\beta$ functions  of SYM and orienti/2f theories
must coincide. The difference emerges only at the $1/N$ level.
From Eq.~(\ref{deltact}) we deduce that
\beq
\Delta \beta_1 = (\beta_1)_{{\rm orienti/2f}} -
(\beta_1)_{{\rm SYM}} = 2N - \frac{3}{N}\,;
\label{ny}
\eeq
cf. Table~\ref{TaT}.

\subsection{Wilsonian $\beta$}
\label{tsp}

To acquaint the reader with the issue,
it is convenient to begin by
outlining how the NSVZ $\beta$ function is obtained in 
SU($N$) SUSY gluodynamics from instanton calculus.\footnote{For a review see 
Refs.~\cite{15,Shifman:1999mv}. {\em Note that here we limit ourselves to
one and two loops.}}
We will rephrase the standard argument
in a slightly different form, which will, hopefully,
make the presentation somewhat simpler.

Let us introduce a small mass term for the gluino field, $m_{\rm adj}$;
then the integration over the fermion moduli can be carried out explicitly.
The instanton measure reduces to a product of an integral over 
the bosonic moduli,
$ \rho^{-5} \,d\rho\, d^4x_0\, d\Omega$, times a prefactor
\beq
M_{\rm UV}^{3N}\,  m_{\rm adj}^N \, \left(\frac {1}{g^2_0}\right)^{2N}
\exp\left(-\frac{8\pi^2}{g^2_0}
\right)\,.
\label{mex1}
\eeq
Here $\rho$ is the instanton radius, $x_0$ its center, while
$\Omega$ parametrizes the instanton color orientation
in SU($N$). Furthermore,
$M_{\rm UV}$ is the ultraviolet (Pauli-Villars) cut-off,
$g^2_0$ and $m_{\rm adj}$ are  the gauge coupling and
the gluino mass, respectively,  normalized at $M_{\rm UV}$.
Supersymmetry endows the pre-factor
(\ref{mex1}) with a crucial  property:
the bosonic and fermionic non-zero modes
completely cancel each other since the instanton
background field conserves two out of four supercharges.
As usual, the zero modes are special.
The fermion zero modes are represented in Eq.~(\ref{mex1}) by
$m_{\rm adj}^N$, while the boson ones by $(g^2_0)^{-2N}$. 
The prefactor (\ref{mex1})
is  renormalization-group invariant (RGI).
This means that a concerted variation of $M_{\rm UV}$
and $g^2_0$ must leave it intact. For what follows
we also need to know that
the renormalization-group running of $m_{\rm adj}$
coincides with that of $g^2$.
Differentiating Eq.~(\ref{mex1}) with respect to $\ln M_{\rm UV}$,
equating the result to zero, and recalling that
${\partial\alpha_0}/{\partial\ln   M_{\rm UV}} = \beta (\alpha_0)$, 
we arrive at the NSVZ $\beta$ function.

It is quite obvious that ignoring the zero modes
(i.e. $m_{\rm adj}^N\, (g^2_0)^{-2N}$) in the instanton prefactor
would result in the vanishing of $\beta_1$ and higher coefficients
of the $\beta$ function. Thus, in the instanton derivation,
the non-vanishing $\beta_1$
in SUSY gluodynamics is a manifestation of the zero-mode effect.
From the perturbative calculation side,  the very same result
emerges as an 
anomaly \cite{holom-anom,wvnv,Arkani-Hamed:1997mj}.

Now, what changes if we pass to 
a daughter theory belonging to the class~(\ref{mex7})?
Let us first examine examples, and then we will build from them a general formula.

\vspace{3mm}

\underline{$N^2$-$QCD$}

\vspace{2mm}

\begin{figure}
\epsfxsize=6cm
\centerline{\epsfbox{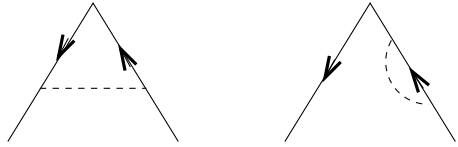}}
\caption{
The one-loop anomalous dimension
of the bifermion operator (or the mass term). The solid line depicts a
fermion, the dashed line a gluon.}
\label{msone}
\end{figure}

\noindent
Again, we introduce a small mass term 
$m_{\rm fund}$, one and the same for all flavors.
The  running law of $m_{\rm fund}$ no more coincides
with that of $g^2$ (which was the case for 
SUSY gluodynamics). However, we can readily find this law
at one loop, which is sufficient for the two-loop $\beta$ function.
The anomalous dimension of the relevant bifermion operator is determined by the diagrams of Fig. \ref{msone}.

Actually, we do {\em not} have to calculate these graphs. 
All we have to do is to  compare the color factors, which 
are determined by $C(R)$ in the corresponding representation.
Namely, $C_{\rm adj}=N$ and $C_{\rm fund}=(N/2)(1-N^{-2})$.
Inspecting the graphs of Fig.~\ref{msone} we readily
conclude that
\beq
m_{\rm fund} \sim \left(g^2\right)^{C_{\rm fund}/N}
\sim\left(g^2\right)^{(1/2)(1-N^{-2})}\,,
\label{mex2}
\eeq
where $\sim$ means ``runs as."
The running law (\ref{mex2}) is valid at one loop.

Let us {\em falsely} assume that Eq.~(\ref{mex2}) is the only modification --- in this way we ignore the non-zero mode contribution
due to the absence of  cancellation, that was
inherent in SUSY gluodynamics, 
in non-supersymmetric daughters.
At one loop the cancellation will still be operative;
the condition $T_{\rm eff} =N/2$ takes care of this.
At the two-loop level the non-zero modes do contribute.
By focusing on the zero-mode-related part
of the $\beta$ function, we will separate the Wilsonian
and anomaly-related parts in $\beta_1$.

The zero-mode-induced instanton prefactor in $N^2$-QCD  is
\begin{eqnarray}
&& 
M_{\rm UV}^{3N}\,  m_{\rm fund}^N \, \left(\frac {1}{g^2_0}\right)^{2N}
\exp\left(-\frac{8\pi^2}{g^2_0}
\right) 
\nonumber\\[4mm]
\sim
&&
M_{\rm UV}^{3N}\,  \left(g^2\right)^{2\,T_{\rm fund}\,
C_{\rm fund}/N}\, \left(\frac {1}{g^2_0}\right)^{2N}
\exp\left(-\frac{8\pi^2}{g^2_0}
\right) ,
\label{mex3}
\end{eqnarray}
where we used the fact that the power of $m_{\rm fund}$
is in fact $2T$ (cf. Eq.~(\ref{mex1})). 
Performing the same procedure as in SUSY gluodynamics (RGI!),
we would get a ``zero-mode-$\beta$" at two loops
\beq
\mbox{zero-mode-}\beta =  -\frac{3N\alpha^2}{2\pi}
\left[ 1+\frac{\alpha}{2\pi}\left(2N-\frac{2}{N}\, CT
\right)
\right]\,.
\label{mex4}
\eeq
This result implies, in turn, that
\beq
(\beta_1)_{\rm anom} = (\beta_1)_{\rm SYM} - 6\Delta (CT)\,,
\label{beta1anom}
\eeq
where 
\beq
\Delta (CT) \equiv (CT)_{\rm fund} - (CT)_{\rm SYM}\,.
\label{opredd}
\eeq

\vspace{3mm}

\underline{$Orienti/2f$}

\vspace{2mm}

\noindent
In addition to $m_{\rm fund}$, we introduce a small mass term
$m_A$ for the two-index antisymmetric Dirac spinor
that is present in the theory at hand. Given the 
corresponding value of $C(R)$
and the anomalous dimension it entails (through the diagrams of
Fig.~\ref{msone}) we arrive at the conclusion
\beq
m_A \sim \left(g^2\right)^{C_{A}/N}\sim 
 \left(g^2\right)^d\,,\qquad d \equiv
\frac{(N-2)(N+1)}{N^2}\,,
\label{mex11}
\eeq
The zero-mode instanton prefactor
takes the form
\begin{eqnarray}
&&
M_{\rm UV}^{3N}\,  m_A^{N-2} m_{\rm fund}^2 \, \left(\frac {1}{g^2_0}\right)^{2N}
\exp\left(-\frac{8\pi^2}{g^2_0}
\right)
\nonumber\\[4mm]
\sim 
&&
M_{\rm UV}^{3N}\,  \left(g^2\right)^{2T_AC_{A}/N}m_{\rm fund}^2 \, \left(\frac {1}{g^2_0}\right)^{2N}
\exp\left(-\frac{8\pi^2}{g^2_0}
\right) .
\label{mex12}
\end{eqnarray}
Exploiting the RGI property of Eq.~(\ref{mex12}) we 
readily obtain the orienti/2f ``zero-mode-$\beta$,"
\beq
\mbox{zero-mode-}\beta =  -\frac{3N\alpha^2}{2\pi}
\left[ 1+\frac{\alpha}{2\pi}\left(N+2 -\frac{3}{N^2}
\right)
\right]\,,
\label{mex13}
\eeq
which leads to exactly the same formula as 
in Eq.~(\ref{mex4}),
\beq
(\beta_1)_{\rm anom} = (\beta_1)_{\rm SYM} - 6\Delta (CT)\,,
\label{beta1anomm}
\eeq
with
\beq
\Delta (CT) \equiv \sum_{A,\,\,\rm fund}
C(R)T(R) - (CT)_{\rm SYM}\,.
\label{opreddd}
\eeq

\vspace{3mm}

\underline{{\em Master formulae}}

\vspace{2mm}

\noindent
Summarizing our considerations, we will make here a general statement
that is
valid for any daughter theory with $T_{\rm eff} =N/2$.
In any theory satisfying the above condition
the total second coefficient of the $\beta$ function
is
\beq
(\beta_1)_{\rm D} = (\beta_1)_{\rm SYM} - 2\Delta (CT)\,,
\label{tscbf}
\eeq
where D stands for daughter, while the Wilsonian part
\beq
(\beta_{1\,\,\rm Wilsonian})_{\rm D} = 4\,\Delta (CT)\,.
\label{wscbf}
\eeq
Here
\beq
\Delta (CT) =\sum_R C(R)T(R)_{\rm D} - CT_{\rm SYM}\,.
\label{dede}
\eeq
In the $N^2$-QCD and orienti/2f models
\beq
\Delta (CT) = \left\{
\begin{array}{ll}
-\frac{N^2}{4}-\frac 14\,,\\[4mm]
-N+\frac{3}{2N}\,,
\end{array}
\right.
\label{lfo}
\eeq
respectively.

\newpage


\end{document}